\documentclass[fleqn,usenatbib]{mnras}

\usepackage{mathptmx}

\usepackage[T1]{fontenc}
\usepackage{ae,aecompl}

\usepackage{graphicx}	
\usepackage{amsmath}	
\usepackage{amssymb}	
\usepackage{upgreek}	
\usepackage{booktabs}
\usepackage{comment}

\newcommand{\civ}{C\,{\sc iv}}
\newcommand{\ha}{H$\alpha$}
\newcommand{\hb}{H$\beta$}
\newcommand{\mgii}{Mg\,{\sc ii}}

\newcommand{\nii}{[N\,{\sc ii}]}
\newcommand{\msol}{M$_{\odot}$}
\newcommand{\mbh}{$M_{\rm BH}$}

\newcommand{\luv}{$L_{1350}$}

\newcommand{\lopt}{$L_{5100}$}
\newcommand{\led}{$L/L_{\rm Edd}$}
\newcommand{\gt}{$t_\mathrm{grow}$}
\newcommand{\rp}{P$_{500\, \rm MHz}$}

\title[Black hole masses of radio-loud quasars in CARLA]{VLT/SINFONI study of black hole growth in high redshift radio-loud quasars from the CARLA survey}

\author[Marinello et al.]{M. Marinello$^{1,2,3},$\thanks{E-mail: murilo.marinello@gmail.com}
R.A. Overzier$^{1,4}$, H.J.A. R\"{o}ttgering$^{2}$, J.D. Kurk$^{5}$, C. De Breuck$^{6}$, \newauthor
J. Vernet$^{6}$, D. Wylezalek$^{6}$, D. Stern$^{7}$, K.J. Duncan$^{2}$, N. Hatch$^{8}$, N. Kashikawa$^{9}$, \newauthor
Y.-T. Lin$^{10}$, R.S. Nemmen$^{4}$, A. Saxena$^{2,11}$
\\
$^{1}$\,Observat\'{o}rio Nacional/MCTIC, Rua General Jos\'e Cristino, 77, S\~ao Crist\'ov\~ao, Rio de Janeiro, RJ 20921-400, Brazil\\
$^{2}$\,Leiden Observatory, University of Leiden, PO Box 9513, 2300 RA Leiden, The Netherlands\\
$^{3}$\,Laborat\'orio Nacional de Astrof\'isica, Rua Estados Unidos 154, Itajub\'a, MG, 37504-364, Brazil\\
$^{4}$\,Institute of Astronomy, Geophysics and Atmospheric Sciences, University of S\~{a}o Paulo, S\~{a}o Paulo, SP 05508-090, Brazil\\
$^{5}$\,Max-Planck-Instit\"{u}t f\"{u}r Extraterrestrische Physik, Giessenbachstrasse, D-85741 Garching, Germany\\
$^{6}$\,European Southern Observatory, Karl-Schwarzschild-Strasse 2, 85748 Garching bei M\"{u}nchen, Germany\\
$^{7}$\,Jet Propulsion Laboratory, California Institute of Technology, Mail Stop 169-221, Pasadena, CA 91109, USA\\
$^{8}$\,School of Physics and Astronomy, University of Nottingham, University Park, Nottingham NG7 2RD, United Kingdom\\
$^{9}$\,Department of Astronomy, School of Science, The University of Tokyo, 7-3-1 Hongo, Bunkyo-ku, Tokyo, 113-0033, Japan\\
$^{10}$\,Institute of Astronomy and Astrophysics, Academia Sinica, Taipei 10617, Taiwan\\
$^{11}$\,INAF-Osservatorio Astronomico di Roma, Via Frascati 33, I-00040 Monteporzio (RM), Italy
}

\date{Accepted XXX. Received YYY; in original form ZZZ}

\pubyear{2019}

\hypersetup{draft}

\begin{document}
\label{firstpage}
\pagerange{\pageref{firstpage}--\pageref{lastpage}}
\maketitle

\begin{abstract}
We present VLT/SINFONI observations of 35 quasars at $2.1<z<3.2$, the majority of which were selected from the Clusters Around Radio-Loud AGN (CARLA) survey. CARLA quasars have large \civ-based black hole masses (\mbh\ $>10^{9}$ \msol) and powerful radio emission (\rp $>$ 27.5 W\,Hz$^{-1}$). We estimate \ha-based \mbh, finding a scatter of 0.35 dex compared to \civ. We evaluate several recipes for correcting \civ-based masses, which reduce the scatter to 0.24 dex. The radio power of the radio-loud quasars is at most weakly correlated with the interconnected quantities \ha-width, \lopt\ and \mbh, suggesting that it is governed by different physical processes. However, we do find a strong inverse correlation between \civ\ blueshift and radio power linked to higher Eddington ratios and \lopt. Under standard assumptions, the BH growth time is longer than the cosmic age for many CARLA quasars, suggesting that they must have experienced more efficient growth in the past. If these BHs were growing from seeds since the epoch of reionization, it is possible that they grew at the Eddington limit like the quasars at $z\sim6-7$, and then continued to grow at the reduced rates observed until $z\sim2-3$. Finally, we study the relation between \mbh\ and environment, finding a weak positive correlation between \mbh\ and galaxy density measured by CARLA.
\end{abstract}



\section{Introduction}
\label{intro}

The Clusters Around Radio-Loud Active Galactic Nuclei (AGN) project (CARLA) was a 400 h \textit{Spitzer} snapshot survey that targeted 419 powerful radio-loud AGN (radio galaxies and quasars) with $L_{\rm 500\ MHz}>27.5$ W Hz$^{-1}$ in the redshift range $1.3<z<3.2$. The selection was done at a rest-frame frequency of 500 MHz estimated by interpolating between the flux densities at 1.4 GHz from the NRAO VLA Sky Survey (NVSS) and 74 MHz from the VLA Low-frequency SKy Survey (VLSS), and ensured that the type 1 and 2 sources have comparable radio luminosities. The CARLA sample was further limited to contain equal fractions of type 1 and type 2 sources. The main goal of CARLA was to systematically investigate the environments of radio galaxies and quasars \citep{galametz12,wylezalek13,wylezalek14}. By selecting galaxies on the basis of relatively red ([3.6] -- [4.5]) colors, the CARLA survey successfully identified excesses of galaxies in the fields around many of the radio-loud AGN. The luminosity function of the galaxies in the overdensities derived by \citet{wylezalek14} showed that the overdense regions have the expected luminosity function of a (proto-)cluster at the redshift of the radio-loud AGN (RLAGN). In a pilot study of 48 radio galaxies, \citet{galametz12} showed that radio galaxies tend to lie, on average, in overdense regions consistent with clusters and proto-clusters of galaxies. In one of the densest fields identified in the CARLA study, \citet{cooke15} found that massive galaxies assembled their stars faster compared to the field. Recently, \citet{noirot16} spectroscopically confirmed two structures around powerful $z=2$ radio galaxies from CARLA. Using \textit{Hubble Space Telescope} (\textit{HST}) slitless spectroscopy they found 8 star-forming members in the field of MRC\,2036--254 and 8 star-forming galaxies (and 2 AGN) in the field of B3\,0756+406. The spectroscopic follow-up of CARLA overdensities has since been expanded to yield 16 galaxy structures at $1.8<z<2.8$ in 20 of the densest CARLA fields \citep{noirot18}. Although a proper interpretation of the CARLA results must await more complete spectroscopic follow-up, the results are consistent with numerous previous studies that found strong evidence for overdense environments and (proto-)clusters associated with powerful radio galaxies at high redshift \citep[see][for a review]{overzier16}.

The local $\mathrm{M_{BH}}-\sigma_*$ relation combined with the galaxy stellar-to-halo mass relation suggests that there is a connection, albeit an indirect one, between the mass of Supermassive Black Holes (SMBHs) and their large-scale environments \citep{ferrarese00,gebhardt00,kormendy13}. 
The CARLA sample offers a unique chance to investigate at what redshifts these relations may have been established. 
For instance, \citet{nesvadba11} analyzing a small sample of HzRGs with very massive BHs found an offset of 0.6\,dex with respect to the $\mathrm{M_{BH}}-\sigma_*$ relation. Moreover, the dense environments found around a significant fraction of the RLAGN targeted by CARLA could be particularly conducive to galaxy merging or halo gas accretion rates that are atypical for ordinary galaxies in the early universe. Although it is difficult to directly measure BH masses for the (type 2) radio galaxies in the CARLA sample, BH masses are available for the (type 1) radio-loud quasars in CARLA. Measurements based on the width of the \civ\ line from the Sloan Digital Sky Survey (SDSS) indicate that CARLA quasars have very massive BHs \citep[log(\mbh/\msol)$>$8.5;][]{hatch14}. It will therefore be interesting to study whether the BH masses correlate with any other properties of the quasar hosts or their large-scale (cluster) environments provided by CARLA. \citet{hatch14} found a mild trend for more massive SMBHs to be located in denser environments. However, for CARLA sources located at high redshift, most of the BH mass estimates in that study were based on the \civ\ line, which is known to be problematic with large scatter and systematic offsets \citep{shen08}. Increasing the accuracy of the BH mass determinations will be a crucial first step if one wants to study any correlations involving the BH mass \citep{uchiyama18}. Therefore, an accurate measurement of the BH mass is necessary before we can properly assess possible correlations between the SMBHs and other properties of their hosts and environments.

Reverberation mapping (RM) is a powerful technique for probing the inner parsec of type-1 quasars \citep[see][for a review]{peterson14}. Mapping the delay time between the variability of the continuum and the emission line response to this variation allows us to estimate fundamental parameters, such as the BH mass and the size of the broad line region (BLR) \citep{kaspi00}. However, RM observations are time-consuming and hence delay times can only be obtained for a relatively small sample of (nearby) sources. Alternatively, BH masses can be estimated from single epoch (SE) spectra under the assumption of virial motions of the BLR gas, 
\mbh~$\propto R_{\rm BLR} v_\mathrm{gas}^2/G$. The RM radius-luminosity ($R-L$) relation shows that the continuum luminosity at 5100 \AA\ (\lopt) can be used as a proxy for the R$_{\rm BLR}$, while the full width at half maximum (FWHM) of a BLR line, usually \hb, is related to the velocity dispersion of the emitting gas \citep{kaspi00}. This technique was extended to other lines, such as \ha\ and \mgii\ in the rest-frame optical and ultraviolet (UV), respectively, and is widely used to determine BH masses from SE quasar spectra for quasars over a wide range of redshifts \citep{shen11}.

\citet{vestergaard06} presented a RM analysis of nearby AGN observed in the UV and \hb, and found an empirical relation that can be used to estimate BH masses based on \civ. With the increase of RM data this relation was later updated using a larger sample of 25 nearby AGN \citep{park13}. Using a similar approach, \citet{wang09} obtained an empirical relation between \hb\ and \mgii.  The resulting \mgii--\mbh\ calibration is considered to be very reliable, and has been applied to large samples of quasars \citep{shen11,marziani13}. Another reliable line that is frequently used is \ha. \citet{greene05} showed that the FWHM and luminosity of \ha\ are good analogs of the FWHM(\hb) and \lopt.

Despite the success of SE BH mass determinations, the technique also has its limitations. Depending on the redshift of the quasars, rest-frame UV or optical lines may not be available. For instance, \ha\ is available in optical spectra only at $z<\,0.4$, while \hb\ can be detected only up to $z\sim0.8$. At $0.8<z<2.2$ the BH mass can be estimated reliably using the \mgii\ line detected in optical spectra, while at $z>2.2$ only the \civ\ line can be used. Furthermore, the scaling relation for SE BH mass determinations relies on the tightness of the $R-L$ relation, and has additional scatter when using lines other than \hb. While results based on \ha, \hb, and \mgii\ are usually consistent with those provided by RM results, and hence provide a good estimate for the BH mass, the \civ\ line is affected by several non-gravitational broadening effects making this line the least reliable of all \citep{shen08}. Therefore, in order to obtain more accurate BH masses, NIR observations are required.
For instance, $K$-band spectroscopy is able to probe \ha\ at $1.9<z<2.7$, \hb\ at $4.0<z<5.0$, and \mgii\ at $6.8<z<7.5$.

Several studies have attempted to improve the \civ-based estimates by determining empirical relations between the main observables used in the SE BH mass estimate, e.g., FWHM and continuum luminosity. For instance, \citet{park17} obtained RM for an updated sample of AGN to improve the \civ\ BH mass estimator using FHWM(\civ) and the luminosity at 1350 \AA\ (\luv) as BLR size proxy. \citet{runnoe13} proposed a correction to the \civ\ BH mass estimator using the peak ratio between Si\,{\sc iv}+O\,{\sc iv}] and \civ. \citet{assef11} used a different approach, taking the ratio between \lopt\ and \luv\ as an additional parameter in the SE BH mass determination. In yet another approach, \citet{denney12} constructed a correction factor based on the ratio of the FWHM and $\sigma$ of the line. Recently, \citet{coatman17} used NIR spectra for a large sample (230 quasars) in order to obtain a reliable relation between the \mbh\ estimated from \ha\ (and \hb) and that obtained from \civ. Despite all these attempts to improve the BH mass estimators based on \civ, the Balmer lines remain the most reliable for BH mass determination \citep[e.g.][]{kaspi00,greene05,vestergaard06}.

In this paper we use observations from the Spectrograph for INtegral Field Observations in the Near Infrared \citep[SINFONI;][]{eisenhauer03} on the Very Large Telescope (VLT) to probe the optical rest-frame spectra of 35 high redshift ($z>2.2$) radio-loud quasars from the CARLA survey. We use \ha\ to estimate the BH masses, and compare with the estimates from \civ\ and the various recipes that have been suggested to correct the \civ-based measurements for non-virial contributions. We use the new \mbh\ to study its relation to several other physical parameters of the CARLA quasars, such as accretion rate, luminosity, radio power, and growth time. Finally, we exploit the fact that we now have a large sample of quasars for which both accurate BH masses and a measurement of the local environment exists from the CARLA project in order to investigate a possible relation between \mbh\ and local environment.

This paper is organized as follows. In Section~2 we describe the selection of the sample targeted with SINFONI, the observations and the data reduction. In Section~3 we present a description of the techniques used for the BH mass determination. In Section~4 we present redshifts and BH masses estimated from \ha, and compare the results with values obtained using several methods from the literature that were designed to correct estimates based on \civ\ for non-gravitational effects. Using the updated BH masses we then estimate the Eddington ratio (\led) and the BH growth time. In Section~5 we study the correlations between the \mbh\ and the luminosity, FWHM(\ha), radio power, \led, and growth time. Finally, we use the accurate measurements of \mbh\ based on \ha\ to study the relation  between the masses of the BHs and the Mpc-scale environment of their host galaxies using the galaxy surface density measurements from the CARLA survey. We give a summary of the results and concluding remarks in Section~6. In this paper, we adopt a $\Lambda$CDM cosmology with $H_0$\,=\,70\,km\,s$^{-1}$\,Mpc$^{-1}$, $\Omega_M$\,=\,0.3 and $\Omega_{\Lambda}$\,=\,0.7. All magnitudes and colors are given in the AB photometric system.

\section{Sample and Observations}
\label{data}

\subsection{Sample selection}
\label{sec:sample}

Starting from the original full CARLA sample, a subset of 30 quasars was selected for follow-up observations. The redshift range of all but one of the targets is $z=2.1-2.6$, and the \ha\ line falls within the wavelength range 2.0--2.4 $\upmu$m. One target (SDSS J094113+114532) has a higher redshift of $z=3.19$, for which we observe \hb\ at 2.04 $\upmu$m. In our analysis we assume that the FWHM of this line is roughly equivalent to that of \ha\ measured for the other objects \citep{vestergaard06,coatman17}. This subset covers the full range of parameter space of the CARLA sample (i.e., UV luminosities of 
$45.5 < \mathrm{log}(L_{\mathrm UV} / \mathrm{[erg~s^{-1}]}) < 47.5$, \mbh$_\mathrm{,CIV} = 10^{8.5-10.5}$ \msol, and radio power of $27.6 < \mathrm{log} (\mathrm{P_{500~MHz}} / \mathrm{[W~Hz^{-1}]}) < 29.2$). For completeness, we include in parts of our analysis 5 additional quasars that were selected from SDSS purely on the basis of their very high (\civ-based) BH masses, even though they are radio-quiet and thus not part of the CARLA sample. Furthermore, as the type of analysis performed in this paper is typically limited to just the radio-loud quasars in CARLA and not the radio galaxies, the latter  occasionally show (in $\sim$20\% of the cases) broad line region \ha\ lines \citep{nesvadba11}. We thus add to our analysis a small sample of six of such broad line radio galaxies from the sample of \citet{nesvadba11}. Three of these radio galaxies are part of the CARLA sample, while the other three match the selection criteria of the CARLA survey but were not in the sample. Our final sample thus consists of 30 radio-loud quasars, 5 radio-quiet quasars and 6 broad line radio galaxies at $z\sim2-3$.  

\begin{figure}
\includegraphics[width=\columnwidth]{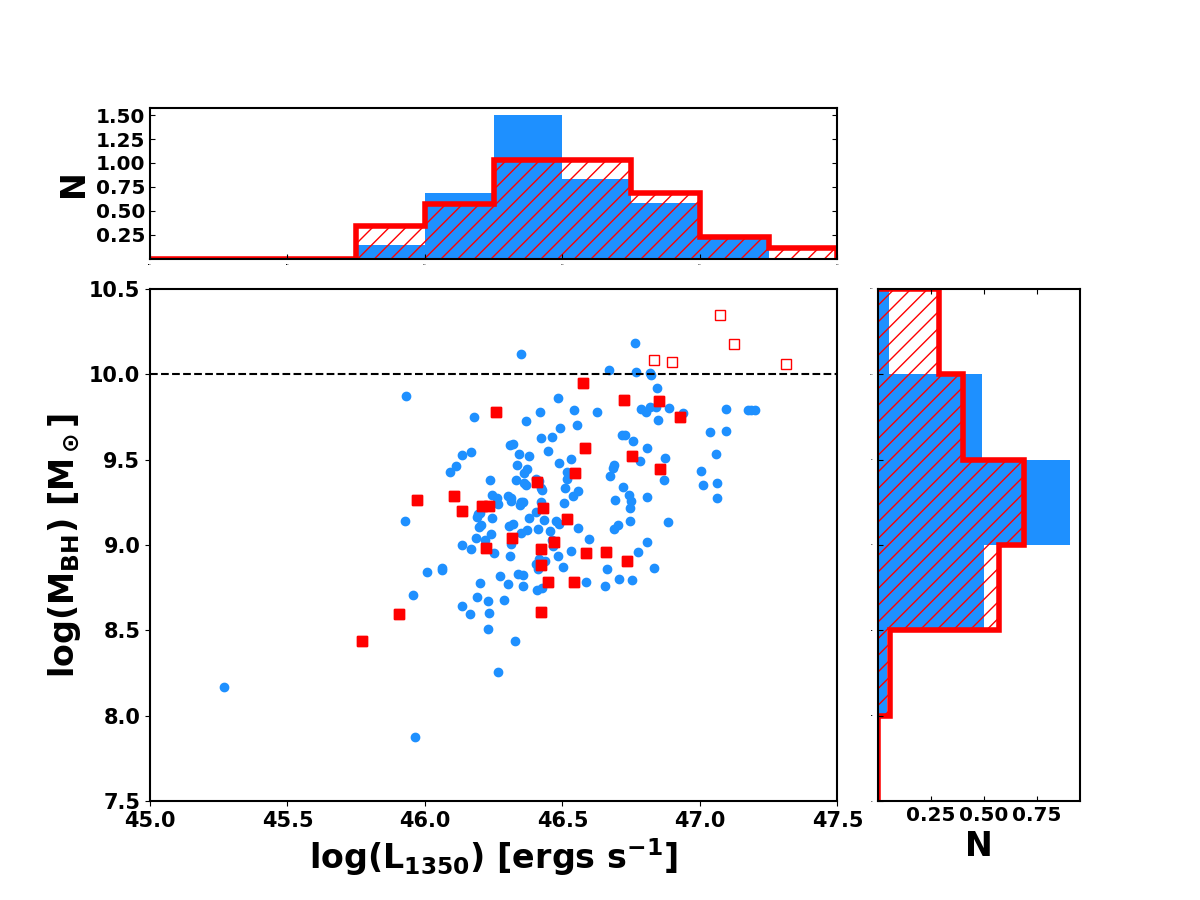}\\
\\
\\
\includegraphics[width=\columnwidth]{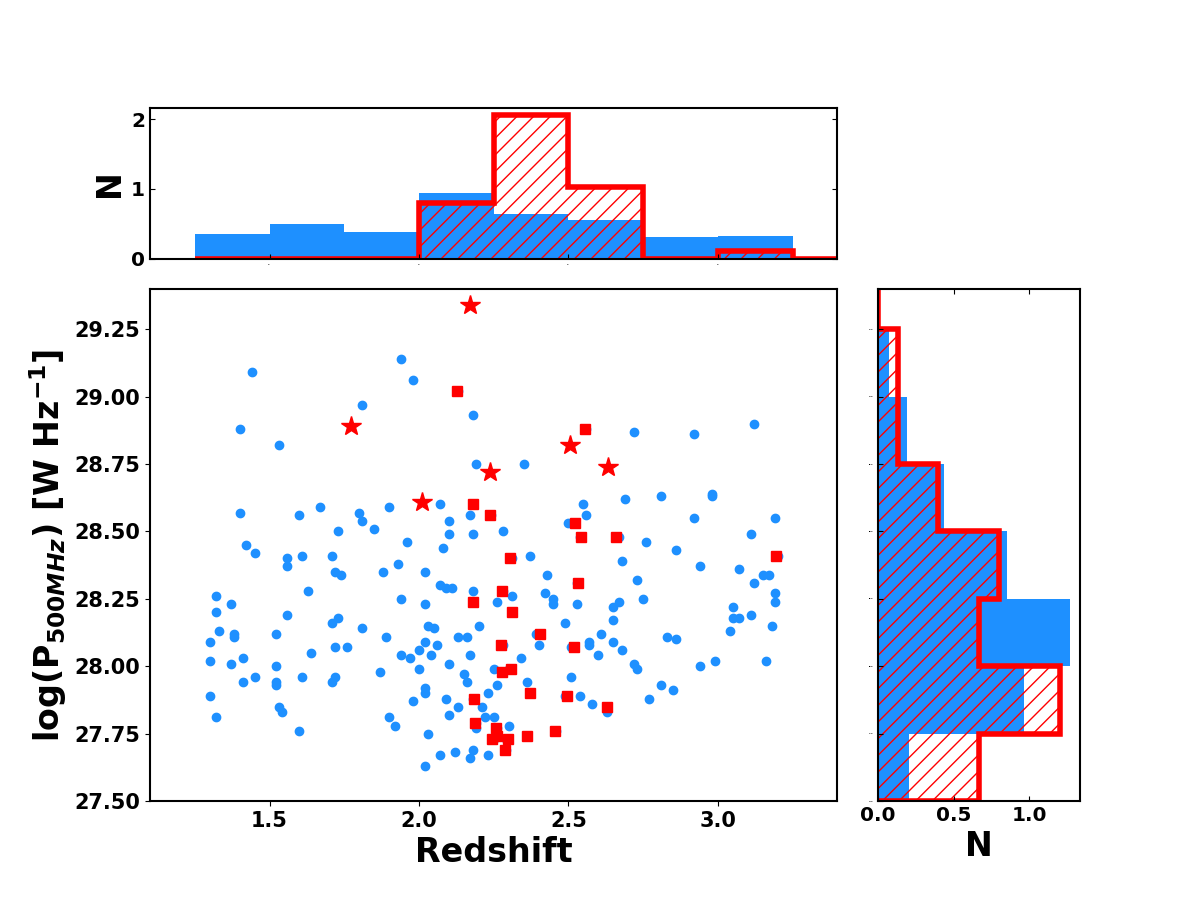}
\caption{Sample parameters distribution showing redshift versus radio power (bottom panel) and UV luminosity versus \civ-based BH mass (top panel). The full CARLA quasar sample is indicated with blue circles. CARLA sources targeted with SINFONI are indicated with filled red squares. The 5 SDSS radio-quiet quasars that are not part of CARLA are indicated with the open red squares. The 6 broad-line radio galaxies from CARLA with BH mass measurements from \citet{nesvadba11} are indicated by the red stars in the bottom panel (\civ-based BH masses are not available for this sample). The right and top histograms show that our SINFONI targets (red hatched histogram) form a fair subsample of the full CARLA sample (blue shaded histogram) in radio power, UV luminosity and BH mass, even though the SINFONI sample covers a more limited range in redshift.}  
\label{fig:sample}
\end{figure}

Figure~\ref{fig:sample} shows the sources selected for this work compared to the full sample of $\sim200$ CARLA quasars in redshift, radio power, UV luminosity and \mbh\  parameter space. The figure shows that the SINFONI subsample, while only $\sim$10\%\ the size of the full CARLA sample, is representative of the full sample in radio power, UV luminosity and BH mass and covers its central redshift range. The radio galaxies from \citet{nesvadba11} are concentrated near the highest radio luminosities and the highest \mbh. 

A list of the targets and relevant observational information in our quasar sample is given in Table~\ref{tab:obs-info}. Table \ref{tab:rgs} summarizes the main parameters determined for the six broad line radio galaxies from \citet{nesvadba11} that were used in our analysis. The available parameters include \ha-based BH mass estimates and Eddington ratios that were determined in a similar fashion to the analysis performed on the CARLA quasars in this paper. 

\subsection{Observations}

We used SINFONI at UT4 of the VLT at the European Southern Observatory (ESO) to observe a sample of 35 high-redshift quasars. The purpose of the project was to obtain NIR spectra probing the rest-frame optical region around \ha\ in order to obtain more accurate BH masses for quasars selected from the CARLA survey and the SDSS. 
The data were taken over 6 years (2009--2015) as part of programs 095.B-0323(A), 094.B-0105(A), 093.B-0084(A), 092.B-0565(A), 091.B-0112(A), 090.B-0674(A), and 089.B-0433(A) (PI: Jaron Kurk). Except for 089.B-0433(A), all programs targeted radio-loud quasars from CARLA. The 089.B-0433(A) targets were selected from SDSS on the basis of their very high (\civ-based) BH masses but are not part of the CARLA sample (see Sect. \ref{sec:sample}). 

We used the SINFONI $K$-band (1.95--2.45 $\upmu$m) with the spatial scale of $125\times250$ mas (without adaptive optics). This setup delivers a spatial aperture of 8\arcsec$\times$8\arcsec\ with a spectral resolution of $\sim$400 km s$^{-1}$, allowing a robust sky subtraction. The observational strategy includes on-source dithering (ABBA pattern) in order to optimize the exposure time on the source and sky subtraction, and a telluric standard star for telluric band removal and flux calibration. We took 14 to 16 individual exposures of 300\,s each, resulting in total integration times of 4200-4800\,s. 

\subsection{Data reduction}

The data reduction was performed using the SINFONI pipeline v2.9.0 \citep[][]{modigliani07} with the \textit{Reflex} interface. The pipeline creates master calibration images and applies them to the observations, starting with a non-linearity map constructed from the flat-field frames. A master dark frame is created from the dark frames and the hot pixels are mapped. Individual flat-field images are combined into a master flat. Optical distortion and slitlet distances are computed, and the dispersion solution for the wavelength calibration is determined from arc spectra. Finally, individual exposures of the quasar and standard star are stacked. We performed an extra sky background removal step by determining a median stack of the pixels surrounding the source and combining them to create a residual sky map. This map was then subtracted from the source spectrum to remove background residuals remaining in the final data cube (these residuals can be seen in the form of spikes in the uncorrected spectrum).

We determined the spatial location of each quasar using a 2-dimensional Gaussian fit, and extracted the spectrum from a 5\,$\sigma$ pixel aperture around this position\footnote{The median change in FWHM(\ha) is about 1\% when using a more conservative aperture of 3\,$\sigma$.}. Because this program was designed to operate in poor weather conditions, no effort was made to obtain spatially resolved information from the data cubes. We modeled and subtracted Paschen lines from the telluric standard spectrum and the removal of telluric absorption was performed by dividing the quasar spectrum by a scaled telluric template derived from the standard star using the task \textit{Telluric} in the IRAF \textit{onedspec} package. Finally, the spectrum was flux calibrated using the observed standard star. For quasars with photometry in the $K$-band available from UKIDSS or 2MASS, we scaled the final spectrum to have the same flux as expected based on the photometry. For the remaining 7 quasars for which no previous $K$-band photometry exists, we estimated the absolute scale of the continuum using a template fitting technique described in Section 3.3. Because we are interested only in the properties of the emission lines (i.e., flux and FWHM), we fitted and subtracted the underlying continuum near \ha. We modeled the continuum using a power law, fitting the regions free of emission lines between rest-frame 6250--6400\AA\ and 6600--6750\AA\ (redshifted using the redshifts from the SDSS catalog). The resulting final spectra around \ha\ are shown in Figs. \ref{fig:ha_fit1}, \ref{fig:ha_fit2} and \ref{fig:ha_fit3} of the Appendix.

\begin{table*}
\begin{scriptsize}
\centering
\caption{Sample Information.}
\label{tab:obs-info}
\begin{tabular}{lrrccccccccccc} 
\hline
\hline
SDSS Name    & R.A. & Dec.    & $z^a$ & $T_{exp}$   & log(P$_{\rm 500\ MHz}$) & $u^\prime$ & $g^\prime$    & $r^\prime$  & $i^\prime$ & $z^\prime$ &  $J$   &  $H$    & $K$   \\
& (J2000)    & (J2000)    &  & (s) & (W\,Hz$^{-1}$)   & (mag) & (mag) & (mag) & (mag) & (mag) & (mag)& (mag) & (mag) \\
\hline
\multicolumn{14}{l}{Radio-loud quasars from CARLA}\\
\hline
J012514$-$001828 &01:25:17.1  &$-$00:18:29 &2.279054 &4200 &28.28 &19.06  &18.31  &18.36  &18.27  &18.04  &16.94  &16.72  &15.91\\
J082707+105224   &08:27:06.4  &10:52:24    &2.278000 &4800 &27.98 &19.75  &18.94  &18.90  &18.71  &18.39  &$-$    &$-$    &$-$  \\
J090444+233354   &09:04:44.3  &23:33:54    &2.258323 &4200 &27.77 &18.06  &17.40  &17.30  &17.00  &16.73  &15.76  &15.25  &14.21\\
J092035+002330   &09:20:35.8  &00:23:31    &2.493972 &4800 &27.89 &19.30  &18.51  &18.48  &18.46  &18.18  &17.22  &16.74  &15.94\\
J094113+114532   &09:41:13.5  &11:45:32    &3.193793 &4800 &28.41 &21.39  &19.41  &19.27  &19.33  &19.35  &18.39  &18.12  &17.41\\
J102429$-$005255 &10:24:29.5  &$-$00:52:55 &2.556504 &4200 &28.88 &19.01  &18.33  &18.28  &18.23  &17.93  &16.92  &16.37  &15.59\\
J104257+074850   &10:42:57.6  &07:48:51    &2.660536 &4800 &28.48 &18.81  &17.76  &17.50  &17.32  &17.17  &18.18  &17.47  &16.61\\
J110344+023209   &11:03:44.5  &02:32:10    &2.517125 &4200 &28.07 &19.35  &18.53  &18.44  &18.33  &18.01  &16.98  &16.37  &15.49\\
J111857+123441   &11:18:57.3  &12:34:42    &2.125651 &4800 &29.02 &18.87  &18.50  &18.49  &18.33  &18.13  &17.59  &16.98  &16.10\\
J112338+052038   &11:23:38.1  &05:20:38    &2.183412 &4800 &27.88 &19.65  &19.20  &19.09  &18.77  &18.50  &17.66  &17.04  &16.09\\
J115901+065619   &11:59:01.7  &06:56:19    &2.186956 &4800 &27.79 &20.49  &19.75  &19.29  &18.93  &18.60  &17.66  &16.95  &15.97\\
J120301+063441   &12:03:01.0  &06:34:42    &2.180930 &4800 &28.24 &21.15  &19.74  &18.91  &18.39  &18.04  &16.99  &16.42  &15.51\\
J121255+245332   &12:12:55.8  &24:53:32    &2.373100 &4800 &27.90 &19.53  &19.10  &19.02  &19.02  &18.84  &$-$    &$-$    &$-$  \\
J121911$-$004345 &12:19:11.2  &$-$00:43:46 &2.296210 &4200 &27.73 &18.27  &17.89  &17.94  &17.98  &17.90  &17.13  &16.68  &16.08\\
J122836+101841   &12:28:36.9  &10:18:42    &2.303086 &4800 &28.40 &19.43  &18.79  &18.72  &18.58  &18.31  &17.24  &16.55  &15.78\\
J133932$-$031706 &13:39:32.6  &$-$03:17:06 &2.311469 &4200 &28.20 &19.19  &18.66  &18.62  &18.52  &18.30  &17.28  &16.69  &15.93\\
J140445$-$013021 &14:04:45.9  &$-$01:30:22 &2.520401 &4200 &28.53 &18.98  &18.29  &18.19  &18.10  &17.91  &16.91  &16.36  &15.45\\
J141906+055501   &14:19:06.8  &05:55:02    &2.287456 &4200 &27.69 &19.78  &19.21  &19.18  &19.12  &18.89  &18.33  &17.93  &16.96\\
J143331+190711   &14:33:31.9  &19:07:12    &2.360114 &4200 &27.74 &19.74  &18.85  &18.84  &18.72  &18.45  &$-$    &$-$    &$-$  \\
J145301+103617   &14:53:01.5  &10:36:17    &2.275276 &4200 &28.08 &19.97  &19.30  &19.29  &19.11  &18.77  &17.89  &17.38  &16.41\\
J151508+213345   &15:15:08.6  &21:33:45    &2.245700 &4800 &27.73 &19.30  &18.48  &18.32  &18.14  &17.85  &16.81  &16.21  &15.28\\
J153124+075431   &15:31:24.1  &07:54:31    &2.455450 &4200 &27.76 &19.82  &19.21  &19.10  &19.06  &19.10  &18.37  &17.77  &17.37\\
J153727+231826   &15:37:27.7  &23:18:26    &2.259551 &4800 &27.74 &19.83  &19.38  &19.02  &18.79  &18.59  &$-$    &$-$    &$-$  \\
J153925+160400   &15:39:25.1  &16:04:00    &2.542400 &4200 &28.48 &20.33  &19.48  &19.23  &19.17  &19.01  &$-$    &$-$    &$-$  \\
J154459+040746   &15:44:59.4  &04:07:46    &2.182000 &4800 &28.60 &18.75  &18.33  &18.26  &18.04  &17.81  &17.17  &16.53  &15.73\\
J160016+183830   &16:00:17.0  &18:38:30    &2.404757 &4200 &28.12 &19.68  &18.85  &18.77  &18.60  &18.28  &$-$    &$-$    &$-$  \\
J160154+135710   &16:01:54.5  &13:57:11    &2.237000 &4800 &28.56 &19.11  &18.49  &18.38  &18.18  &17.96  &17.01  &16.20  &15.64\\
J160212+241010   &16:02:12.6  &24:10:11    &2.530514 &4200 &28.31 &19.66  &18.79  &18.71  &18.58  &18.22  &17.16  &16.54  &15.70\\
J230011$-$102144 &23:00:11.7  &$-$10:21:44 &2.306749 &4200 &27.99 &18.77  &18.35  &18.27  &18.30  &18.16  &$-$    &$-$    &$-$  \\
J231607+010012   &23:16:07.2  &01:00:13    &2.629261 &4200 &27.85 &18.85  &18.32  &18.13  &18.05  &17.93  &17.15  &16.22  &16.03\\
\hline
\multicolumn{14}{l}{Non-CARLA quasars having $\mathrm{M_{BH}}$(\civ) $>10^{10}$ \msol}\\
\hline
J005814+011530   &00:58:14.3  &01:15:30    &2.519759 &4200 & $-$  &18.71  &17.86  &17.66  &17.66  &17.53  &16.64  &16.11  &15.45\\
J081014+204021   &08:10:14.6  &20:40:21    &2.506104 &4200 & $-$  &17.79  &17.31  &17.28  &17.27  &17.11  &16.16  &15.67  &14.86\\
J115301+215117   &11:53:01.6  &21:51:18    &2.367374 &4200 & $-$  &17.21  &16.67  &16.69  &16.63  &16.43  &15.54  &15.03  &14.26\\
J130331+162146   &13:03:31.3  &16:21:47    &2.276900 &4200 & $-$  &18.80  &18.21  &18.05  &18.01  &17.84  &17.05  &16.44  &15.61\\
J210831$-$063022 &21:08:31.5  &$-$06:30:23 &2.348147 &4800 & $-$  &17.90  &17.38  &17.20  &17.16  &17.03  &16.42  &15.77  &15.01\\
\hline
\hline
\multicolumn{14}{l}{$^a$ Redshifts are from SDSS.}\\
\end{tabular}
\end{scriptsize}
\end{table*}

\begin{table*}
\begin{scriptsize}
\centering
\caption{Broad-line radio galaxies from \citet{nesvadba11}.}
\label{tab:rgs}
\begin{tabular}{lrrccccccc} 
\hline
\hline
SDSS Name        & R.A.         & Dec.       & $z^a$     & log($\mathrm{P_{radio}}$) &  Flux (\ha)    & FWHM (\ha)            &  log(\mbh/\msol) (\ha) & \led\ & $t_{growth}/t(z)$ \\
                 & (J2000)      & (J2000)    &           & (W\,Hz$^{-1}$)            &  (10$^{-15}$ erg s$^{-1}$ cm$^{-2}$)      &  (km\,s$^{-1}$)      &        &        &   \\
\hline
MRC0156--252     &   01:58:33.5 & --24:59:32 & 2.01143   & 28.61$^\dagger$           &   17.5   &  12436           &  10.0   & 0.05    & 3.82   \\
MRC1017--220 	 &   10:19:49.0 & --22:19:58 & 1.77356   & 28.89$^\ddagger$          &   5.3    &  12006           &  9.70   & 0.02    & 6.90   \\
TXS1113--178 	 &   11:16:14.5	& --18:06:22 & 2.23851   & 28.72$^\dagger$           &   14.0   &  11063           &  9.95   & 0.05    & 3.62   \\
MRC1138--262 	 &   11:40:48.3 & --26:29:09 & 2.17201   & 29.34$^\dagger$           &   19.0   &  14900           &  10.3   & 0.02    & 9.28   \\
MRC1558--003 	 &   16:01:17.3	& --00:28:46 & 2.50488   & 28.82$^\ddagger$          &   8.3    &  12425           &  10.0   & 0.04    & 5.17   \\
MRC2025--218 	 &   20:27:59.0	& --21:40:57 & 2.63094   & 28.74$^\ddagger$          &   5.5    &  8023.6          &  9.48   & 0.08    & 2.46   \\
\hline
\hline
\multicolumn{10}{l}{$^a$ Redshifts based on \ha\ taken from \citet{nesvadba11}.}\\
\multicolumn{10}{l}{$^\dagger$ Radio luminosity at 325 MHz taken from \citet{debreuck00}.}\\
\multicolumn{10}{l}{$^\ddagger$ Radio luminosity at 500 MHz taken from \citet{wylezalek13}.}\\
\end{tabular}
\end{scriptsize}
\end{table*}

\section{Analysis}\label{analysis}

\subsection{H$\alpha$ Line Fitting}\label{hafit}

\ha\ is a recombination line that can be produced both in the BLR and in the NLR \citep[e.g., see][]{stern13}. Several authors have claimed the existence of multiple broad components of this line \citep[e.g.,][]{sulentic99}, suggesting that there are different emitting regions inside the BLR producing the observed emission. Surrounding \ha\ are the lines of the \nii\ doublet emitted by the NLR. Because forbidden lines in powerful radio-loud AGN can be broadened by shocks to velocities as high as 1500\,km\,s$^{-1}$ \citep{debreuck01}, we include a narrow component with FWHM$_{\rm narrow}<$1500\,km\,s$^{-1}$ and force the broad component to have FHWM$_{\rm broad}>$1500\,km\,s$^{-1}$. Taking into account these possible components, we fit \ha\ by solving for the parameters of a multi-Gaussian fitting function that minimizes $\chi^2$.

\begin{figure*}
\includegraphics[width=\textwidth]{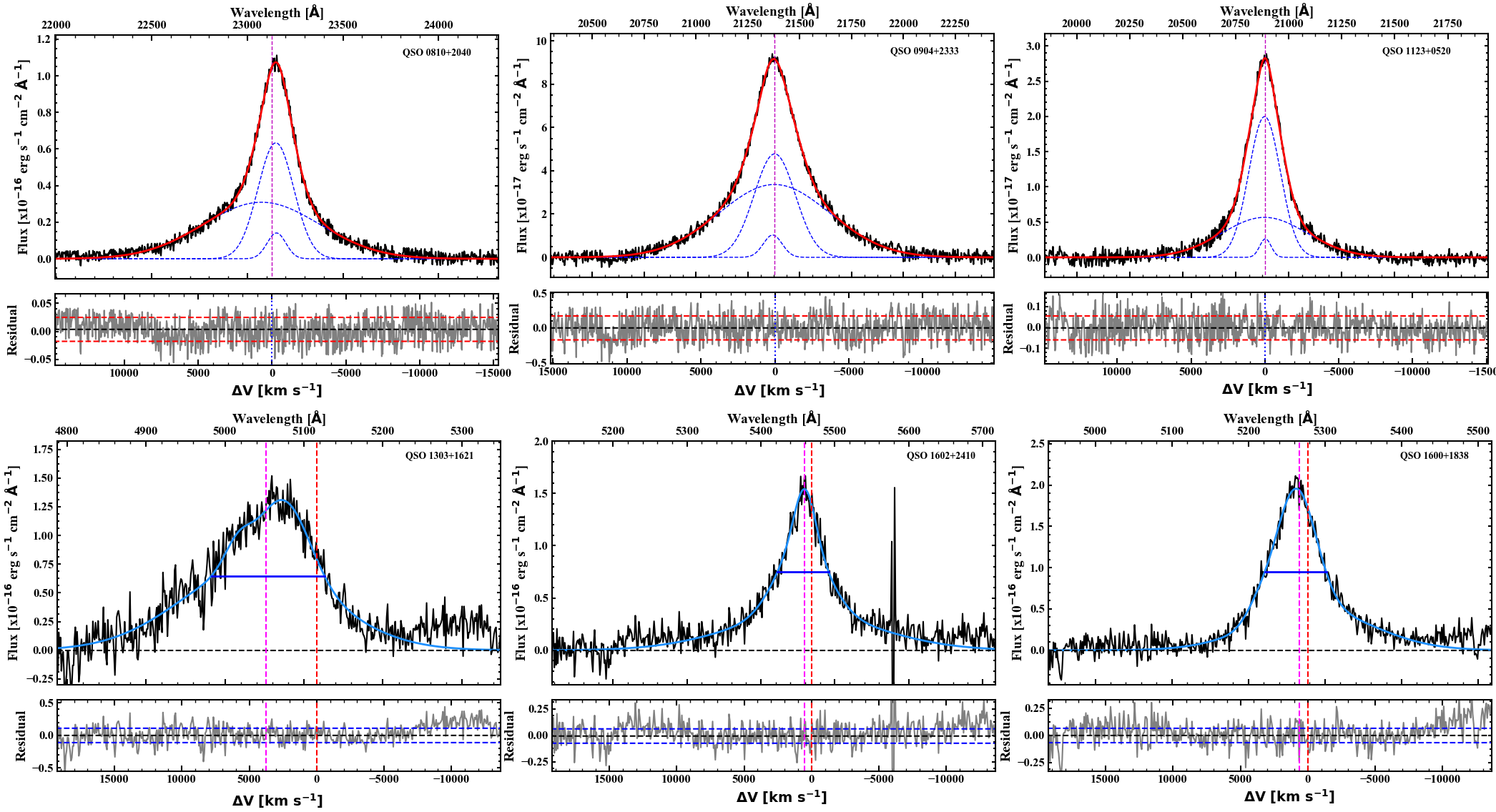}
\caption{Examples of the line fitting routine applied to the \ha\ (top row of panels) and \civ\ line profiles (bottom row of panels). The spectra and the residuals of the subtracted best-fit model are shown in black and gray, respectively. The bold red lines in the top panels show the best-fit, the dashed blue lines show the individual components of the model, and the vertical dashed magenta line shows the centroid of \ha. The bottom panels show examples of \civ\ in the SDSS spectra of the same quasars. The blue lines in the lower panels show the best-fit model, and the red and magenta vertical dashed lines show the central wavelength of \civ$\lambda$1549 as expected based on \ha\ and the actual centroid of the \civ\ profile, respectively. The horizontal blue line shows the FWHM of \civ. Dashed red and blue horizontal lines in the rms panels show the 1-$\sigma$ level for \ha\ and \civ, respectively.}
\label{fig:linefit}
\end{figure*}

We used one Gaussian for each of the lines of the \nii\ doublet, up to two broad Gaussians for the BLR component of \ha, and one Gaussian for the narrow component of \ha. However, in none of our quasars did we detect \nii\ emission with a peak flux larger than 3$\times$ the rms noise measured in the line-free continuum. Since no \nii\ detections were found, we fitted the whole line complex with just one Gaussian for the NLR \ha\ component and (at most) two Gaussians for the BLR \ha\ component. We selected just one broad Gaussian function when one of the two broad components was below 3$\sigma$ of the continuum. Again, if the narrow component found was weaker than the 3$\sigma$ threshold we ignored this component. We found a narrow component in 8 out of the 35 sources in our sample having a relevant contribution to \ha. From the best fit we estimated the FWHM, flux and velocity dispersion (second moment of the line) of the total broad component of \ha, i.e., the summed broad components of the line (hereafter broad component of \ha) \citep{peterson04}. This approach is similar to \citet{shen11}, and reproduces well the observed line profiles of \ha\ in all our quasars. The errors on the parameters were obtained by fitting 10,000 monte carlo realizations of the data. The results for the summed broad components of \ha\ are summarized in Columns~2-5 of Table~\ref{tab:fitting}. An example of the best fit profiles can be seen in the top panels of Figure~\ref{fig:linefit}. The results of these fits are shown in Figs. \ref{fig:ha_fit1}, \ref{fig:ha_fit2} and \ref{fig:ha_fit3} of the Appendix.

In order to obtain accurate measurements of the blueshift of \civ\ (see Section 3.2), we need to estimate accurately the systemic redshift of the quasars. \citet{coatman16} show that redshifts can be reliably obtained from the centroid of \ha\ defined as the intersection point that separates half of the flux of the line. Using the results from the summed broad components of the \ha\ fits we estimated the systemic redshifts for each source (Column~5 in Table~\ref{tab:fitting}).

\subsection{C\,{\sc iv} Line Fitting}
\label{civfit}

In order to measure the properties of the \civ\ line, we first remove the underlying continuum. We fit a power law function to the line free regions at rest-frame 1400--1450\AA\ and 1700--1705\AA, corrected for the Doppler shift using the redshifts estimated from \ha\ in the previous section \citep[similar to][]{shen11}. We found a median value for the power law index of -1.2, which is consistent with the mean value for a composite quasar spectrum from SDSS/DR3 \citep{vandenberk01}. We subtracted the power law from each SDSS spectrum, resulting in a pure emission line spectrum around \civ. Different from \ha, the profile of \civ\ exhibits a much more complex structure. Several physical phenomena such as winds and outflows can cause non-gravitational broadening of this line resulting in an often asymmetrical profile \citep[e.g.,][]{denney12,coatman16}. To model this line we used a similar approach as \citet{shen11}, who considered a line profile composed of three Gaussians. While other authors have used a 6th-order Gauss-Hermite function to model \civ, these approaches are consistent within 10\% \citep{assef11,coatman16}. 

We model \civ\ with three Gaussians which provide a good fit among the variety of the line profiles observed in the sample, including blue asymmetries and absorption features. To fit \civ\ we perform two rounds of fitting. First we apply a 20-pixel boxcar filter to produce a smoothed spectrum. An initial fit is performed on the smoothed spectrum and any regions above and below 3$\sigma$ (i.e., strong noise spikes and absorption features) are masked. This mask was applied to the original spectrum and the final fit was performed. 

\begin{figure}
\includegraphics[width=\columnwidth]{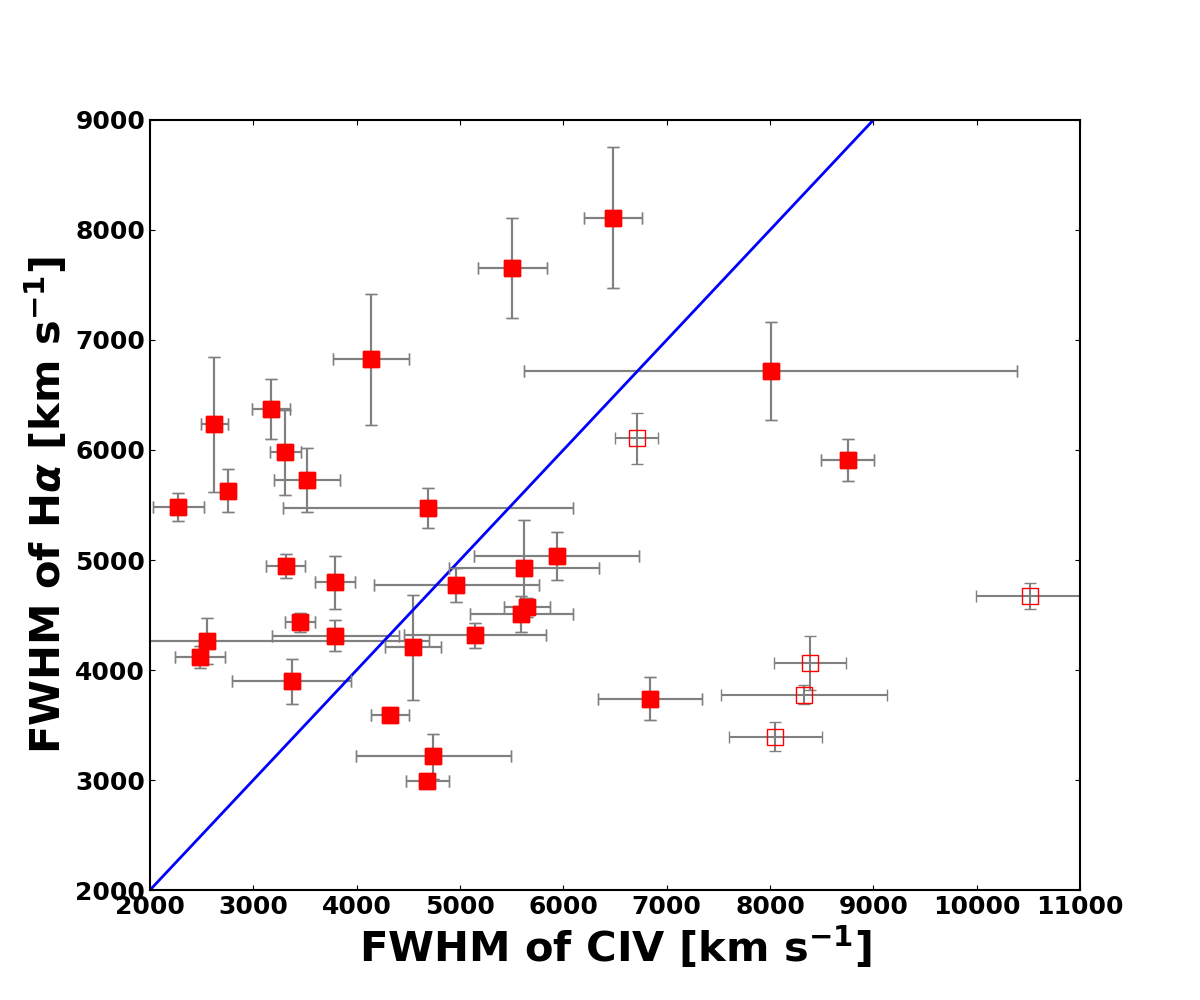}
\caption{FWHM of \civ\ and \ha. The large scatter observed in the plot suggests a poor correlation between the width of these two lines. Filled squares are the CARLA sources in our sample. Open squares show the 5 radio-quiet SDSS quasars in our sample that are not part of the CARLA survey. The blue line shows the unitary line. The large scatter suggests that different broadening mechanisms, other than gravitational broadening, may be acting on \civ.}
\label{fig:fwhm-civ-ha}
\end{figure}

The bottom row of panels of Figure~\ref{fig:linefit} shows examples of the best fit profiles, and all fits are shown in Figs. \ref{fig:civ_fit1}, \ref{fig:civ_fit2} and \ref{fig:civ_fit3} of the Appendix. The flux and FWHM were estimated from the total fit of the profile. The \civ\ blueshift, defined as BS$_{\rm CIV}=(\lambda_c - 1549.48)/c$, relative to the measured line centroid\footnote{The line centroid, $\lambda_{\rm c}$, is the wavelength that separates half of the flux of the line}, $\lambda_c$, was also estimated. A positive value of BS$_{\rm CIV}$ means that the centroid of \civ\ is shifted blueward of the expected rest-frame wavelength of the line as determined from \ha. The best fit model parameters including the blueshift are summarized in Table~\ref{tab:fitting}. Figure~\ref{fig:fwhm-civ-ha} shows a comparison between the FWHM determined from \ha\ and \civ. The large scatter observed between the two values suggests that mechanisms, other than gravitational broadening, likely contribute to the \civ\ line profile.

\subsection{Monochromatic Luminosity}
\label{sedfit}

In this work, our goal is to estimate the BH mass using \ha\ and compare it with the estimates based on \civ\ using the various methods that have been proposed to correct \civ-based measurements for non-gravitational contributions. The last piece of information needed for these BH mass determinations is the rest-frame optical continuum luminosity, which is a proxy for the radius of the line emitting region in the BLR through the $R-L$ relation \citep{kaspi05,bentz06}. The \civ-based BH mass relies on the FWHM of \civ\ and the continuum luminosity at 1350 \AA\ (L$_{1350}$). Deriving \luv\ is straightforward if the spectrum is well flux-calibrated. We calculate \luv\, the monochromatic flux at 1350 \AA, using the power law obtained in the previous section (Column~3 of Table~\ref{tab:fitting2}).

Similarly, the luminosity at 5100 \AA\ (\lopt) is necessary for the BH mass estimate using the \ha\ (or \hb) line. Since our spectra do not cover the region around 5100 \AA\ we used a similar approach as \citet{hewett06}, fitting a model spectral energy distribution (SED) simultaneously to the spectrum and the available photometric data. This is an accurate method for estimating the shape of the SED and successfully reproduces the observed magnitudes of quasars with a precision better than 0.1 mag. \lopt\ is calculated from the best fit SED model. This method is the same as used by \citet{coatman16} to obtain the continuum luminosities for more than 200 high redshift quasars. We constructed a simple parametric quasar SED template consisting of a reddened power-law and a Balmer continuum. We used the reddening law for quasars derived in \citet{zafar15}. The Balmer continuum was forced to have an integrated flux of 10$\%$ of the power-law in the region blueward of the Balmer edge (3646 \AA) \citep[as in][]{hewett06}. Emission lines were added to the template using the values listed for the SDSS composite from \citet{vandenberk01}, which includes all broad and narrow lines and the Fe\,{\sc ii} pseudo-continuum. Each of the parameters were varied, and the resulting templates were used to compute synthetic  magnitudes. The best fit model was found by comparing the modeled $g,r,i,z,J,H,K$ magnitudes to those given by the SDSS \citep[DR7][]{schneider10} and UKIDSS surveys using a $\chi^2$ minimization routine. We excluded the $u$ filter from the fitting process because it is not sufficiently covered by the SDSS spectrum. Finally, the luminosity at 5100 \AA\ was measured from the continuum component of the best fit SED. Overall, this method accurately reproduced the observed fluxes to $\sigma<0.1$ mag per band. However, for a couple of quasars the $H$-magnitude was above this threshold (but still lower than 0.15 mag), and we note that a 0.15 mag error in magnitude translates to a 12$\%$ flux error. 

\begin{figure}
\includegraphics[width=\columnwidth]{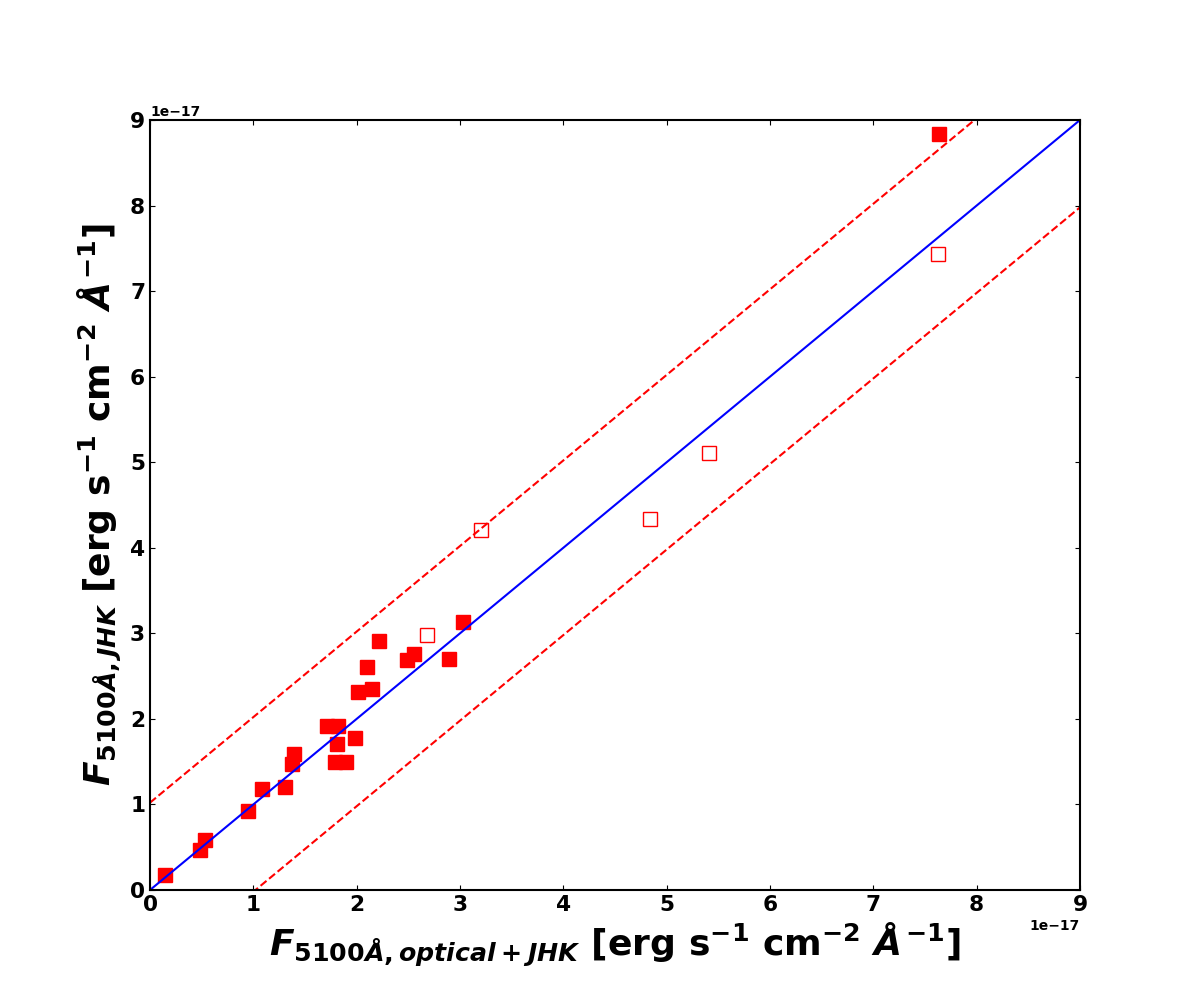}
\caption{Measured flux densities at 5100 \AA\ from the SED fitting performed in Section~3.3. The abscissa shows the values obtained from fits using the full optical+NIR photometry, while the ordinate shows the results without the $J,H,K$ photometry. The values are consistent within a standard deviation of 18$\%$ (red dashed lines). The blue line is the unitary line. CARLA quasars are indicated with filled red squares. The 5 radio-quiet SDSS quasars with the open red squares. 
See the text of Section 3.3 for details.}
\label{fig:fluxcomp}
\end{figure}

For seven quasars, NIR photometry was not available. For these sources we therefore fit the SED template only to the SDSS optical magnitudes. In order to test if these results still reliably predict the luminosity at 5100 \AA, we performed a test by fitting all quasars for which $J,H,K$ data does exist first with and then without the NIR magnitudes. The results are shown in Figure~\ref{fig:fluxcomp}. The difference in the fluxes at 5100 \AA\ determined for these objects using the two methods has a standard deviation of 18$\%$. We used this value as the typical error for the seven sources for which NIR photometry is not available.

\begin{figure}
\includegraphics[width=\columnwidth]{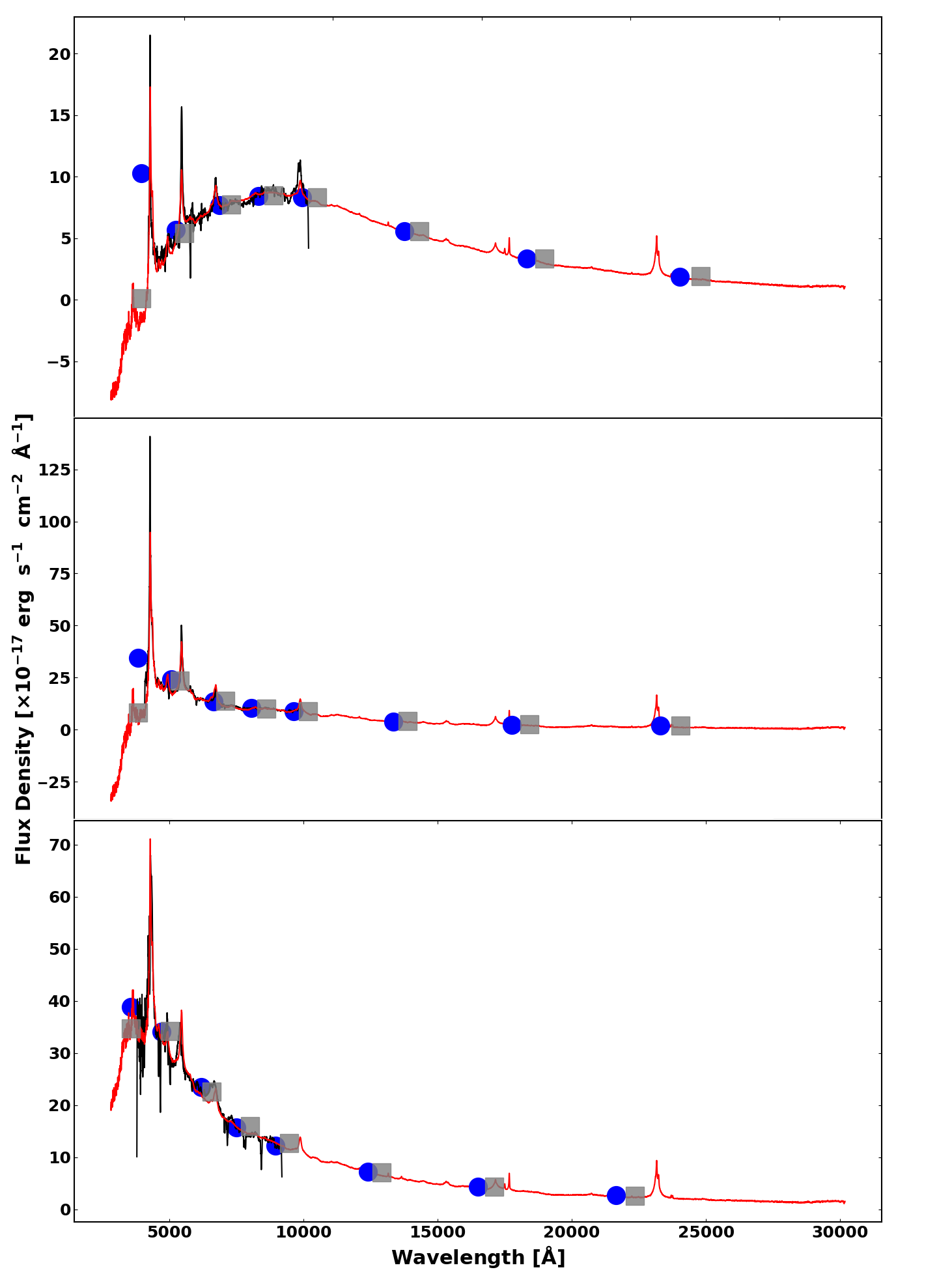}
\caption{Examples of the SED template fit to optical spectra and the observed magnitudes. The blue and green dots, respectively, show the observed and modeled $u,g,r,i,z,J,H,K$ magnitudes. The best-fit template and the observed SDSS spectra are shown in red and black, respectively. See the text of Section 3.3 for details.}
\label{fig:sedfit}
\end{figure}

\begin{figure}
\includegraphics[width=\columnwidth]{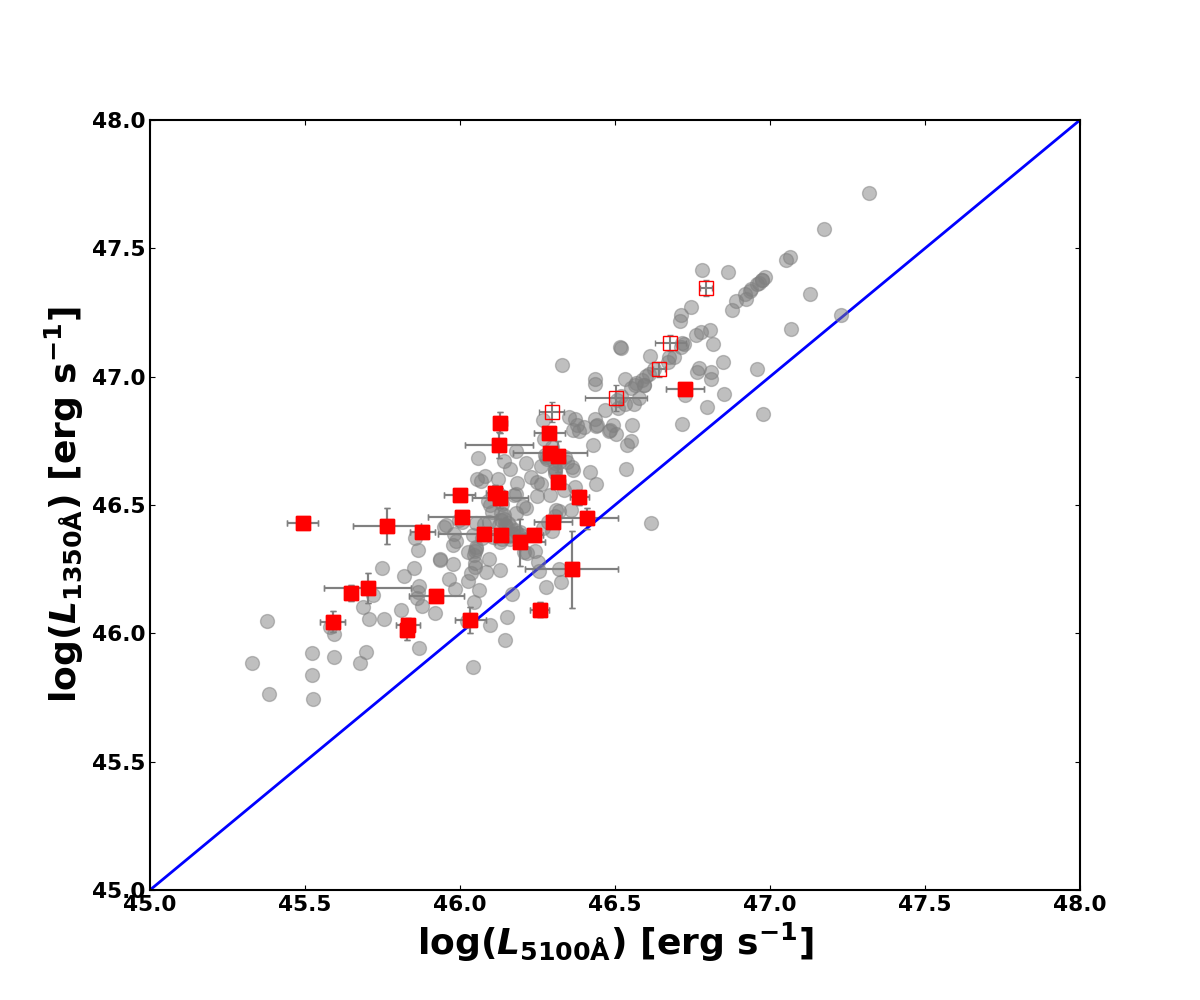}
\caption{Rest-frame optical ($L_{5100}$) versus ultraviolet ($L_{1350}$) luminosities. The red filled squares show the values for \lopt\ and \luv\ for the SINFONI targets studied in this paper that belong to the CARLA sample. Red open squares show the 5 radio-quiet SDSS quasars that are not in CARLA. The blue line shows the unitary line. The sources from \citet{coatman17} are shown in grey circles. See the text of Section 3.3 for details.}
\label{fig:lum-5100-1350}
\end{figure}

Figure~\ref{fig:sedfit} shows some examples of the SED fits obtained. The values found for \lopt\ are given in Column~4 of Table~\ref{tab:fitting2}. \citet{coatman17} presented a compilation of more than 230 quasars observed in the NIR with the goal of rehabilitating the \civ-based method for estimating BH masses. Their sample is composed of quasars at $1.5<z<4$, with \civ-blueshifts up to 5000\,km\,s$^{-1}$, luminosities of $45.5<\mathrm{log}(L_{5100})<48.0$, and $8.5<\mathrm{log}(M_\mathrm{BH})<10.5$. The main difference between the CARLA sample and the \citet{coatman17} sample is that the latter is dominated ($\sim90\%$) by radio-quiet sources, whereas our sample is predominantly radio-loud (86\% of the sources). Comparision between these two  samples could thus provide powerful insight into the link between radio activity in quasars and other observable quantities. We plot the values obtained for the rest-frame optical and UV luminosities in Figure~\ref{fig:lum-5100-1350} together with the luminosities of all quasars in the \citet{coatman17} sample. Our values are consistent with their results and with the slope ($\alpha=1.044$) obtained by \citet{shen11} for the SDSS sample.

\begin{table*}
\begin{scriptsize}
\caption{Results of the \ha\ and \civ\ line fitting.}
\label{tab:fitting}
\begin{tabular}{lrrcccrrcrr} 
\hline
\hline
SDSS Name            & \multicolumn{5}{c}{\ha} & \multicolumn{5}{c}{\civ} \\
\cmidrule(lr){2-6}\cmidrule(lr){7-11}
  & 			Flux$^a$  			& FWHM  		   & $\sigma_{l}$   	& $z$        & $S/N$ & Flux$^a$ 		& FWHM   	   	 	  & $\sigma_{l}$   		   & Blueshift      & $S/N$  \\
  &    					  			& (km\,s$^{-1}$)   & (km\,s$^{-1}$) 	&            &       &          		& (km\,s$^{-1}$)      & (km\,s$^{-1}$)         & (km\,s$^{-1}$) & \\
\hline
\multicolumn{11}{l}{Radio-loud quasars from CARLA}\\
\hline
J012514$-$001828 & 8.24$\pm$0.16		& 4433$\pm$87 	   & 3126$\pm$61	    & 2.279 & 50 	 & 37.08$\pm$1.36   & 3450.8$\pm$142 	  & 4278$\pm$157 		   & -0.39   	    & 27  \\
J082707+105224   & 8.97$\pm$0.29		& 4768$\pm$154 	   & 3046$\pm$98	    & 2.284 & 30 	 & 17.97$\pm$0.69   & 4965.1$\pm$801 	  & 4450$\pm$170 		   & 235.75  	    & 26  \\
J090444+233354   &35.68$\pm$0.66		& 4569$\pm$84 	   & 3151$\pm$58	    & 2.257 & 53 	 & 53.24$\pm$1.69   & 5647.8$\pm$223 	  & 3736$\pm$118 		   & 464.15  	    & 31  \\
J092035+002330   & 1.31$\pm$0.10		& 8113$\pm$638 	   & 3430$\pm$270	    & 2.489 & 12 	 & 24.59$\pm$1.39   & 6484.7$\pm$280 	  & 4476$\pm$253 		   & 189.49  	    & 17  \\
J094113+114532$^\dagger$   & 0.96$\pm$0.09		& 6232$\pm$609 	   & 2633$\pm$257	    & 3.193 & 10 	 & 11.69$\pm$0.53   & 2621.9$\pm$132 	  & 3704$\pm$168 		   & -57.58  	    & 22  \\
J102429$-$005255 & 3.92$\pm$0.13		& 5630$\pm$197 	   & 2377$\pm$83	    & 2.556 & 28 	 & 28.65$\pm$0.95   & 2757.2$\pm$66 	  & 3614$\pm$120 		   & 221.03  	    & 30  \\
J104257+074850   & 3.11$\pm$0.27		& 6823$\pm$596 	   & 2822$\pm$246	    & 2.659 & 11 	 &  9.15$\pm$0.73   & 4136.9$\pm$369 	  & 4241$\pm$339 		   & 147.42  	    & 12  \\
J110344+023209   & 6.04$\pm$0.26		& 5032$\pm$217 	   & 3344$\pm$144	    & 2.512 & 23 	 & 21.81$\pm$1.25   & 5935.3$\pm$800 	  & 4080$\pm$234 		   & -0.53   	    & 17  \\
J111857+123441   &36.77$\pm$2.38		& 5978$\pm$388 	   & 2522$\pm$163	    & 2.126 & 15 	 & 30.74$\pm$0.92   & 3310.8$\pm$154 	  & 4034$\pm$120 		   & 30.23   	    & 33  \\
J112338+052038   & 6.77$\pm$0.12		& 2989$\pm$56 	   & 2061$\pm$38	    & 2.182 & 52 	 & 13.48$\pm$0.69   & 4684.8$\pm$207 	  & 3565$\pm$184 		   & 519.25  	    & 19  \\
J115901+065619   & 9.73$\pm$0.31		& 5906$\pm$192 	   & 2492$\pm$81	    & 2.183 & 30 	 &  7.71$\pm$0.97   & 8750.9$\pm$255 	  & 3643$\pm$461 		   & 2819.44 	    & 8   \\
J120301+063441   &14.86$\pm$0.36		& 4120$\pm$101 	   & 2724$\pm$67	    & 2.179 & 40 	 &  8.47$\pm$0.61   & 2482.4$\pm$242 	  & 3200$\pm$233 		   & -37.95  	    & 14  \\
J121255+245332   & 2.69$\pm$0.23		& 4931$\pm$434 	   & 2081$\pm$183	    & 2.405 & 11 	 & 5.510$\pm$0.94   & 5620.9$\pm$722 	  & 3861$\pm$664 		   & 2164.23 	    & 6   \\
J121911$-$004345 &12.78$\pm$0.23		& 3590$\pm$67 	   & 2258$\pm$42	    & 2.304 & 53 	 & 13.56$\pm$0.84   & 4323.5$\pm$182 	  & 3689$\pm$231 		   & 2064.21 	    & 15  \\
J122836+101841   & 6.23$\pm$0.20		& 4310$\pm$141 	   & 1965$\pm$64	    & 2.302 & 30 	 & 12.12$\pm$0.80   & 3793.1$\pm$614 	  & 3500$\pm$232 		   & -100.02 	    & 15  \\
J133932$-$031706 & 7.62$\pm$0.25		& 5471$\pm$182 	   & 3515$\pm$117	    & 2.310 & 29 	 & 16.59$\pm$1.27   & 4691.1$\pm$1403	  & 4765$\pm$364 		   & -79.39  	    & 13  \\
J140445$-$013021 &22.85$\pm$1.11		& 4262$\pm$208 	   & 3675$\pm$179	    & 2.517 & 20 	 & 29.46$\pm$1.02   & 2552.3$\pm$2151 	  & 4534$\pm$157 		   & 168.99  	    & 29  \\
J141906+055501   & 3.20$\pm$0.19		& 7656$\pm$456 	   & 3790$\pm$225	    & 2.294 & 16 	 &  9.73$\pm$0.84   & 5507.6$\pm$336	  & 4103$\pm$356 		   & 187.76  	    & 12  \\
J143331+190711   & 7.07$\pm$0.35		& 5728$\pm$289 	   & 2416$\pm$122	    & 2.358 & 19 	 & 23.10$\pm$0.76   & 3518.4$\pm$321	  & 3884$\pm$129 		   & -133.65 	    & 30  \\
J145301+103617   & 6.49$\pm$0.23		& 4508$\pm$165 	   & 3386$\pm$124	    & 2.276 & 27 	 & 10.19$\pm$0.90   & 5592.1$\pm$497	  & 3833$\pm$341 		   & -273.64 	    & 11  \\
J151508+213345   &13.08$\pm$0.55		& 6372$\pm$269 	   & 2693$\pm$113	    & 2.248 & 23 	 & 28.56$\pm$0.95   & 3171.5$\pm$180	  & 4812$\pm$160 		   & 294.71  	    & 30  \\
J153124+075431   & 1.13$\pm$0.06		& 3896$\pm$205 	   & 1643$\pm$86	    & 2.473 & 18 	 &  6.46$\pm$0.40   & 3369.8$\pm$571	  & 3633$\pm$229 		   & 1083.40 	    & 16  \\
J153727+231826   & 4.35$\pm$0.28		& 6717$\pm$446 	   & 2838$\pm$188	    & 2.265 & 15 	 &  2.61$\pm$0.88   & 8007.8$\pm$2386 	  & 4237$\pm$1431 		   & 1766.47 	    & 3   \\
J153925+160400   & 1.75$\pm$0.11		& 3215$\pm$203 	   & 1349$\pm$85	    & 2.550 & 15 	 &  5.00$\pm$0.72   & 4743.1$\pm$751 	  & 4429$\pm$641 		   & 1227.27 	    & 7   \\
J154459+040746   &26.55$\pm$0.61		& 5480$\pm$126 	   & 2314$\pm$53	    & 2.185 & 43 	 & 19.57$\pm$0.84   & 2275.8$\pm$249 	  & 3655$\pm$158 		   & 317.50  	    & 23  \\
J160016+183830   &20.94$\pm$2.37		& 4205$\pm$476 	   & 1778$\pm$201	    & 2.405 & 8  	 & 19.95$\pm$0.67   & 4543.5$\pm$270 	  & 3322$\pm$111 		   & 547.80  	    & 30  \\
J160154+135710   &11.03$\pm$0.24		& 4942$\pm$108 	   & 3296$\pm$72	    & 2.237 & 45 	 & 17.34$\pm$0.71   & 3312.5$\pm$186 	  & 3500$\pm$144 		   & -119.23 	    & 24 \\
J160212+241010   &19.79$\pm$0.98		& 4796$\pm$238 	   & 2005$\pm$99	    & 2.529 & 20 	 & 16.06$\pm$0.68   & 3789.9$\pm$194 	  & 4133$\pm$176 		   & 431.34  	    & 23 \\
J230011$-$102144 & 4.05$\pm$0.10		& 4313$\pm$113 	   & 2749$\pm$72	    & 2.318 & 38 	 &  8.92$\pm$1.00   & 5145.6$\pm$683 	  & 3919$\pm$443 		   & 1315.68 	    & 8  \\
J231607+010012   & 5.77$\pm$0.30		& 3739$\pm$194 	   & 2472$\pm$128	    & 2.649 & 19 	 &  9.31$\pm$1.32   & 6840.3$\pm$502 	  & 4499$\pm$640 		   & 2797.38 	    & 7  \\
\hline
\multicolumn{11}{l}{Non-CARLA quasars having $\mathrm{M_{BH}}$(\civ) $>10^{10}$ \msol}\\
\hline
J005814+011530   & 9.08$\pm$0.35		& 3391$\pm$132 	   & 2797$\pm$109	    & 2.528 & 25 	 & 17.50$\pm$1.97   & 8050.9$\pm$452 	  & 4303$\pm$486 		   & 3218.35 	    & 9   \\
J081014+204021   &40.27$\pm$0.90		& 3776$\pm$84 	   & 3222$\pm$72	    & 2.524 & 44 	 & 25.51$\pm$1.93   & 8328.7$\pm$807 	  & 4281$\pm$325 		   & 3245.79 	    & 13  \\
J115301+215117   &20.58$\pm$0.78		& 6108$\pm$231 	   & 2573$\pm$97	    & 2.372 & 26 	 & 76.71$\pm$3.97   & 6708.6$\pm$206 	  & 4700$\pm$243 		   & 2568.54 	    & 19  \\
J130331+162146   & 2.94$\pm$0.17		& 4067$\pm$245 	   & 1703$\pm$102	    & 2.301 & 16 	 & 22.74$\pm$1.91   & 8390.7$\pm$348 	  & 4773$\pm$402 		   & 3640.50 	    & 12  \\
J210831$-$063022 &15.07$\pm$0.37		& 4672$\pm$117 	   & 3003$\pm$75	    & 2.376 & 39 	 & 20.27$\pm$1.46   &10511.0$\pm$519	  & 4584$\pm$331 		   & 5312.81 	    & 14  \\
\hline
\hline
\multicolumn{11}{l}{$^a$ Fluxes are given in units of 10$^{-15}$ erg s$^{-1}$ cm$^{-2}$.}\\
\multicolumn{11}{l}{$^\dagger$ Due to the high redshift of this source, the measurements were made using \hb}
\end{tabular}
\end{scriptsize}
\end{table*}

\begin{table}
\begin{center}
\caption{Peak ratio and UV and optical continuum luminosities.}
\label{tab:fitting2}
\begin{tabular}{lccc} 
\hline
\hline
SDSS Name       & Peak Ratio  & \luv           & \lopt          \\
                &    $\lambda1400$                      & (erg s$^{-1}$) & (erg s$^{-1}$) \\
\hline
\multicolumn{4}{l}{Radio-loud quasars from CARLA}\\
\hline
J012514$-$001828 &  0.173 &  46.539$\pm$0.02  & 46.00$\pm$0.05 \\
J082707+105224   &  0.203 &  46.146$\pm$0.02  & 45.92$\pm$0.09 \\
J090444+233354   &  0.378 &  46.952$\pm$0.02  & 46.73$\pm$0.06 \\
J092035+002330   &  0.184 &  46.547$\pm$0.03  & 46.11$\pm$0.03 \\
J094113+114532   &  0.211 &  46.430$\pm$0.02  & 45.49$\pm$0.05 \\
J102429$-$005255 &  0.161 &  46.702$\pm$0.02  & 46.29$\pm$0.12 \\
J104257+074850   &  0.285 &  46.448$\pm$0.04  & 46.41$\pm$0.10 \\
J110344+023209   &  0.380 &  46.529$\pm$0.03  & 46.38$\pm$0.03 \\
J111857+123441   &  0.172 &  46.396$\pm$0.02  & 45.88$\pm$0.04 \\
J112338+052038   &  0.272 &  46.032$\pm$0.03  & 45.83$\pm$0.04 \\
J115901+065619   &  0.195 &  45.650$\pm$0.05  & 46.03$\pm$0.05 \\
J120301+063441   &  0.129 &  45.693$\pm$0.03  & 46.26$\pm$0.03 \\
J121255+245332   &  0.604 &  46.420$\pm$0.07  & 45.77$\pm$0.11 \\
J121911$-$004345 &  0.500 &  46.822$\pm$0.04  & 46.13$\pm$0.02 \\
J122836+101841   &  0.298 &  46.354$\pm$0.09  & 46.19$\pm$0.08 \\
J133932$-$031706 &  0.175 &  46.526$\pm$0.04  & 46.13$\pm$0.09 \\
J140445$-$013021 &  0.172 &  46.778$\pm$0.02  & 46.29$\pm$0.05 \\
J141906+055501   &  0.203 &  46.050$\pm$0.04  & 45.59$\pm$0.04 \\
J143331+190711   &  0.106 &  46.454$\pm$0.02  & 46.01$\pm$0.11 \\
J145301+103617   &  0.317 &  46.010$\pm$0.04  & 45.83$\pm$0.01 \\
J151508+213345   &  0.202 &  46.590$\pm$0.02  & 46.32$\pm$0.02 \\
J153124+075431   &  0.367 &  46.159$\pm$0.03  & 45.65$\pm$0.01 \\
J153727+231826   &  1.010 &  46.250$\pm$0.15  & 46.36$\pm$0.15 \\
J153925+160400   &  0.318 &  46.180$\pm$0.06  & 45.70$\pm$0.14 \\
J154459+040746   &  0.113 &  46.384$\pm$0.02  & 46.13$\pm$0.04 \\
J160016+183830   &  0.241 &  46.386$\pm$0.02  & 46.08$\pm$0.15 \\
J160154+135710   &  0.219 &  46.384$\pm$0.02  & 46.24$\pm$0.03 \\
J160212+241010   &  0.265 &  46.435$\pm$0.02  & 46.30$\pm$0.06 \\
J230011$-$102144 &  0.561 &  46.730$\pm$0.05  & 46.13$\pm$0.11 \\
J231607+010012   &  0.408 &  46.690$\pm$0.06  & 46.32$\pm$0.02 \\
\hline
\multicolumn{4}{l}{Non-CARLA quasars having $\mathrm{M_{BH}}$(\civ) $>10^{10}$ \msol}\\
\hline
J005814+011530   &  0.837 &  46.920$\pm$0.05  & 46.50$\pm$0.10 \\
J081014+204021   &  0.678 &  47.133$\pm$0.03  & 46.68$\pm$0.05 \\
J115301+215117   &  0.514 &  47.346$\pm$0.03  & 46.79$\pm$0.02 \\
J130331+162146   &  0.594 &  46.870$\pm$0.04  & 46.30$\pm$0.04 \\
J210831$-$063022 &  1.002 &  47.032$\pm$0.03  & 46.65$\pm$0.02 \\
\hline
\hline
\end{tabular}
\end{center}
\end{table}

\section{Results}
\label{results}

\subsection{Black Hole Masses}
\label{blackholes}

Single epoch BH masses can be determined under the assumption of virial motions using two parameters, the velocity of the gas and its distance from the BH. The strong $R-L$ relation between the continuum luminosity at 5100 \AA\ and the BLR radius obtained from RM allows us to use \lopt\ as a measure of the BLR radius, while the FWHM of the Balmer lines serve as a proxy for the gas velocity. Both parameters were obtained in Section 3, and we can thus estimate the BH mass using the FHWM of \ha\ and \lopt. The general equation to estimate the BH mass takes the form of equation~\ref{eq:BHmass} \citep{shen11}:

\begin{equation}
\mathrm{log}(M_\mathrm{BH}/M_{\odot})= a + b\times \mathrm{log}(\frac{FWHM}{\mathrm{10^3\,km\,s^{-1}}}) + c\times \mathrm{log}(\frac{\lambda L_{\lambda}}{\mathrm{10^{42}\,erg\,s^{-1}}}).
\label{eq:BHmass}
\end{equation}

\begin{figure*}
\includegraphics[width=\columnwidth]{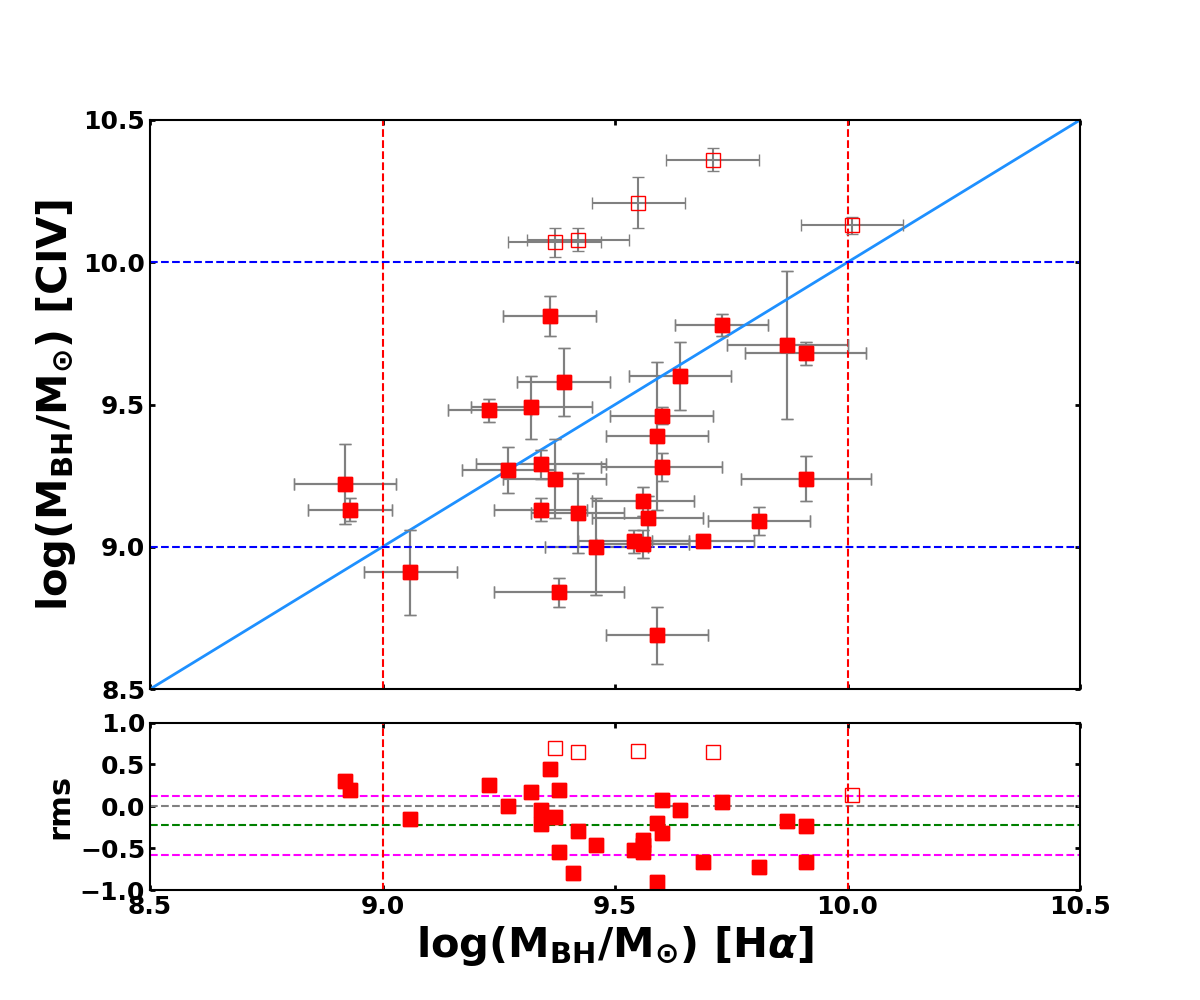}
\includegraphics[width=\columnwidth]{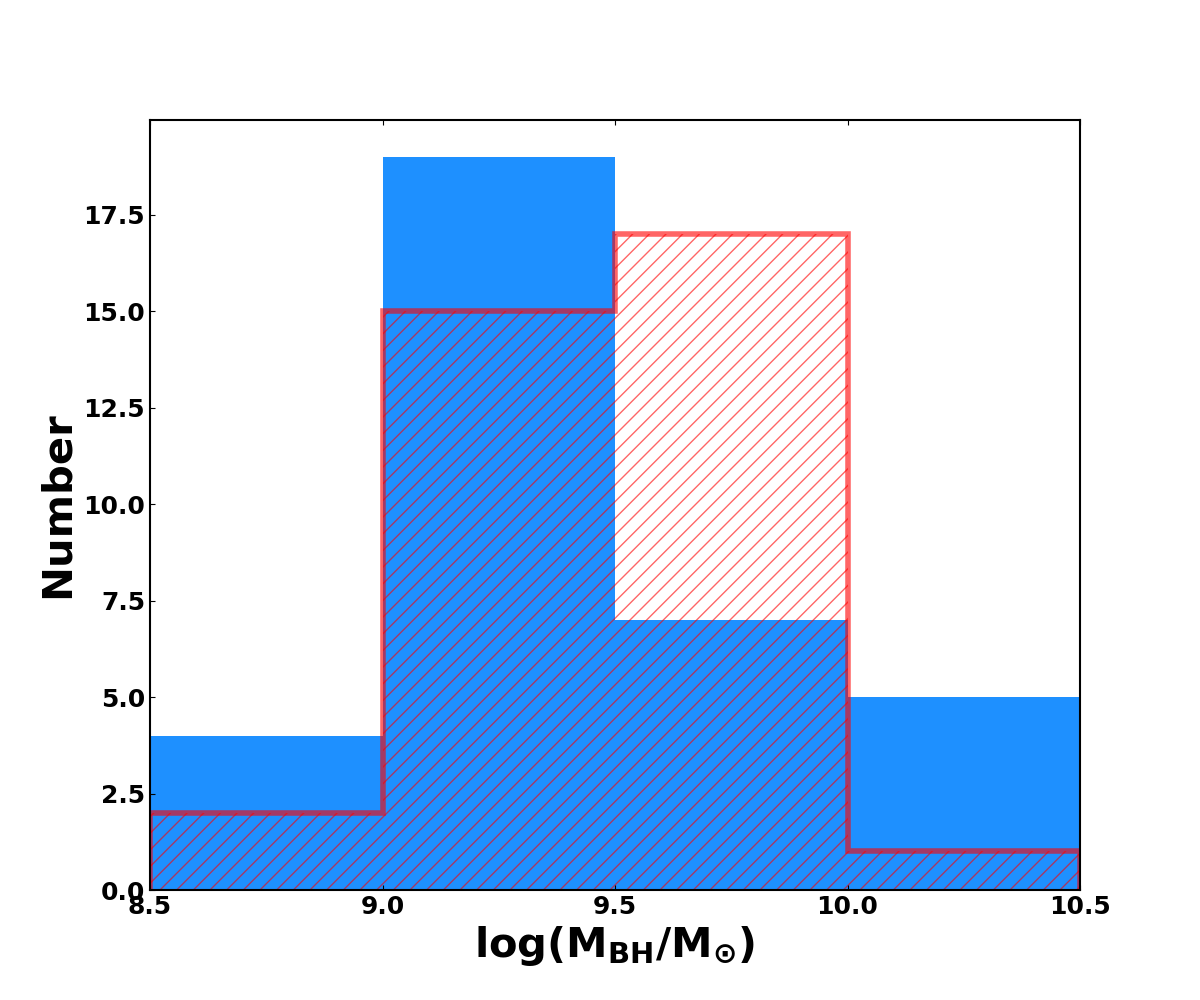}
\caption{BH masses estimated using \ha\ and \civ. Left panel: values obtained using \citet{vestergaard06} for \civ\ and \ha\ (upper left panel) and the difference between log(\mbh) (\civ) and log(\mbh) (\ha) (left bottom panel). 
Filled squares show the CARLA quasars and open squares show the 5 radio-quiet quasars that are not from CARLA. Dotted lines (red for \ha\ and blue for \civ) separate the regions where the \mbh\ are below $10^9$ \msol\ and above $10^{10}$ \msol. Right panel: the distribution of \mbh\ for our sample (blue for \civ\ and red for \ha). While the \ha-based masses are higher on average, the number of sources above $10^{10}$ \msol\ is reduced with respect to that of the \civ-based masses.}
\label{fig:BHM_ha}
\end{figure*}

\citet{shen11} presented BH masses for the entire SDSS quasar sample using the equations derived by \citet{vestergaard06} for \civ\ (at high redshift) and \ha\ (at low redshift). Since the CARLA quasars were selected from the SDSS as well, the BH masses of our quasars were previously determined based on \civ. Using the results from Section~3 we have used the \ha\ line to obtain new BH masses. The new measurements should be relatively free from the type of uncertainties that affect determinations based on \civ. We use Equation~\ref{eq:BHmass} with the parameters $(a,b,c)=(6.91\pm0.02,2.0,0.50)$ as in \citet{vestergaard06}, by converting the FWHM of \ha\ to the FHWM of \hb\ using the empirical relation presented by \citet{coatman17}. Table~\ref{tab:BHM_ha} summarizes the values obtained. We compare these values with those found for \civ\ in Figure~\ref{fig:BHM_ha}. The left panel shows that there is a large scatter between the two estimates, as expected based on previous studies \citep[e.g.,][]{coatman17}. Only considering the radio-loud quasars from CARLA (filled red squares), the \civ-based \mbh\ is systematically lower by 0.23\,dex on average, with a scatter of 0.35\,dex. Despite the \ha-based masses being somewhat higher than those estimated from \civ, it leads to fewer catastrophic outliers. This can be easily seen from the fact that of the 5 non-CARLA radio-quiet quasars that had \civ-based masses in excess of $10^{10}$ $M_\odot$ and that were observed with SINFONI, for only 1 object \ha\ gives a consistent answer (open symbols in Figure \ref{fig:BHM_ha}). The right-hand panel of Figure~\ref{fig:BHM_ha} shows the distribution in \mbh\ estimated from \civ\ and \ha, indicating again that although the \ha-based method leads to masses that are on average somewhat higher, it leads to fewer masses at the very high mass end of $\sim10^{10}$ $M_\odot$. 

Because of the ease of accessibility of the \civ\ line in optical spectra out to very high redshift, its robustness for SE BH mass determinations has been discussed by many authors, leading to novel attempts for providing reliable \mbh-calibrations that are based on this line. With the idea of finding the optimal method for determining the BH masses of the radio-loud quasar population targeted by CARLA, we will test several empirical calibrations that have been proposed for improving the \mbh\ estimate using \civ\ by comparing them with our results obtained using \ha.

\begin{figure*}
\includegraphics[width=\columnwidth]{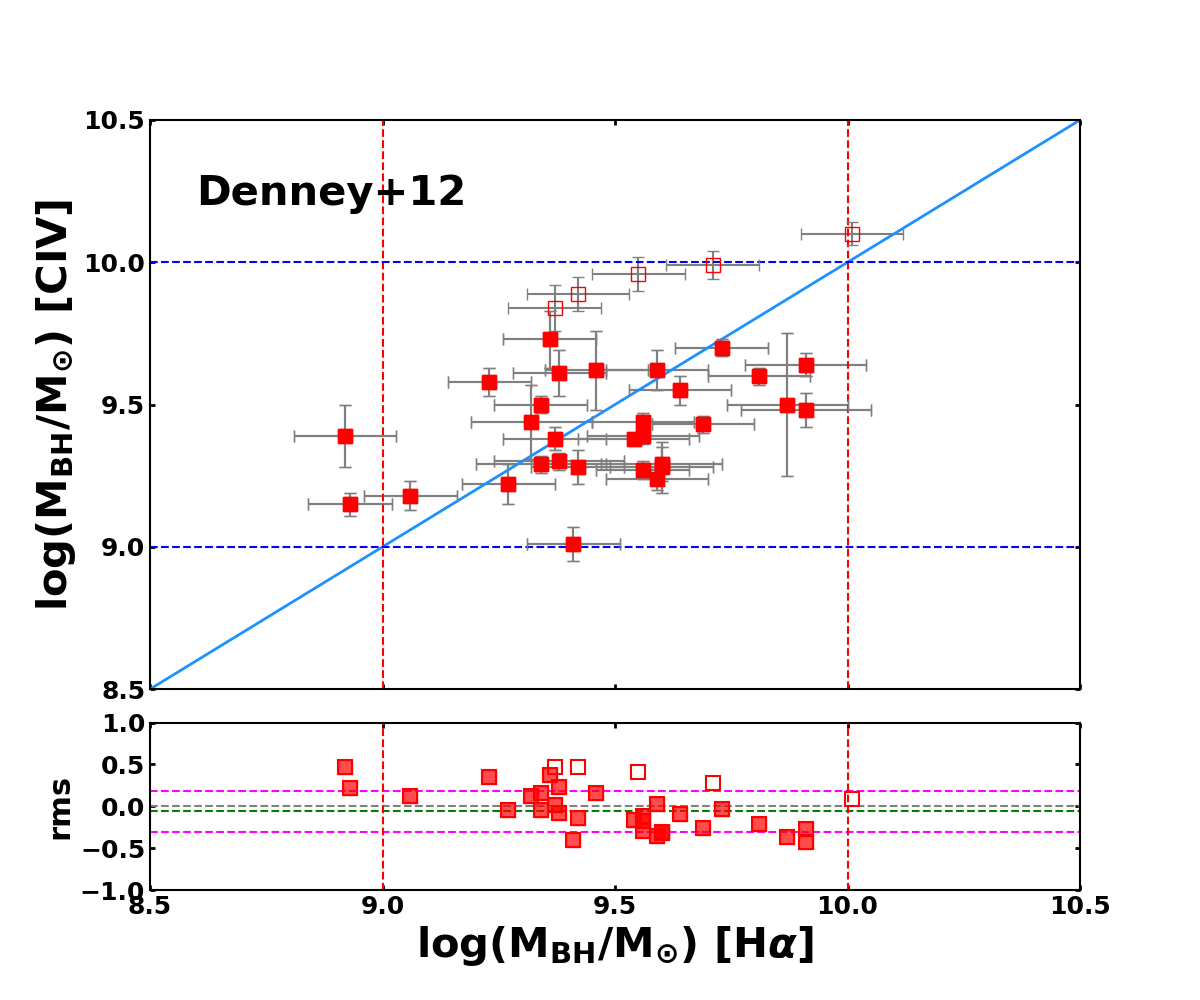}
\includegraphics[width=\columnwidth]{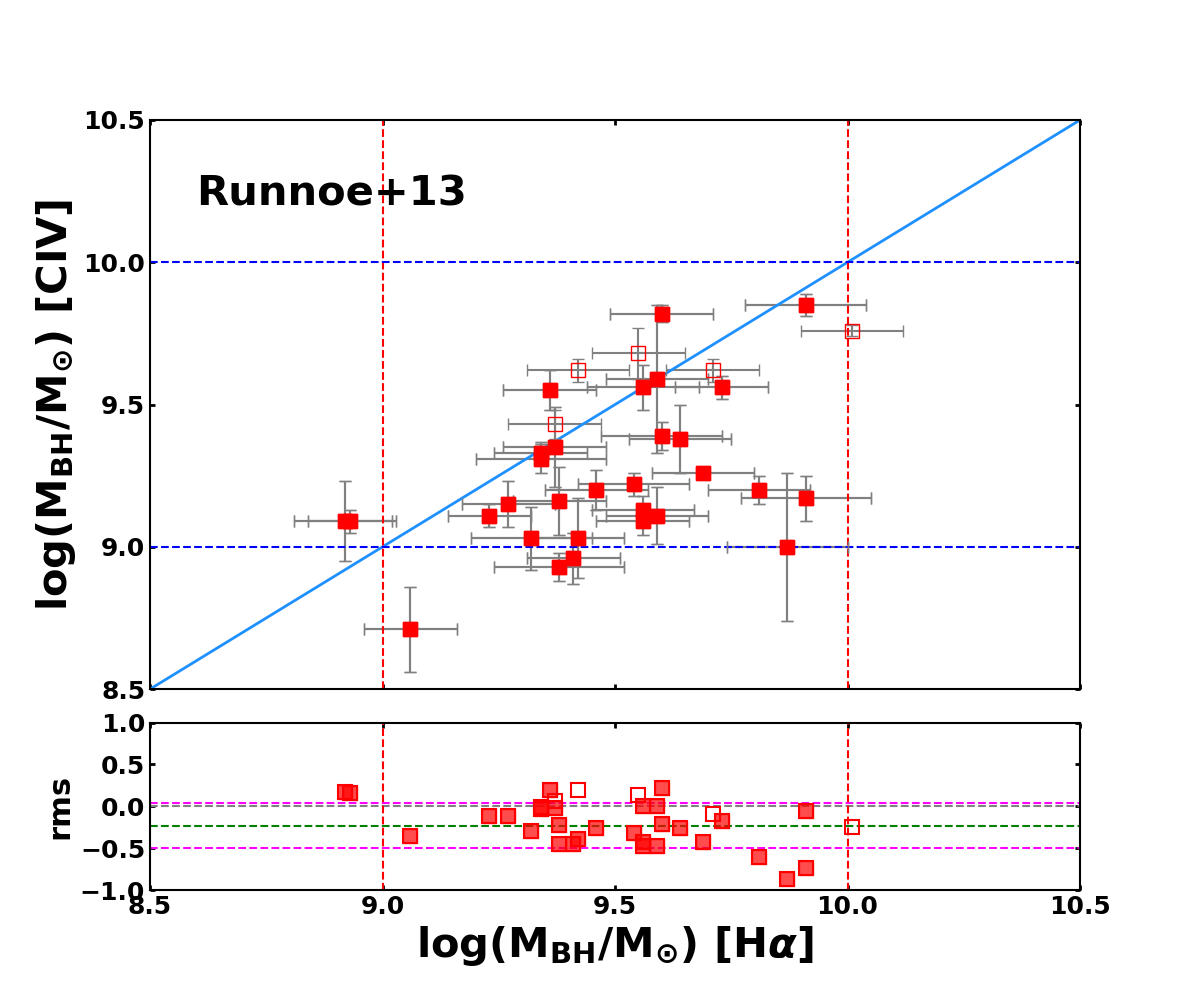}
\includegraphics[width=\columnwidth]{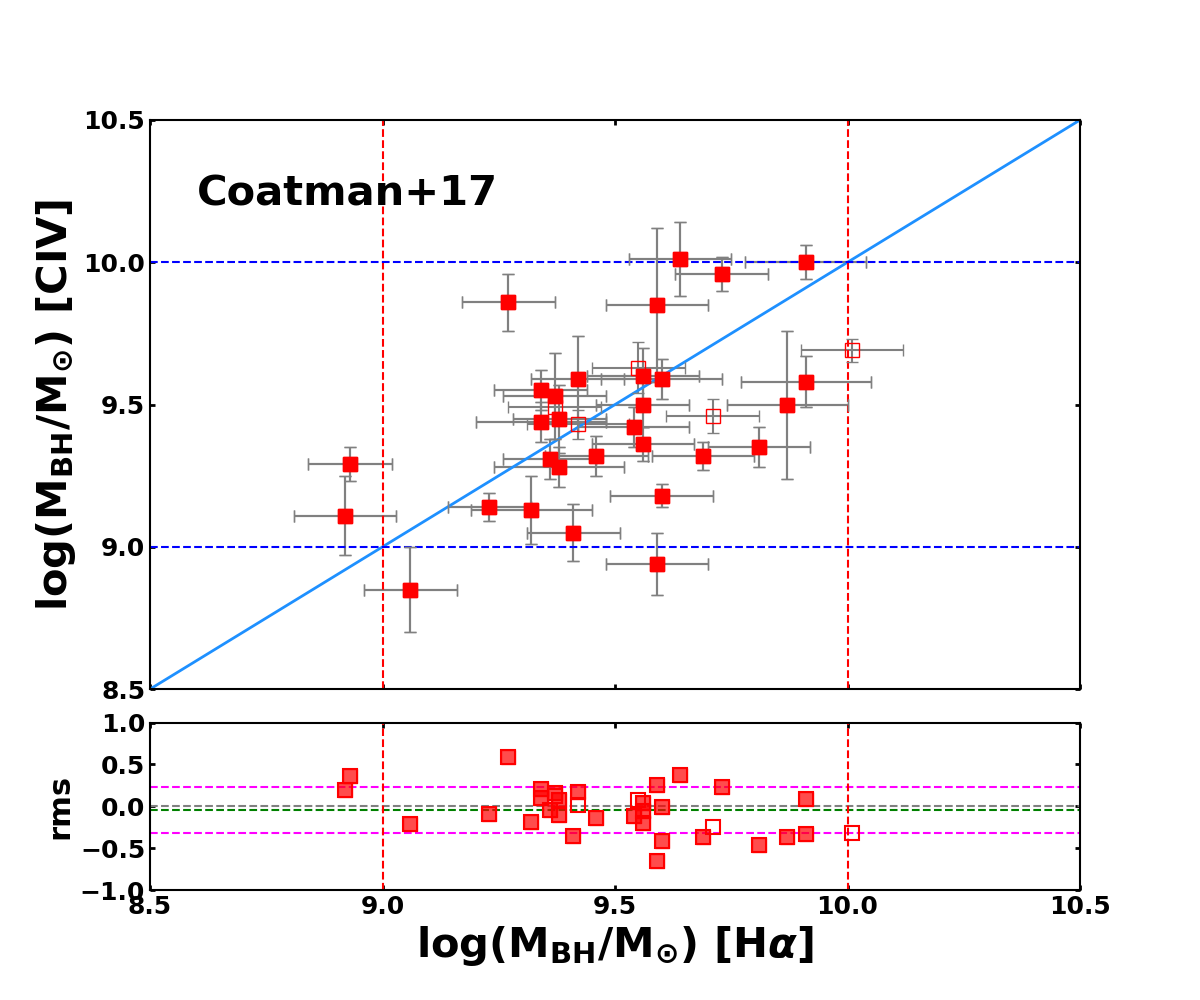}
\includegraphics[width=\columnwidth]{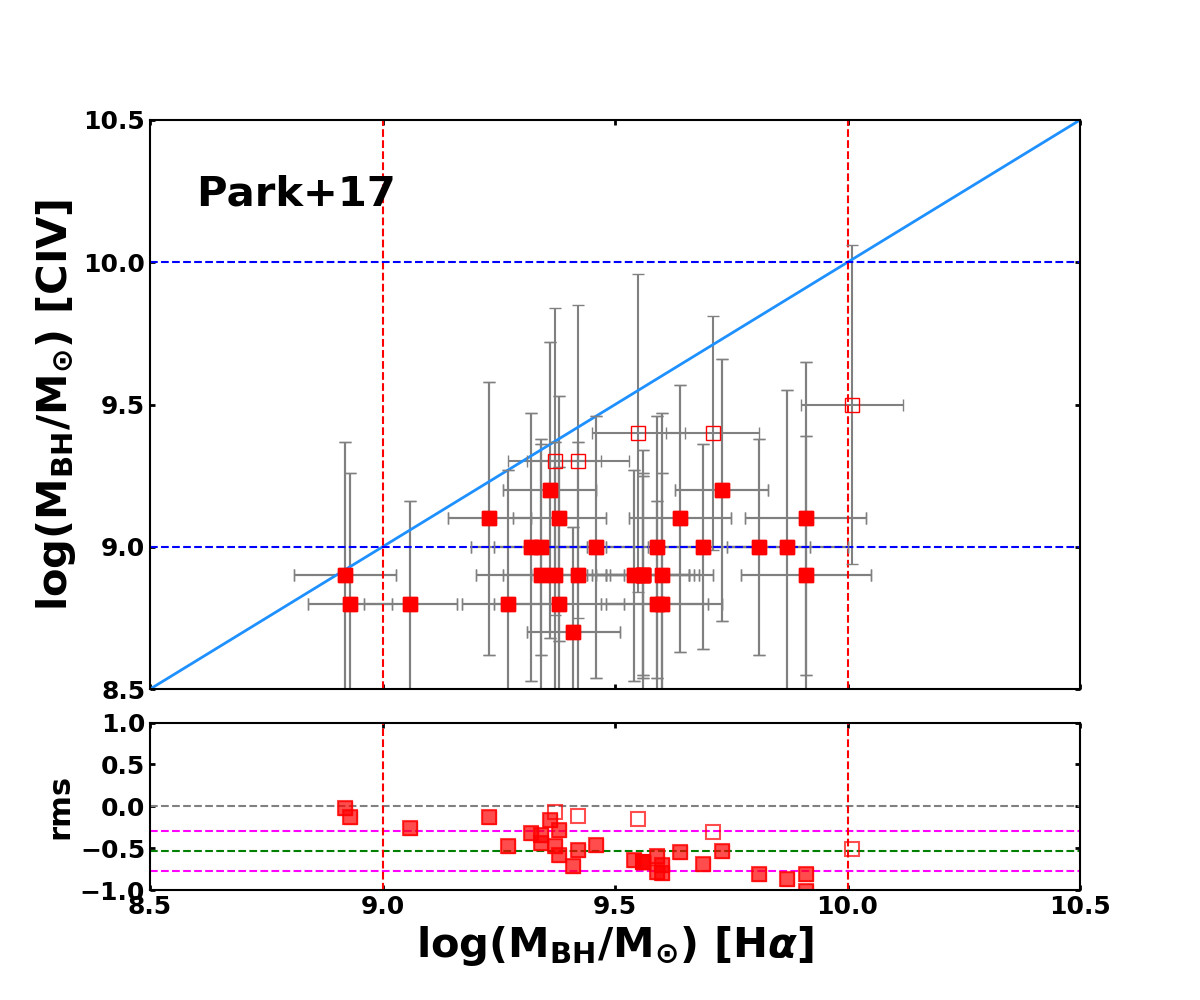}
\caption{Rehabilitation methods for \civ-based \mbh\ estimates compared to our \ha-based estimates. The top-left panel shows the method from \citet{denney12}. The top-right panel shows the method of \citet{runnoe13}. The bottom-left panel shows the method of \citet{coatman17}. The bottom right panel shows the \civ\ SE estimates using the \citet{park17} new RM results. In all panels, filled squares show the radio-loud quasars from CARLA and empty squares the radio-quiet quasars that are not in CARLA. The red and blue dashed lines demarcate the range $9<\mathrm{log}(\mathrm{M}_{\mathrm{BH}}/\mathrm{M}_\odot)<10$. Below each panel we show the absolute differences between the corrected \civ\ values and the corresponding \ha-based values. The results of these calibrations are summarized in Table \ref{tab:rehab}.}
\label{fig:BHM_corr}
\end{figure*}

\citet{denney12} used a combination of RM data with SE BH mass estimates based on \civ\ in order to investigate any offsets. Their results showed that in several cases \civ\ has a non-reverberating component (likely related to outflows) that varies from source to source. Comparing with the \mbh\ estimated from \hb\ they showed that the scatter between the estimates based on these lines correlates with the shape of \civ. They suggested a correction for \mbh\ using \civ\ based on the ratio between the FWHM and dispersion ($\sigma_{\rm l,CIV}$). We compare the \mbh\ corrected by this method with the ones obtained from \ha. The top-left panel of Figure~\ref{fig:BHM_corr} shows the results. The \citet{denney12} method significantly reduces the scatter between \ha\ and the (corrected) \civ\ from 0.35 to 0.24 dex. Also, the average value for \civ\ is now nearly the same as that of \ha, with a --0.06 dex offset. We note that \citet{denney12} developed their method using relatively low luminosity, radio-quiet sources with log(\mbh/\msol) between 7 and 9. Our results obtained here indicate that the method extends to luminous radio-loud quasars with higher BH masses as well.

With the same goal, \citet{runnoe13} suggested that discrepancies between \mbh\ from \civ\ and \hb\ are mainly due to the fact that \civ\ often presents a non-virial component. Using a sample of 85 bright low-to-intermediate redshift quasars, they found that the difference in widths of \civ\ and \hb\ correlates with the peak flux ratio of Si\,{\sc iv}+O\,{\sc iv}]$\lambda$1400 and \civ\ as the former line complex does not vary with the FWHM of \civ. Using this parameter (Peak Ratio $\lambda$1400 in Table \ref{tab:fitting2}) they derived an empirical correction to the virial equation that reduces the scatter from 0.43 to 0.33\,dex in their sample. The top-right panel of Figure~\ref{fig:BHM_corr} shows how the \citet{runnoe13} correction method compares with our estimates based on \ha. This method also reduces the scatter between the \mbh\ estimated from \ha\ and \civ, but with a tendency to underestimate the true (\ha) \mbh. We find that the average BH mass has an offset of --0.24\,dex with a scatter of 0.27\,dex. Note that the scatter is very similar to that obtained using the \citet{denney12} method, but with a larger negative systematic offset. Furthermore, this method depends strongly on a line that is significantly weaker than \civ. This means that in low S/N spectra the correction factor can become difficult to measure. 

\citet{coatman17} presented another possible solution to improve \civ-based BH mass estimators. Analyzing a sample of 230 quasars, they found a strong correlation between the blueshift of \civ\ and the ratio of the FHWMs of \civ\ and \ha. Their results show that in order to estimate the BH mass using the FWHM of \civ\ it is necessary to first correct its width by a parameter that is a function of the \civ\ blueshift. We estimated the BH masses using the \citet{coatman17} method for our sample. The bottom-left panel of Figure~\ref{fig:BHM_corr} shows the corrected BH masses compared with our measurements based on \ha. The results are comparable to those obtained with the \citet{denney12} method. The scatter between the BH masses from \ha\ and \civ\ are reduced to 0.28 dex with a small systematic offset of --0.04. 

Finally, \citet{park17} presented new RM results based on a larger sample than that presented by \citet{park13}. By using a sample of 35 high S/N AGN they derived a new equation for SE BH masses using \civ. The main difference between their results and previous determinations of SE \civ-based BH mass estimates is the value of the exponent in the luminosity term of the equation. Their method reduces the contribution of the luminosity in the virial equation, and as a result the values obtained are smaller compared to the standard \civ-based estimates that follow \citet{vestergaard06}. The bottom-right panel of Figure~\ref{fig:BHM_corr} shows the BH masses estimated by the \citet{park17} method compared with our results using \ha. Although the scatter is indeed reduced (0.24\,dex), we now find a large systematic offset of --0.53\,dex. This is in agreement with the results shown in Figure~11 of \citet{park17}. They point out that their method overpredicts the masses of low mass BHs ($<10^{8.5}$ \msol), and underpredicts those with high mass BHs similar to our CARLA sample ($>10^{8.5}$ \msol). 

\begin{table*}
\centering
\caption{Accuracy of the \civ\ rehabilitation method results as applied to radio-loud quasars from CARLA.}
\label{tab:rehab}
\begin{tabular}{lccc} 
\hline
\hline
Method            &   Mean offset$^a$  &  Scatter$^b$  & Correlation coefficient$^c$\\          
			      &	     (dex)     &    (dex)     &        ($\rho$, $p$)                 \\ 
\hline
Uncorrected (Figure \ref{fig:BHM_ha})      &    --0.23      &     0.35     &  0.16, 0.41  \\ 
\citet{denney12} (Figure \ref{fig:BHM_corr}, top-left) &    --0.06      &     0.24     &  0.35, 0.06 \\    
\citet{runnoe13}  (Figure \ref{fig:BHM_corr}, top-right) &    --0.24      &     0.27     &  0.43, 0.02  \\    
\citet{coatman17} (Figure \ref{fig:BHM_corr}, bottom-left)&    --0.04      &     0.28     &  0.47, 0.01  \\
\citet{park17}  (Figure \ref{fig:BHM_corr}, bottom-right)   &    --0.53      &     0.24     &  0.27, 0.14  \\
\hline
\hline
\multicolumn{4}{l}{$^a$ Mean difference between log(\mbh) (\civ) and log(\mbh) (\ha).}\\
\multicolumn{4}{l}{$^b$ Standard deviation of the log(\mbh) (\civ) -- log(\mbh) (\ha) residuals.}\\
\multicolumn{4}{l}{$^c$ Spearman rank correlation efficient $\rho$ and $p$-value.}\\
\end{tabular}
\end{table*}

An overview of the successfulness of these corrections as applied to the CARLA sample are given in Table \ref{tab:rehab}, where we list the mean difference between log(\mbh) (\civ) and log(\mbh) (\ha), the scatter, and the Spearman rank correlation coefficients. 
We summarize this Section by stating that the \ha-based \mbh\ masses of radio-loud quasars in CARLA are, on average, about 0.2 dex higher compared to those estimated from \civ, while at the same time the number of objects having $\mathrm{M_{BH}}>10^{10}$ \msol\ is reduced from five objects to just one. The systematic offsets could be reduced simply by a change of calibration constants used for each relation, although this will likely be strongly sample-dependent.

Among the various methods proposed to improve the \civ-based BH masses tested, the \citet{denney12} and \citet{coatman17} methods resulted in the best agreement between \civ- and \ha-based methods (i.e., they had the smallest mean offsets and similar scatter), while the \citet{runnoe13} method had the second strongest correlation (after \citet{coatman17}, but with a large systematic offset). 

\subsection{Eddington Ratio}
\label{accretion}

The Eddington luminosity ($\rm L_{Edd}$) is the theoretical maximum luminosity that an object can emit while balancing the outward radiative pressure and inward gravitational force. The Eddington ratio (\led) is an estimate of the accretion efficiency of a BH, and is believed to be the main driving mechanism of the Eigenvector 1 which correlates with most of the features in Type-1 AGN spectra, such as the ratio of Fe\,{\sc ii}/\hb, the soft X-ray slope, and the blueshift of \civ\ \citep{boroson02,marziani01,shen14}.

To calculate the Eddington ratio, we apply a bolometric correction factor, $f_l$, which relates the intrinsic luminosity of the accretion disk to the measured optical luminosity\footnote{This bolometric correction refers to the bolometric correction required to obtain the accretion disk luminosity, and is smaller than the total observed bolometric corrections typically applied to AGN \citep[e.g.][]{elvis94,marconi04,richards06,netzer07}.}. We then divide this bolometric luminosity by the Eddington luminosity for a given BH mass. $f_l$ is known to depend on luminosity, and for our sample ($45.5 <$log(\lopt)$ < 47.5$\,erg\,s$^{-1}$) its range is 5--7 \citep{marconi04}. In this paper we follow the approach of \citet{netzer07}, which assumes $f_l=7.0$. \led\ can then be estimated through:

\begin{equation}L/L_{\rm Edd}=\frac{f_l~L_{5100}}{L_{\rm Edd}} = \frac{7.0~L_{5100}}{(1.5\times10^{38}~M_\mathrm{BH}/M_{\odot})}\label{eq:led}\end{equation}

\begin{figure*}
\includegraphics[width=\textwidth]{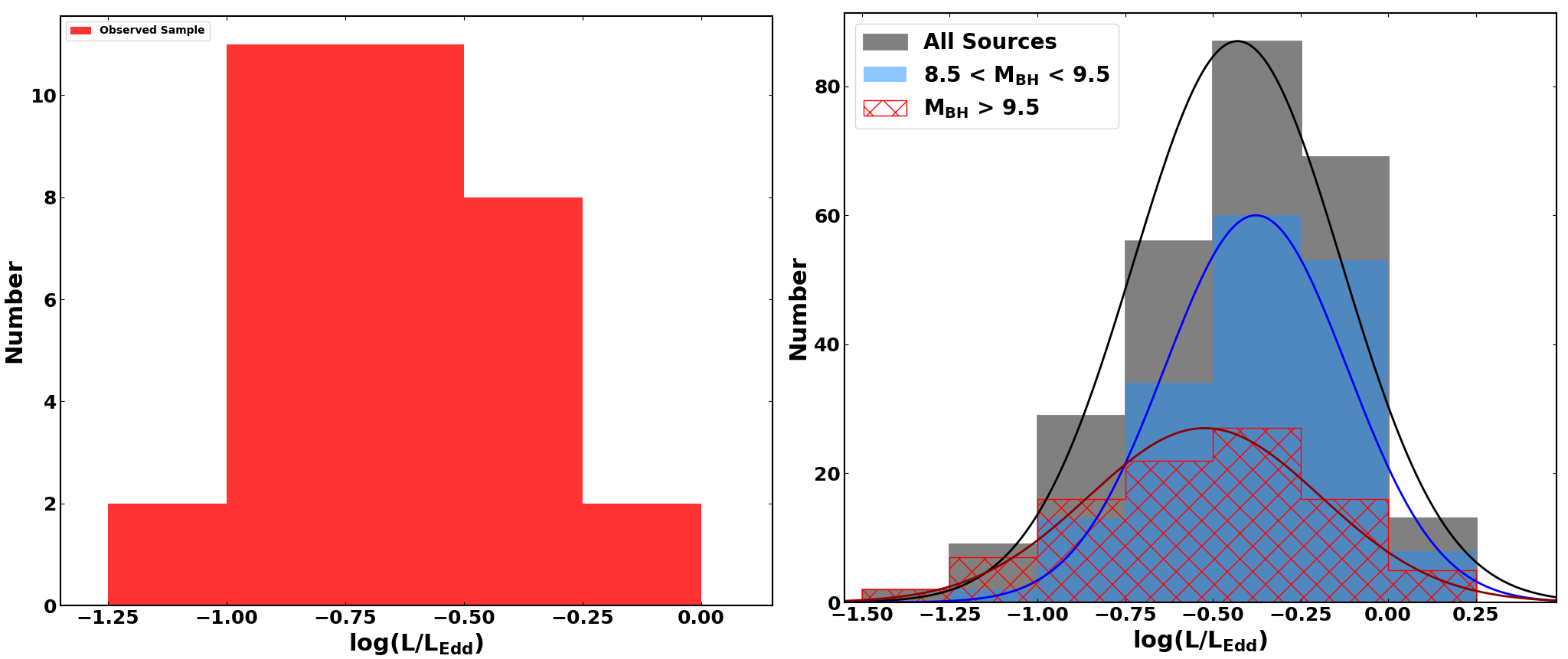}
\caption{Distribution of \led\ for the quasars. Left panel: distribution of \led\ for the CARLA subsample observed with SINFONI. Right panel: Distribution of \led\ for the quasars in our sample combined with quasars from the \citet{coatman17} sample (grey shaded histogram and black line). The red-hatched histogram and red line show the distribution for $M_\mathrm{BH}>10^{9.5}$ \msol, while the blue shaded histogram and blue line show distribution for $M_\mathrm{BH}<10^{9.5}$ \msol.}
\label{fig:led-dist}
\end{figure*}

We use Equation~\ref{eq:led} to estimate \led\ for our sample, as well as for the \citet{coatman17} sample for comparison. The results are shown in Figure~\ref{fig:led-dist}. The left panel shows the distribution of \led\ for the CARLA sample. The values have a relatively broad distribution ($0.26$\,dex) with a mean value of \led$\approx0.2$. In the right panel of Figure~\ref{fig:led-dist} we compare these results to a larger sample by adding to our sample that of \citet{coatman17}, which is composed of both radio-quiet and radio-loud quasars. We split the combined sample into a high (\ha-based) mass (log(\mbh/\msol)$>9.5$) and a low (\ha-based) mass (log(\mbh/\msol)$<9.5$) sample. The lower mass BHs have a higher mean \led$\approx0.4$ (spread of $0.25$\,dex) than the higher mass BHs (mean \led$\approx0.3$ and spread of $0.33$\,dex). The latter distribution is very similar to that found for the CARLA sample shown in the left panel, which is expected because the CARLA sample is exclusively composed of radio-loud quasars with high BH masses. The grey histogram shows the distribution we get when we combine the \citet{coatman17} and CARLA samples (mean \led$\approx0.37$ with a spread of 0.29 dex). This shows that the CARLA quasars have \led\ that are, on average, at most 50\% lower than samples consisting of (mostly) radio-quiet quasars of similar masses. 

\begin{figure}
\includegraphics[width=\columnwidth]{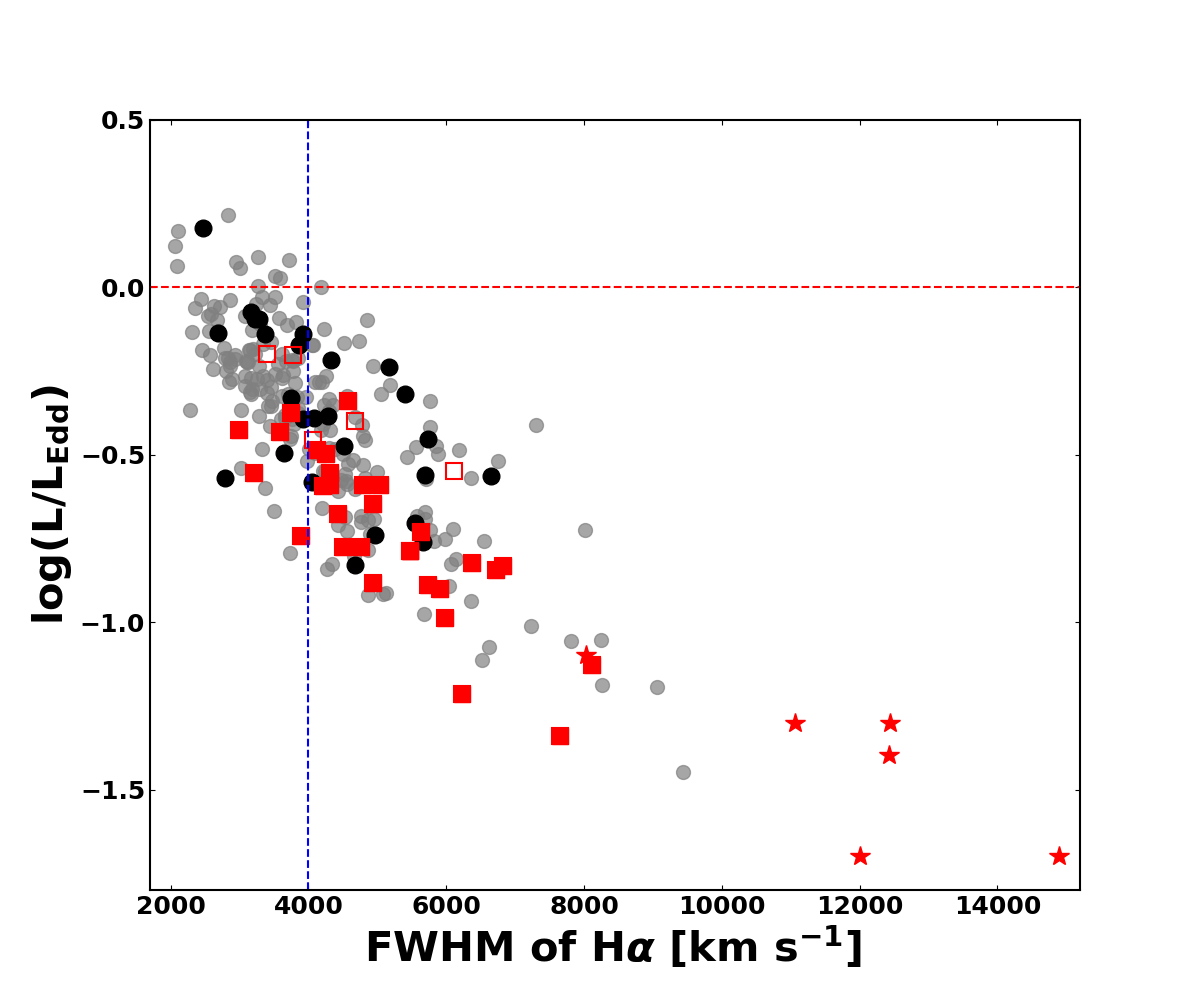}
\includegraphics[width=\columnwidth]{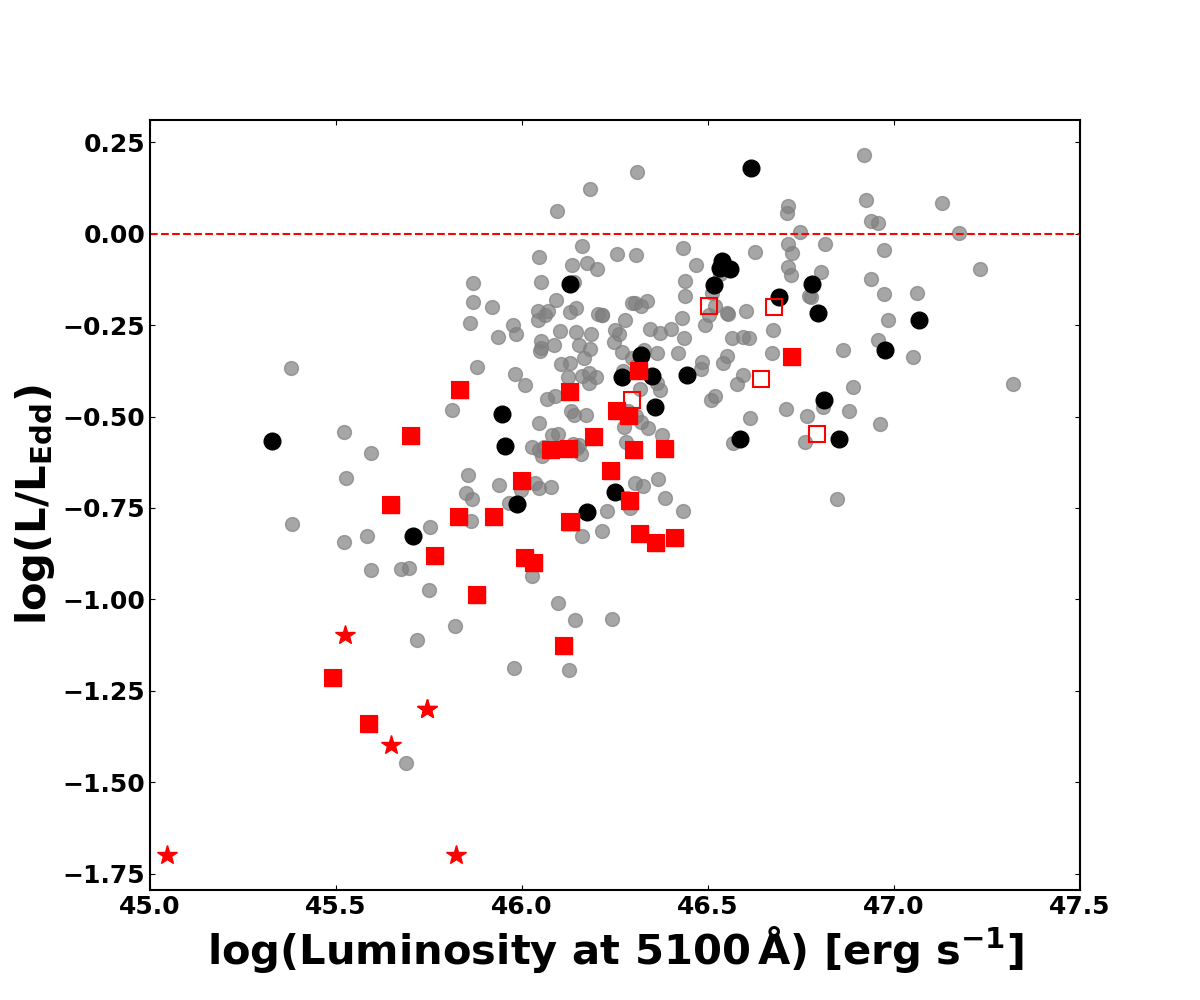}
\includegraphics[width=\columnwidth]{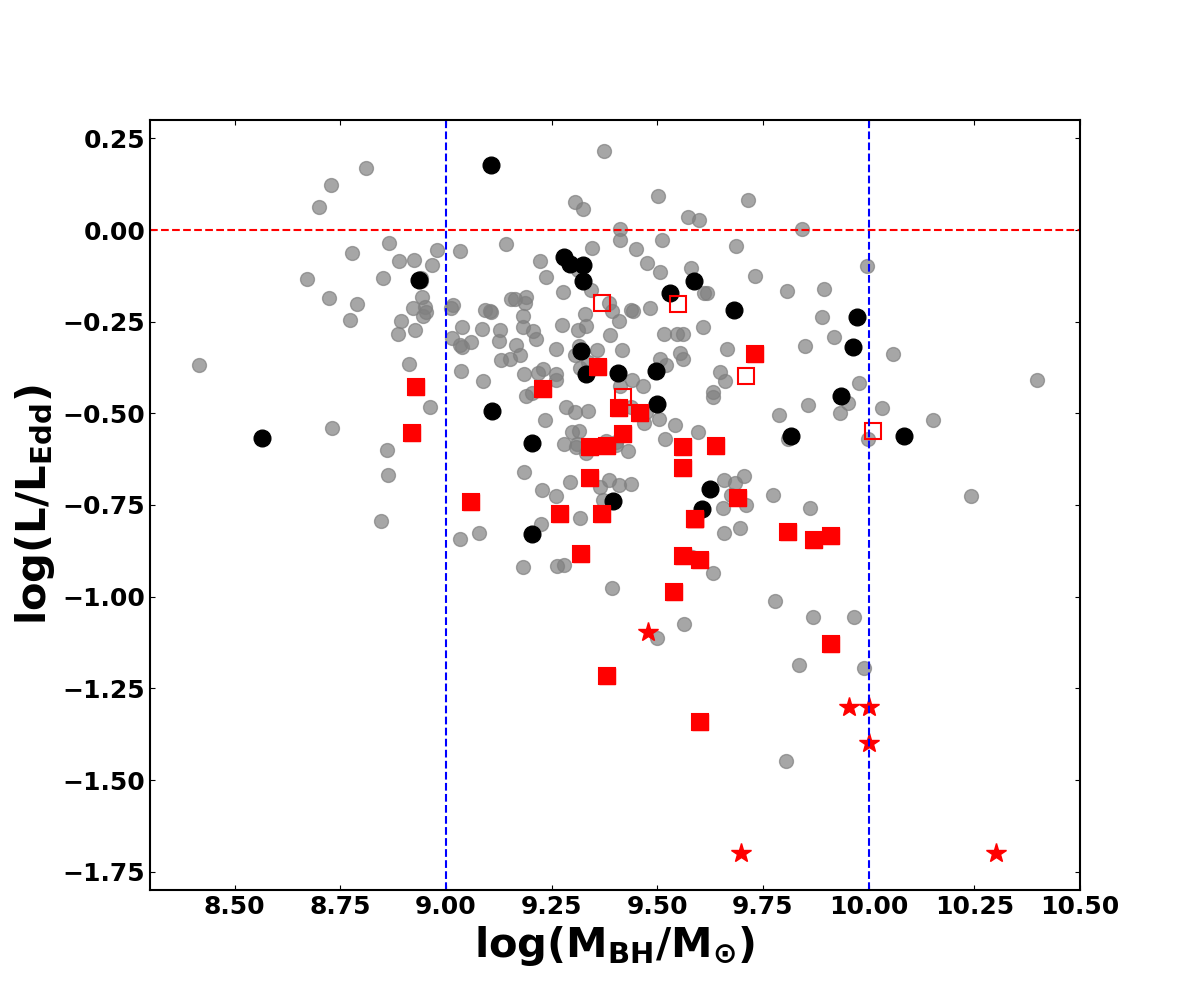}
\caption{Quasar properties as function of Eddington ratio. Panels show (top) the correlation between \led\ and FHWM(\ha), (middle) \led\ and \lopt, and (bottom) \led\ and \mbh. In all panels red filled and open squares, respectively, show the CARLA and non-CARLA quasars from this paper, while the black squares and grey dots show the radio-loud and radio-quiet quasars from \citet{coatman17}, respectively. Red stars are the broad line radio galaxies from \citet{nesvadba11}.}
\label{fig:led-corr}
\end{figure}

In Figure~\ref{fig:led-corr} we plot \led\ as a function of the observables FWHM(\ha) and luminosity (\lopt) and the derived BH mass. We have also added the 6 broad line radio galaxies from \citet{nesvadba11}, and the radio-loud and radio-quiet quasars from \citet{coatman17}. The general trends in the three panels follow readily from Equation 2, in which \led\ is proportional to \lopt\ and inversely proportional to \mbh\ (which itself is proportional to both FWHM(\ha) and \lopt; see Equation 1). The strongest differences are between the radio-quiet objects from \citet{coatman17} on one hand, and the radio-loud quasars from both samples on the other hand. Although the \citet{coatman17} radio-quiet sample contains quasars that are found in the same region of parameter space as the radio-loud objects, it also contains quasars with significantly higher luminosities and Eddington ratios and lower BH masses.   

\subsection{Growth Time}
\label{growth}

\begin{figure}
\includegraphics[width=\columnwidth]{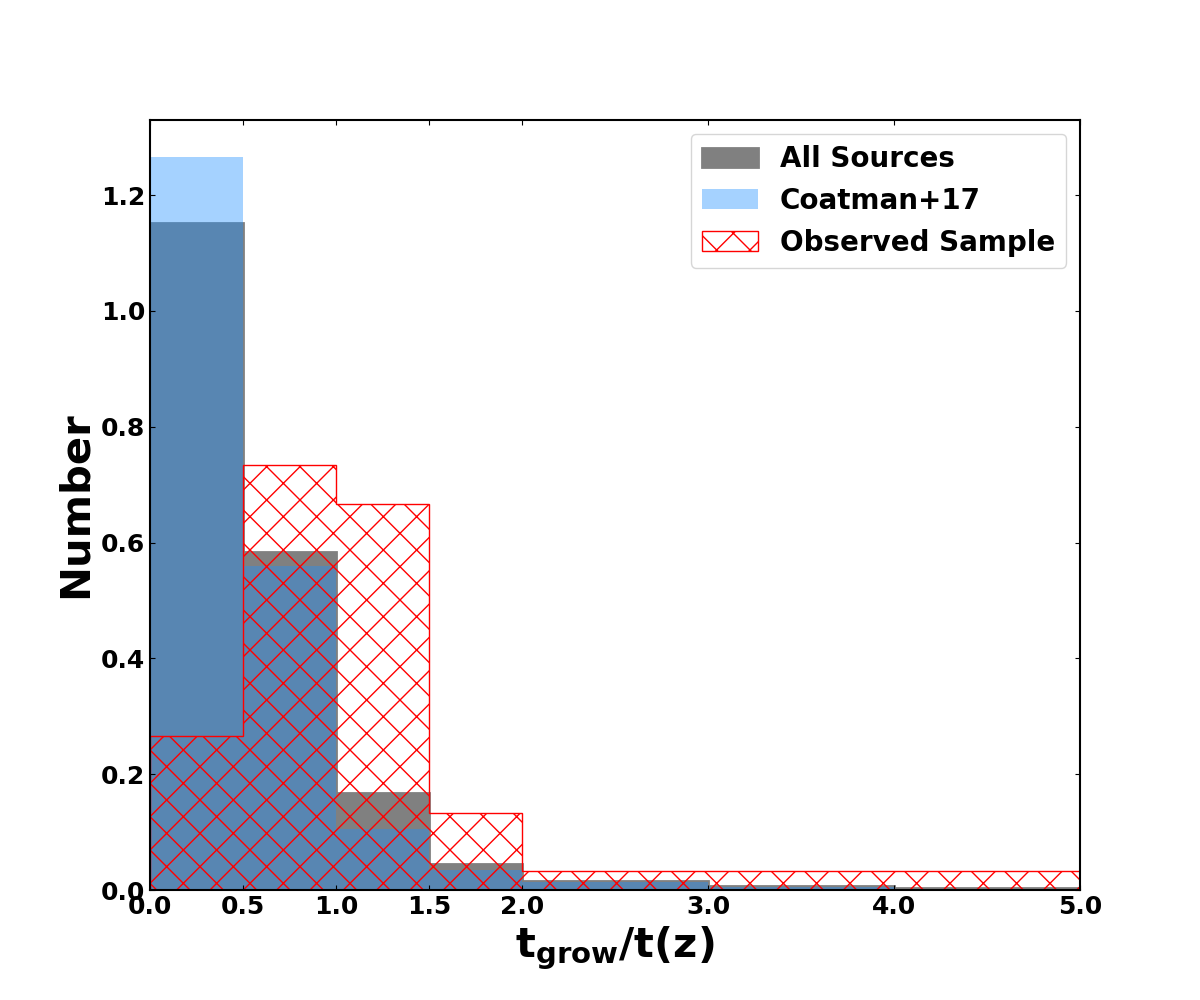}
\caption{Distribution of the quasar growth times. The red dashed histogram shows the values derived for the CARLA sample, the blue solid histogram shows the \citet{coatman17} sample and the grey histogram shows the two samples combined. See Section 4.3 for details.}
\label{fig:gr-dist}
\end{figure}

To estimate BH growth times ($t_\mathrm{grow}$) we first assume that the BH experiences a period of continuous growth during their active phase starting from a seed BH mass. We use the original expression from \citet{salpeter64}:

\begin{equation}
t_\mathrm{grow}= t_{\rm Edd}~\frac{\eta/(1-\eta)}{L/L_{\rm Edd}}~\mathrm{log}\left(\frac{M_\mathrm{BH}}{M_\mathrm{seed}}\right)~\frac{1}{f_\mathrm{active}}~\rm yr, 
\label{eq:growth}
\end{equation}

\noindent
where $t_{\rm Edd}=3.5\times \rm 10^8$\,yr is the Eddington time, $\eta$ is the accretion efficiency, and $f_\mathrm{active}$ is the fraction of the time that the BH is actively accreting. The efficiency depends on BH spin and typical values range an order of magnitude ($\eta=0.04-0.4$) from retrograde to prograde accretion or BH spin of $a=\pm1$. Typical values of $\eta=0.2$ are generally assumed, reflecting a non-zero angular velocity \citep{king04}. The BH seed can have a low mass (M$_{\rm seed}=10^{2-4}$ \msol) when resulting from Population-III stars, or larger (M$_{seed}=10^{4-6}$ \msol) when resulting from direct gas cloud collapse \citep{begelman06}. In this work we follow \citet{netzer07}, assuming $\eta=0.2$, M$_{\rm seed}=10^{4}$ \msol, and $f_\mathrm{active}=1$. Additionally, we calculate the growth time using a lower efficiency, $\eta=0.1$. From these parameters and Equation~\ref{eq:growth} we estimate the growth time for our sample as well as the \citet{coatman17} sample which is dominated by radio-quiet quasars (383 out of 409). The values obtained for \gt\ for our sample are listed in Column~9 of Table~\ref{tab:BHM_ha}, where we divided the values by the age of the universe at each redshift, $t(z)$. We plotted the distribution of growth times in Figure~\ref{fig:gr-dist}. For the assumed efficiency, 15 of our 35 quasars have $t_\mathrm{grow}/t(z)>1$. This indicates that the typical efficiencies or duty cycles assumed are too high for a substantial fraction of our sample. Comparing the two samples in Figure~\ref{fig:gr-dist}, we find that radio-quiet and radio-loud quasars follow somewhat different distributions in \gt, with radio-quiet quasars having, on average, lower growth times compared to radio-loud quasars due to their higher accretion rates.

\begin{figure*}
\includegraphics[height=4.5cm]{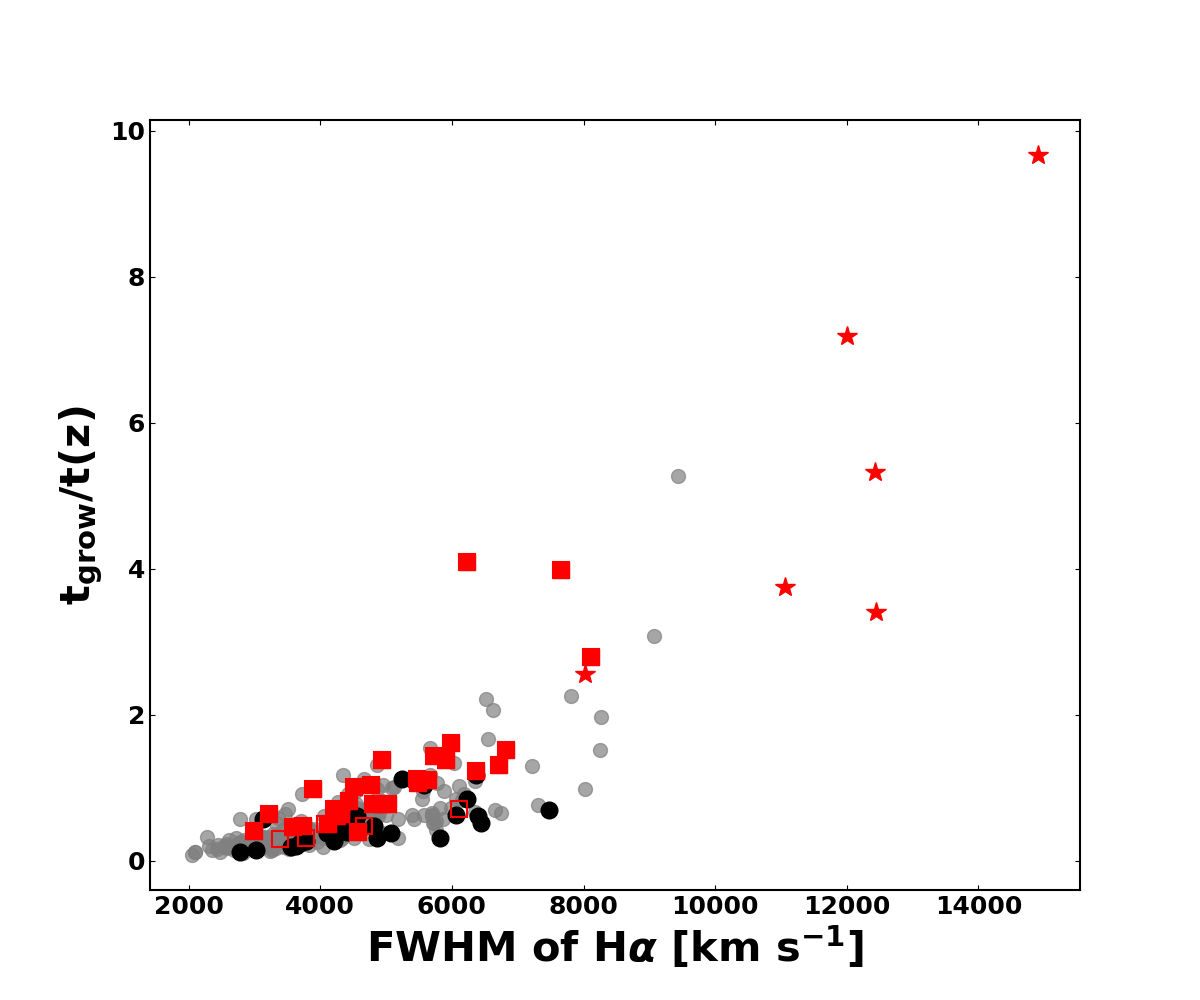}
\includegraphics[height=4.5cm]{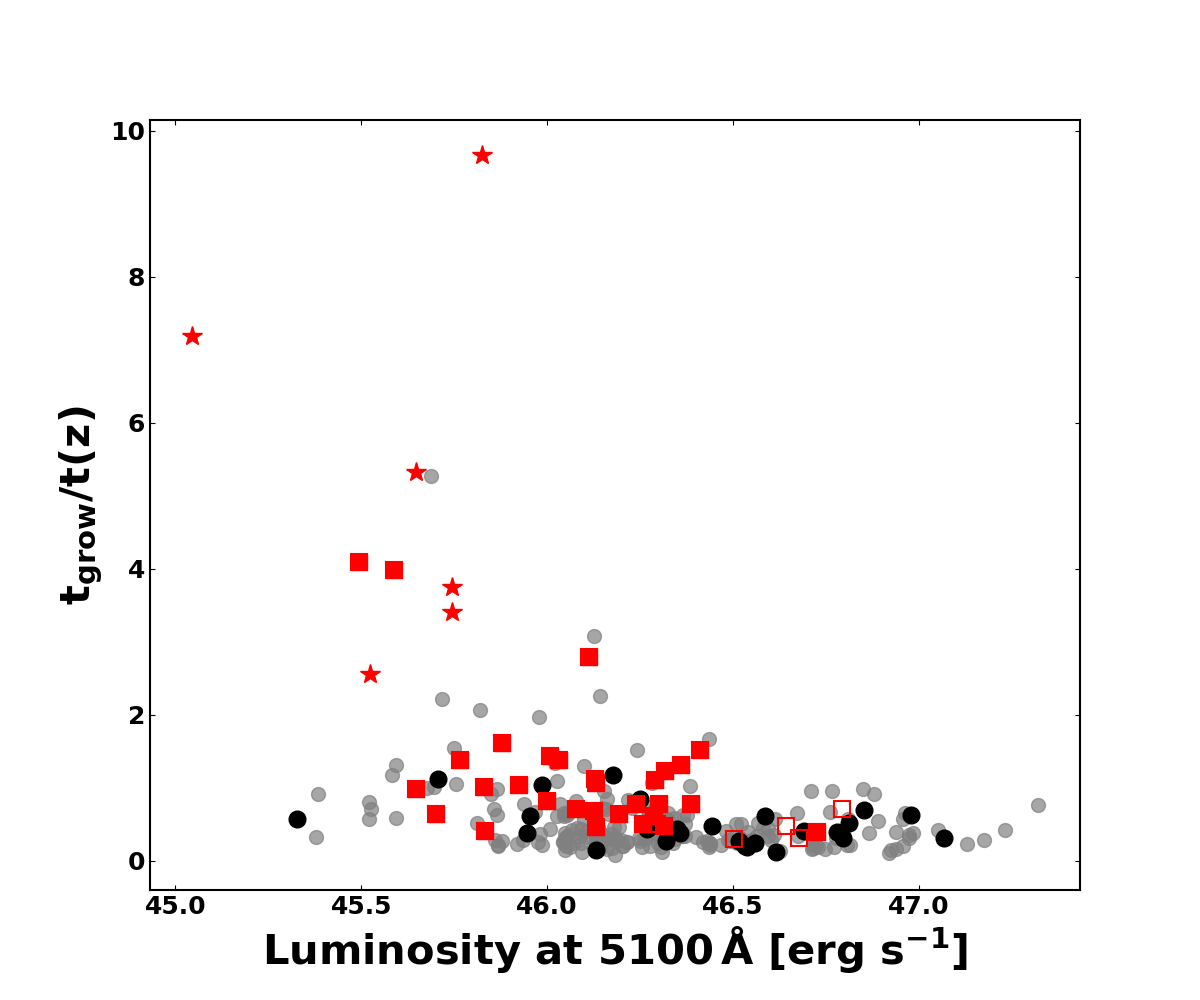}
\includegraphics[height=4.5cm]{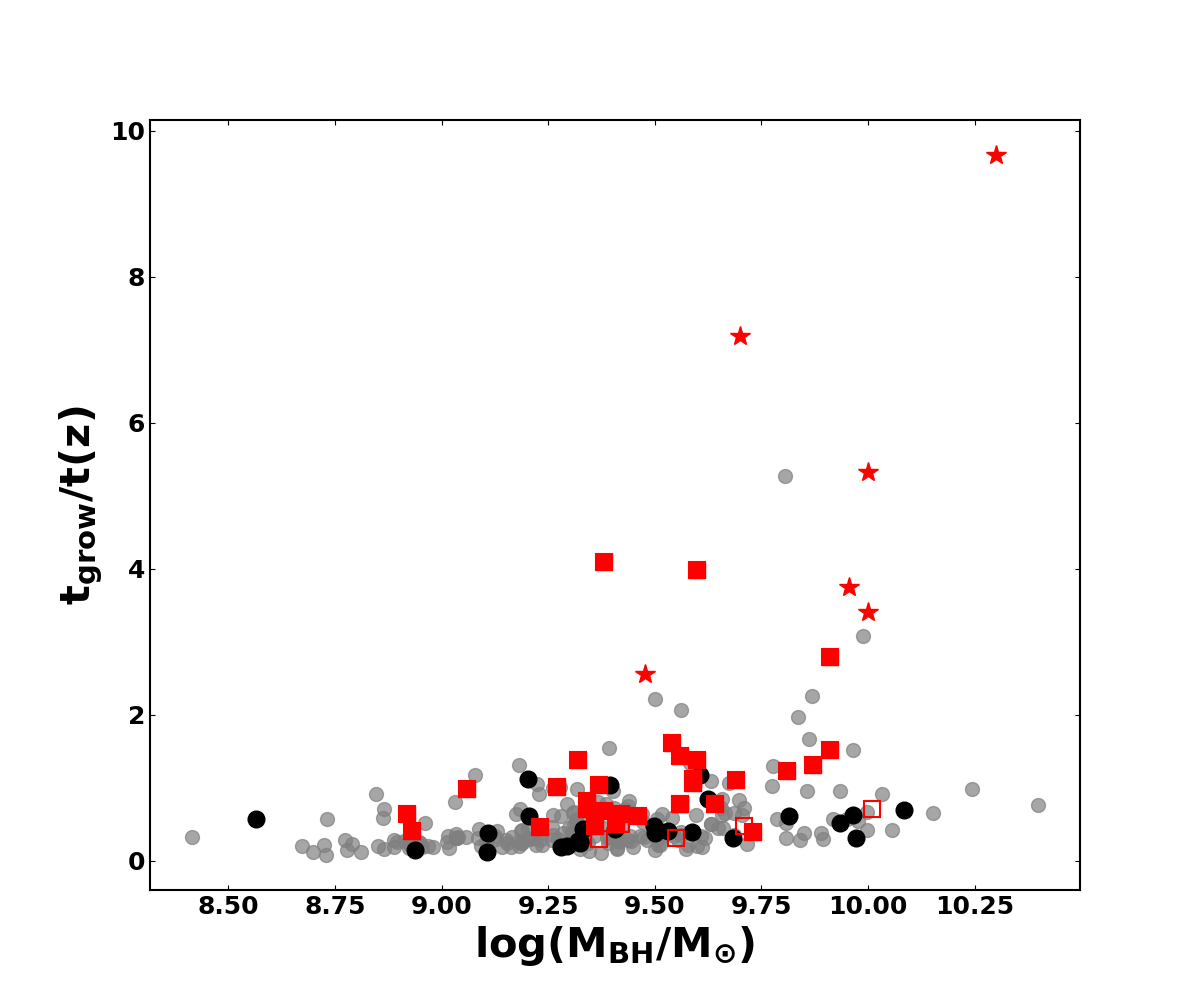}
\caption{Quasar properties as function of the growth time. Panels show \gt\ versus FHWM of \ha\ (left), \gt\ versus \lopt\ (middle), and \gt versus \mbh\ (right). In all panels, red filled squares show the quasars from CARLA, red empty squares show the radio-quiet quasars in our sample that are not from CARLA, and black squares (grey dots) show radio-loud (radio-quiet) quasars from \citet{coatman17}. Red stars are the broad line radio galaxies from \citet{nesvadba11}. See Section 4.3 for details.}
\label{fig:gr-corr}
\end{figure*}

In order to highlight these differences in growth times, we plot \gt$/t(z)$ against FWHM(\ha), optical luminosity, and \mbh\ in Figure~\ref{fig:gr-corr}. Although \gt\ is also a function of FWHM(\ha), \lopt\ and \mbh, and the trends are those that are expected, the normalization by $t(z)$ ensures that the comparison between quasars at different redshifts is done in a fair way. Similar to Figure \ref{fig:led-corr}, the key differences are observed between the radio-loud quasars from CARLA and \citet{coatman17} on one hand, and the radio-quiet quasars from \citet{coatman17} on the other. The latter are found at \gt$/t(z)<1$ in the majority of the cases, whereas the radio-loud quasars exceed the maximum available growth time in more than half of the cases. We will discuss the implications of these results in Section \ref{sec:history}. 

\begin{table*}
\begin{scriptsize}
\centering
\caption{Black Hole masses, Eddington ratios and growth times.}
\label{tab:BHM_ha}
\begin{tabular}{lcccccccc} 
\hline
\hline
SDSS Name            & \multicolumn{1}{c}{\ha} & \multicolumn{1}{c}{\civ} &  \multicolumn{4}{c}{Rehabilitated \civ\ measurements} & & \\
\cmidrule(lr){2-2}\cmidrule(lr){3-3}\cmidrule(lr){4-7}
    & 			log(\mbh/\msol)  & log(\mbh/\msol)  & log(\mbh/\msol)  & log(\mbh/\msol)   & log(\mbh/\msol)   & log(\mbh/\msol) & \led\   & $t_\mathrm{growth}/t(z)$ \\
    &   						 &    				& (Denney+12) 		& (Runnoe+13)       & (Coatman+17)      & (Park+17) 	  &         &    \\
\hline
\multicolumn{9}{l}{Radio-loud quasars from CARLA}\\
\hline  
J012514$-$001828 & 9.34$\pm$0.10  & 9.13$\pm$0.04    & 9.50$\pm$0.03  	 & 9.32$\pm$0.04    & 9.54$\pm$0.07		& 8.95$\pm$0.38   & 0.21    & 0.81 \\
J082707+105224   & 9.36$\pm$0.11  & 9.23$\pm$0.14    & 9.38$\pm$0.04  	 & 9.35$\pm$0.14    & 9.52$\pm$0.15		& 8.85$\pm$0.47   & 0.16    & 1.03 \\
J090444+233354   & 9.73$\pm$0.10  & 9.77$\pm$0.04    & 9.70$\pm$0.03  	 & 9.56$\pm$0.04    & 9.96$\pm$0.06		& 9.24$\pm$0.46   & 0.45    & 0.39 \\
J092035+002330   & 9.91$\pm$0.13  & 9.68$\pm$0.04    & 9.64$\pm$0.04  	 & 9.85$\pm$0.04    & 9.99$\pm$0.06		& 9.09$\pm$0.55   & 0.07    & 2.79 \\
J094113+114532   & 9.37$\pm$0.14  & 8.83$\pm$0.05    & 9.29$\pm$0.03  	 & 8.93$\pm$0.05    & 9.28$\pm$0.07		& 8.84$\pm$0.48   & 0.06    & 4.09 \\
J102429$-$005255 & 9.69$\pm$0.11  & 9.02$\pm$0.02    & 9.43$\pm$0.03  	 & 9.26$\pm$0.02    & 9.31$\pm$0.05		& 8.97$\pm$0.36   & 0.18    & 1.11 \\
J104257+074850   & 9.91$\pm$0.14  & 9.24$\pm$0.08    & 9.47$\pm$0.06  	 & 9.17$\pm$0.08    & 9.57$\pm$0.09		& 8.94$\pm$0.49   & 0.14    & 1.52 \\
J110344+023209   & 9.64$\pm$0.11  & 9.59$\pm$0.12    & 9.55$\pm$0.05  	 & 9.37$\pm$0.12    &10.01$\pm$0.13		& 9.06$\pm$0.47   & 0.25    & 0.78 \\
J111857+123441   & 9.53$\pm$0.12  & 9.01$\pm$0.04    & 9.37$\pm$0.02  	 & 9.22$\pm$0.04    & 9.41$\pm$0.07		& 8.87$\pm$0.37   & 0.10    & 1.61 \\
J112338+052038   & 8.92$\pm$0.09  & 9.12$\pm$0.04    & 9.15$\pm$0.04  	 & 9.08$\pm$0.04    & 9.28$\pm$0.06		& 8.78$\pm$0.46   & 0.37    & 0.40 \\
J115901+065619   & 9.60$\pm$0.11  & 9.68$\pm$0.03    & 9.28$\pm$0.09  	 & 9.81$\pm$0.03    & 9.18$\pm$0.04		& 8.93$\pm$0.57   & 0.12    & 1.37 \\
J120301+063441   & 9.41$\pm$0.10  & 8.60$\pm$0.09    & 9.00$\pm$0.06  	 & 8.96$\pm$0.09    & 9.04$\pm$0.10		& 8.67$\pm$0.37   & 0.32    & 0.50 \\
J121255+245332   & 9.31$\pm$0.13  & 9.49$\pm$0.11    & 9.44$\pm$0.13  	 & 9.02$\pm$0.11    & 9.13$\pm$0.12		& 9.00$\pm$0.47   & 0.13    & 1.37 \\
J121911$-$004345 & 9.23$\pm$0.09  & 9.47$\pm$0.04    & 9.58$\pm$0.05  	 & 9.11$\pm$0.04    & 9.14$\pm$0.05		& 9.12$\pm$0.48   & 0.36    & 0.46 \\
J122836+101841   & 9.41$\pm$0.10  & 9.11$\pm$0.14    & 9.27$\pm$0.06  	 & 9.02$\pm$0.14    & 9.59$\pm$0.15		& 8.88$\pm$0.47   & 0.27    & 0.63 \\
J133932$-$031706 & 9.58$\pm$0.11  & 9.39$\pm$0.26    & 9.62$\pm$0.07  	 & 9.58$\pm$0.26    & 9.85$\pm$0.27		& 9.01$\pm$0.46   & 0.16    & 1.12 \\
J140445$-$013021 & 9.45$\pm$0.11  & 8.99$\pm$0.17    & 9.62$\pm$0.14  	 & 9.19$\pm$0.07    & 9.31$\pm$0.07		& 8.99$\pm$0.46   & 0.31    & 0.61 \\
J141906+055501   & 9.59$\pm$0.13  & 9.27$\pm$0.05    & 9.28$\pm$0.06  	 & 9.38$\pm$0.05    & 9.58$\pm$0.07		& 8.83$\pm$0.46   & 0.04    & 3.98 \\
J143331+190711   & 9.56$\pm$0.12  & 9.10$\pm$0.08    & 9.39$\pm$0.03  	 & 9.56$\pm$0.08    & 9.59$\pm$0.10		& 8.91$\pm$0.36   & 0.12    & 1.43 \\
J145301+103617   & 9.27$\pm$0.10  & 9.27$\pm$0.08    & 9.22$\pm$0.07  	 & 9.14$\pm$0.08    & 9.86$\pm$0.10		& 8.81$\pm$0.47   & 0.16    & 1.00 \\
J151508+213345   & 9.80$\pm$0.11  & 9.08$\pm$0.05    & 9.59$\pm$0.03  	 & 9.20$\pm$0.05    & 9.34$\pm$0.07		& 8.95$\pm$0.38   & 0.15    & 1.22 \\
J153124+075431   & 9.05$\pm$0.10  & 8.90$\pm$0.15    & 9.18$\pm$0.05  	 & 8.70$\pm$0.15    & 8.85$\pm$0.15		& 8.77$\pm$0.36   & 0.18    & 0.98 \\
J153727+231826   & 9.87$\pm$0.13  & 9.70$\pm$0.26    & 9.47$\pm$0.25  	 & 8.97$\pm$0.26    & 9.45$\pm$0.26		& 9.00$\pm$0.55   & 0.14    & 1.31 \\
J153925+160400   & 8.92$\pm$0.11  & 9.21$\pm$0.14    & 9.38$\pm$0.11  	 & 9.09$\pm$0.14    & 9.11$\pm$0.14		& 8.85$\pm$0.47   & 0.27    & 0.63 \\
J154459+040746   & 9.58$\pm$0.11  & 8.68$\pm$0.10    & 9.24$\pm$0.04  	 & 9.11$\pm$0.10    & 8.93$\pm$0.11		& 8.79$\pm$0.36   & 0.16    & 1.05 \\
J160016+183830   & 9.33$\pm$0.14  & 9.28$\pm$0.05    & 9.28$\pm$0.03  	 & 9.31$\pm$0.05    & 9.43$\pm$0.07		& 8.94$\pm$0.46   & 0.25    & 0.71 \\
J160154+135710   & 9.55$\pm$0.10  & 9.01$\pm$0.05    & 9.27$\pm$0.03  	 & 9.08$\pm$0.05    & 9.50$\pm$0.08		& 8.87$\pm$0.35   & 0.22    & 0.77 \\
J160212+241010   & 9.56$\pm$0.11  & 9.15$\pm$0.05    & 9.43$\pm$0.03  	 & 9.13$\pm$0.05    & 9.35$\pm$0.06		& 8.92$\pm$0.44   & 0.25    & 0.77 \\
J230011$-$102144 & 9.38$\pm$0.10  & 9.58$\pm$0.12    & 9.60$\pm$0.08  	 & 9.15$\pm$0.12    & 9.45$\pm$0.12		& 9.12$\pm$0.43   & 0.25    & 0.68 \\
J231607+010012   & 9.35$\pm$0.10  & 9.80$\pm$0.07    & 9.72$\pm$0.10  	 & 9.54$\pm$0.07    & 9.31$\pm$0.07		& 9.16$\pm$0.52   & 0.42    & 0.47 \\
\hline
\multicolumn{9}{l}{Non-CARLA quasars having $\mathrm{M_{BH}}$(\civ) $>10^{10}$ \msol}\\
\hline
J005814+011530   & 9.36$\pm$0.10  &10.06$\pm$0.05    & 9.84$\pm$0.08  	 & 9.42$\pm$0.05    & 9.49$\pm$0.06		& 9.30$\pm$0.54   & 0.63    & 0.30 \\
J081014+204021   & 9.54$\pm$0.10  &10.21$\pm$0.09    & 9.95$\pm$0.06  	 & 9.68$\pm$0.09    & 9.62$\pm$0.09		& 9.40$\pm$0.56   & 0.62    & 0.31 \\
J115301+215117   &10.01$\pm$0.11  &10.13$\pm$0.03    &10.10$\pm$0.04 	 & 9.75$\pm$0.02    & 9.68$\pm$0.04		& 9.45$\pm$0.56   & 0.28    & 0.71 \\
J130331+162146   & 9.41$\pm$0.11  &10.07$\pm$0.04    & 9.89$\pm$0.06  	 & 9.61$\pm$0.04    & 9.42$\pm$0.05		& 9.29$\pm$0.55   & 0.35    & 0.50 \\
J210831$-$063022 & 9.70$\pm$0.10  &10.35$\pm$0.04    & 9.99$\pm$0.05  	 & 9.62$\pm$0.04    & 9.46$\pm$0.06		& 9.41$\pm$0.41   & 0.39    & 0.47 \\
\hline
\hline
\end{tabular}
\end{scriptsize}
\end{table*}

\subsection{Trends with Radio Power}

Radio-loud quasars are known to harbor SMBHs that lie at the high-mass end of the mass function (\mbh$>10^8$ \msol), up to three orders of magnitude higher compared to typical radio-quiet quasars \citep{laor00,mclure01}. This correlation between radio-loudness and BH mass suggests that the properties of the SMBH play some role in the formation of the jet \citep{gu01,lacy01,chiaberge15}. Moreover, radio-loud AGN are more strongly clustered than radio-quiet AGN, suggesting that they are found in more massive halos \citep{overzier03,retana17}. 
This implies that radio-loudness alone is already a good indicator of the BH mass in any quasar sample \citep{retana17}. Our sample is characterized by powerful radio-loud quasars ($P_{\rm 500MHz}>10^{27.5}$ W Hz$^{-1}$) and extremely massive BHs (\mbh$>10^9$ \msol). In this section we investigate if the radio activity correlates with any of the other properties of the sample derived in the previous sections.

\begin{figure*}
\begin{center}
\includegraphics[height=5cm]{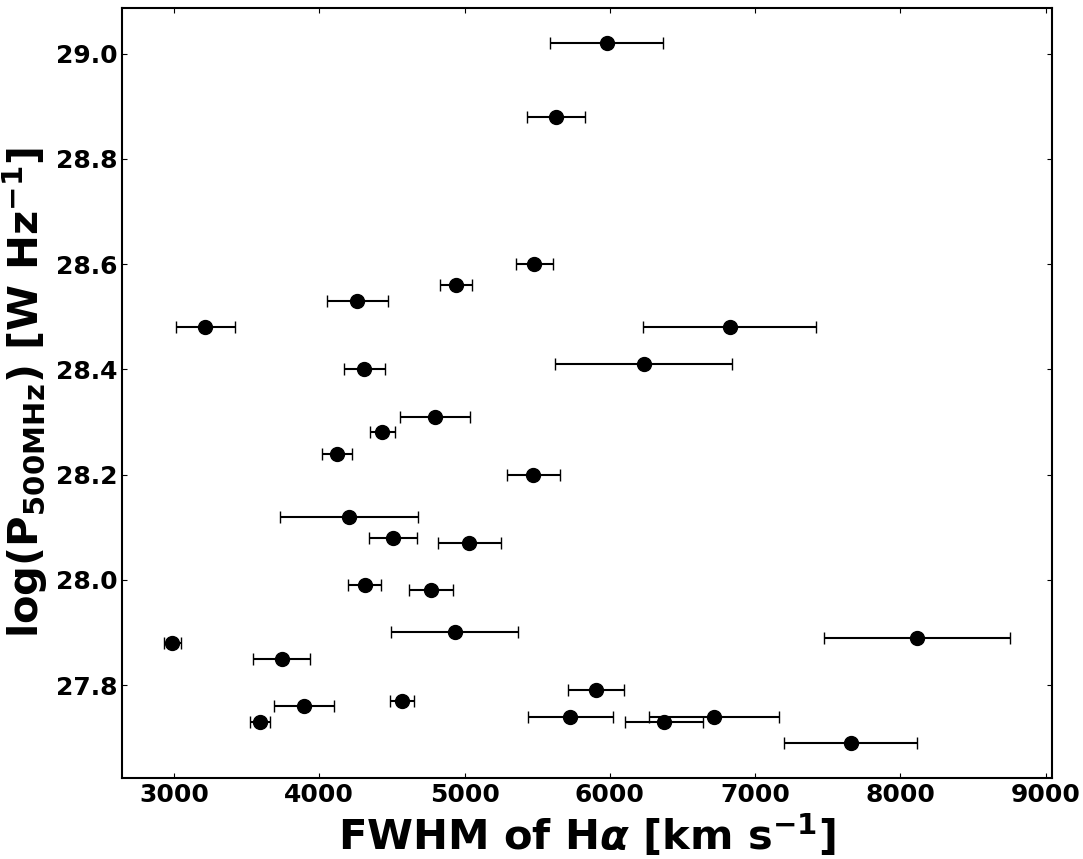}
\includegraphics[height=5cm]{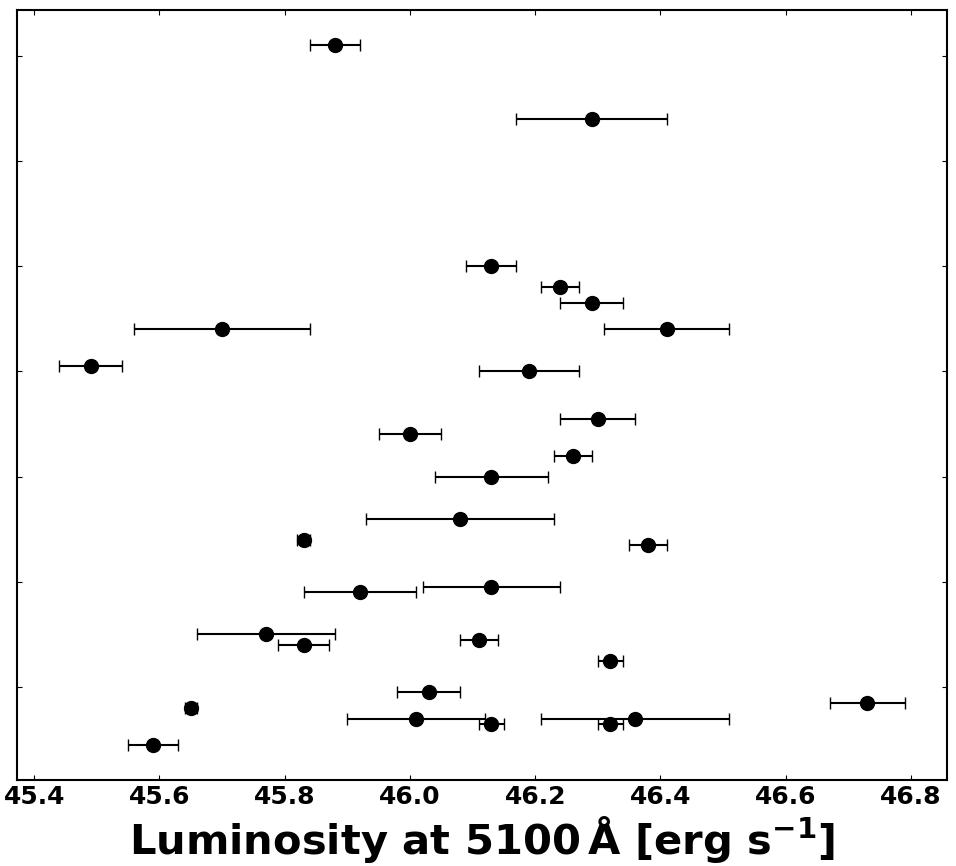}
\includegraphics[height=5cm]{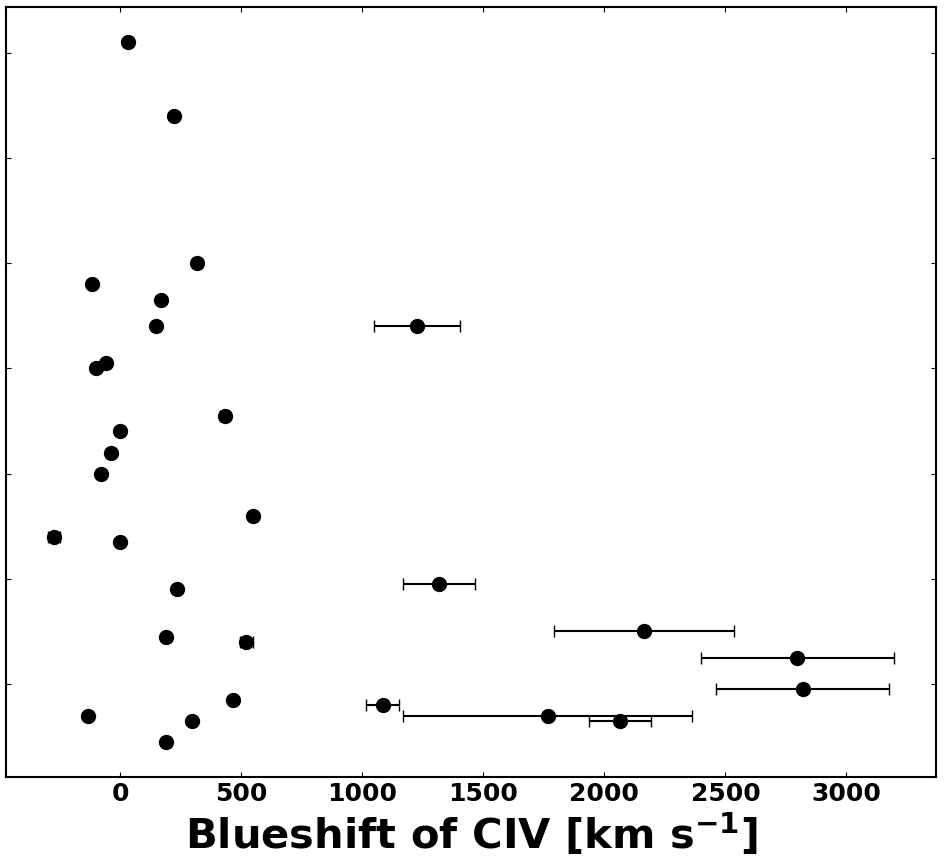}\\
\vspace{5mm}
\includegraphics[height=5cm]{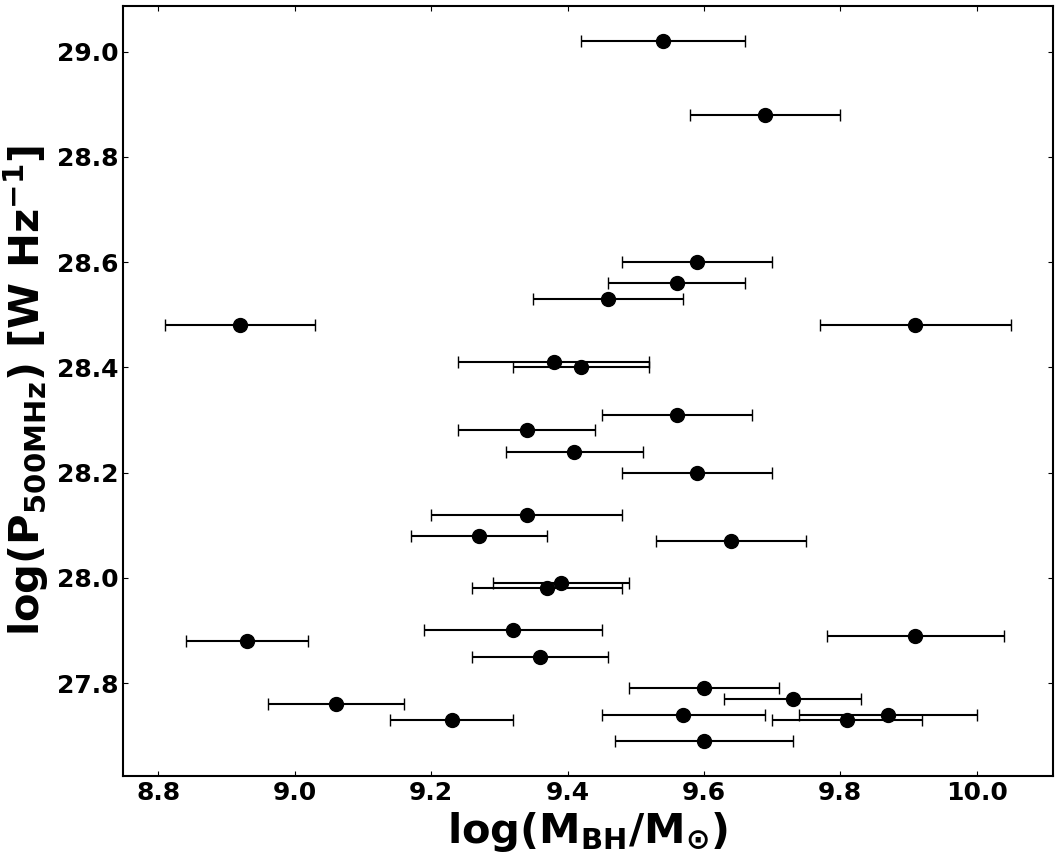}
\includegraphics[height=5cm]{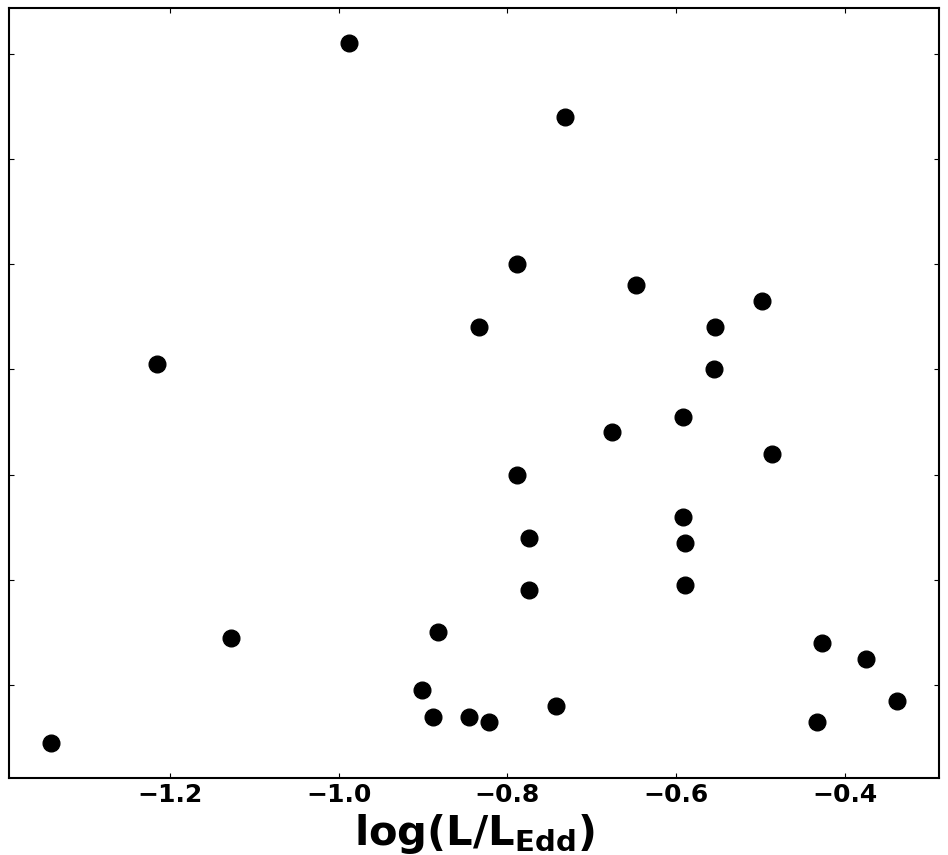}
\includegraphics[height=5cm]{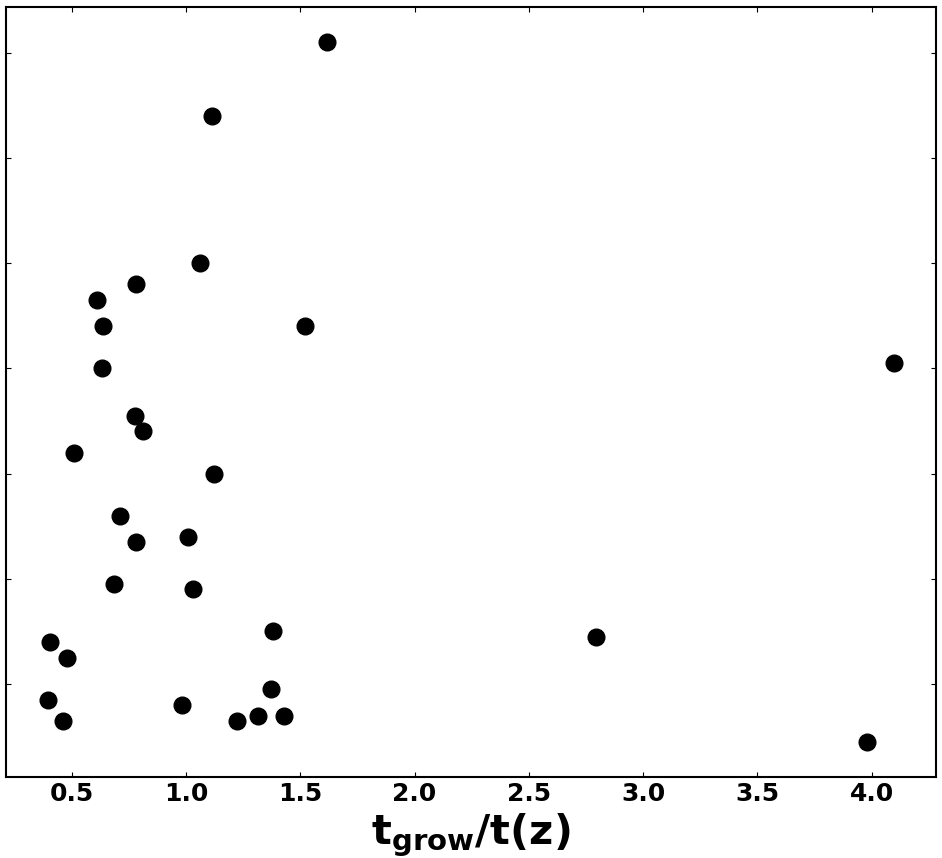}
\end{center}
\caption{Radio power versus other properties derived in this work for radio-loud quasars from CARLA. Top row of panels show \rp\ versus FWHM of \ha\ (left), optical luminosity (middle), and \civ\ blueshift (right). Bottom row of panels show \rp\ versus \mbh\ (left), \led\ (middle), and \gt\ (right).}
\label{fig:radio-corr}
\end{figure*}

We plot in Figure~\ref{fig:radio-corr} radio power versus FWHM(\ha), \lopt, blueshift of \civ, \mbh, \led, and \gt. We do not find any significant correlations between most of these properties and \rp, but we note that our sample spans only about one order of magnitude in radio power, optical luminosity and BH mass and is thus not very well suited to search for such correlations. We do not find a correlation between radio power and BH mass for the radio-loud CARLA quasars, but it must be noted that the range of radio luminosities in our sample is quite small compared to the dynamic range of radio-loud AGN in general (i.e., typically from $10^{24}$ to $>10^{29}$ W Hz$^{-1}$). It has previously been shown that there is a minimum threshold \mbh\ above which radio emission can be triggered \citep{laor00,magliocchetti04}.  Other recent work has suggested a strong correlation between the fraction of galaxies that are radio-loud and the mass of the BHs, but this relation is primarily driven by the correlation between radio-loudness and stellar mass \citep[][]{sabater19}. \citet{xiong13} found a strong correlation between \mbh\ and radio luminosity at 5 GHz for 97 radio-loud quasars at $z=0-2.2$. Their sample consisted of quasars having lower radio luminosities but equally massive BHs compared to the CARLA sample studied here. If the BH mass measurements of \mbh$\sim10^{10}$ \msol\ found for the small sample of very powerful broad line radio galaxies by \citet{nesvadba11} are representative for the population of radio galaxies as a whole, this would indeed suggest that the most powerful radio emission is associated with the most massive BHs. Although radio-quiet quasars exist with similarly high BH masses (e.g., Table \ref{tab:BHM_ha}), this could be explained by the episodic activity of the radio jet formation. 

Because \led\ and \gt\ are both properties of quasars that depend on the entire past history of accretion, it is perhaps not surprising that these properties do not correlate strongly with the (currently observed) radio power. The fact that the optical luminosity and accretion rate do not correlate with radio power further suggests that radio power has no influence on what happens on scales of the accretion disk. This does not mean that radio power is not a good proxy for the strength of possible feedback effects associated with the radio jets in these quasars on larger scales. \citet{hardcastle19}, for example, found a strong correlation between the radio power and kinetic luminosity density in a large sample of radio-loud AGN with radio luminosities up to $\sim10^{26}$ W\,Hz$^{-1}$ \citep[see also][]{sabater19}.

The strongest trend we find is a lack of large \civ\ blueshifts at the highest radio powers. The largest \civ\ blueshifts ($>$1500 km s$^{-1}$) are exclusively found for the lowest power ($<10^{28}$ W Hz$^{-1}$) radio sources (top-right panel of Figure~\ref{fig:radio-corr}). This general trend is furthermore consistent with our finding that the 5 radio-quiet quasars that are not in CARLA have some of the highest \civ\ blueshifts (Table \ref{tab:fitting}). It is likely that this trend is mainly driven by a correlation between the magnitude of the blueshift of \civ\ and Eddington ratio, as shown in Figure~\ref{fig:bs-led}. Although there is a large amount of scatter in the size of the blueshift at any given Eddington ratio, the median blueshift increases with Eddington ratio for radio-quiet and radio-loud objects alike (the median blueshift is about 2000 km s$^{-1}$ for \led\ of $\sim1$, while it is about 0 km s$^{-1}$ for \led\ of $\sim0.1$). It is well-known that larger \civ\ blueshifts tend to occur in brighter and higher \led\ quasars \citep{richards11}. Because high redshift quasar samples are furthermore skewed toward higher luminosities compared to low redshift quasar samples \citep[e.g.][]{mazzucchelli17}, this could explain the large blueshifts observed in the CARLA sample and is consistent with the fact that we see the strongest blueshifts for the objects with the lowest radio luminosities, as these also tend to have the highest optical luminosities. Although it is likely that the radio core luminosities would correlate with optical luminosity, the CARLA sample was constructed at low radio frequencies where the emission from the jet/lobes dominates rather than from the core.

One way to interpret these observations is that as the mass accretion rate onto the SMBH (i.e., the Eddington ratio \led) increases, the power and velocity of the wind (the latter traced by the \civ\ blueshift) gets stronger, possibly due to the increased radiation pressure which drives the mass-loss from the accretion disk \citep[e.g.][]{proga98,zubovas12}. At the same time, an increase in the Eddington ratio leads to a change in the structure of the accretion disk, making it thinner and thereby suppressing the formation of a relativistic jet \citep[e.g.][]{meier01,avara16}. Such jet suppression as due to a change in the accretion state is widely observed in stellar mass BHs \citep[e.g.][]{fender12}. The upper right panel of Figure~\ref{fig:radio-corr} shows that near-zero blueshifts occur at all radio powers, while the largest blueshifts ($>1500$\,km\,s$^{-1}$) are exclusively found for the lowest radio power sources. This result, combined with the correlation between blueshift and \led\ shown in Figure~\ref{fig:bs-led}, suggests a scenario in which as the accretion rate increases, the outflow as traced by the \civ-blueshift becomes stronger (perhaps due to an increase in radiation pressure), leading to a quenching of the relativistic radio jets. Further discussion of this topic is outside the scope of this work.

\section{Discussion}

\subsection{\civ\ versus \ha-based mass estimates}

Among the methods used to estimate the \mbh\ of quasars from SE spectra, the most reliable are those based on the Balmer lines for at least three reasons: (i) most RM studies were done using the continuum luminosity at 5100 \AA\ and \hb; (ii) the Balmer lines are `well behaved' lines, showing highly symmetrical profiles; and (iii) the Balmer lines are located in a spectral region with low contamination of other lines. In Section~4.1 we used $K$-band spectroscopy of CARLA quasars to estimate \mbh\ using the \ha\ line, thereby obtaining new estimates that should be significantly more reliable than the previous ones that were based on \civ. We found that the BH masses obtained using \ha\ are significantly different from those based on \civ. One important check that needs to be performed on \ha-based masses is to make sure they are not affected by double peaked emission lines. \citet{jun17} warn that lines with a FWHM in excess of 8000\,km\,s$^{-1}$ could be indicative of double-peaked lines which could bias the \mbh\ estimates. Their analysis of 26 SDSS quasars at $0.7<z<2.5$ with extremely high BH masses (\mbh$>10^{9.5}$ \msol) showed that in seven quasars a double peak was present (five of which had FWHM$>8000$\,km\,s$^{-1}$). In our sample, four out of 35 quasars have FHWM(\ha)$>8000$\,km\,s$^{-1}$, but we do not find any evidence for double peaks based on the high level of symmetry of the lines. 

\begin{figure}
\includegraphics[width=\columnwidth]{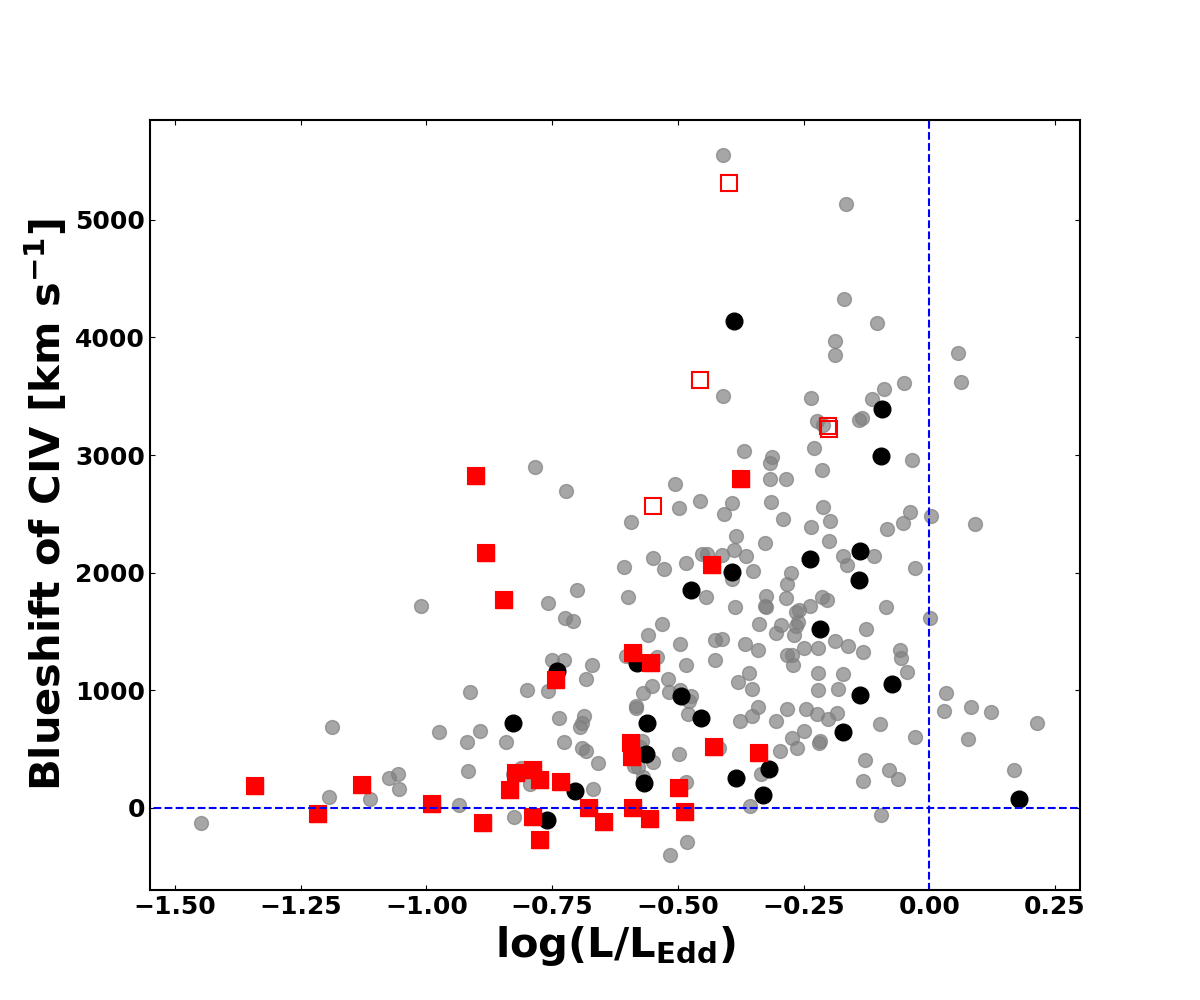}
\caption{Blueshift of \civ\ as a function of Eddington ratio. Red filled and empty squares, respectively, show the CARLA and non-CARLA quasars observed in this work. Black circles (grey circles) show the radio-loud (radio-quiet) quasars from \citet{coatman17}. Vertical and horizontal blue dashed lines indicate the Eddington limit and a zero blueshift of \civ, respectively.}
\label{fig:bs-led}
 \end{figure}

The asymmetric shape of \civ\ is one of the main factors causing the problems in the \civ-based methods. The presence of a non-reverberating component of \civ\ would imply that not all of the flux observed comes from the virialized region. \citet{denney12} show that this component is often present, and their correction is based on the correlation between the non-variable component of the line and the \civ\ shape profile. The non-reverberating component of \civ\ is also associated with outflows and winds originating from the very central region of the AGN \citep{bachev04,baskin05}. \citet{coatman16} found for a small sample of quasars that the large blueshift of \civ\ is present in sources accreting around the Eddington limit. We therefore plot in Figure~\ref{fig:bs-led} the blueshift of \civ\ as a function of the Eddington ratio. There is a weak correlation ($\rho=0.40$ with $p=0.002$) between these quantities, showing that the non-variable component of \civ\ may indeed be associated with a non-reverberating component possibly related to feedback from the AGN. 

Among the various methods for \civ\ rehabilitation that we tested in Section 4.1 (see Table \ref{tab:rehab}), the \citet{denney12} and \citet{coatman17} methods best reproduced the BH masses estimated from \ha\ for our sample of radio-loud quasars. However, as pointed out by \citet{park17} (see their Figure~11), the \citet{coatman17} method has the potential to overestimate \mbh\ when the \civ\ blueshift is negative. \citet{coatman17} warn about extrapolating their method in the negative blueshift regime, since their sample does not cover a large dynamic range. However, both methods have the advantage over, for example, the \citet{runnoe13} correction method, because the $S/N$ of the 1400 \AA\ line complex is usually faint in high-redshift sources and negative \civ\ blueshifts are furthermore rare. 

Other factors may contribute to the scatter between the \civ\ and Balmer methods such as differences in the mass scale relationships for each line \citep{mcgill08}. Another common source of error is contamination of \hb. Most studies focus on the problems with \civ\ since \hb\ has been studied in detail in RM studies. However, this line has its own problems such as a strong narrow line component (although not important in our sample), host galaxy contamination, and a strong underlying Fe\,{\sc ii} pseudo-continuum \citep{bentz09,park12}. These problems are worse in low $S/N$ spectra \citep{assef11}.

\begin{figure*}
\includegraphics[width=0.48\textwidth]{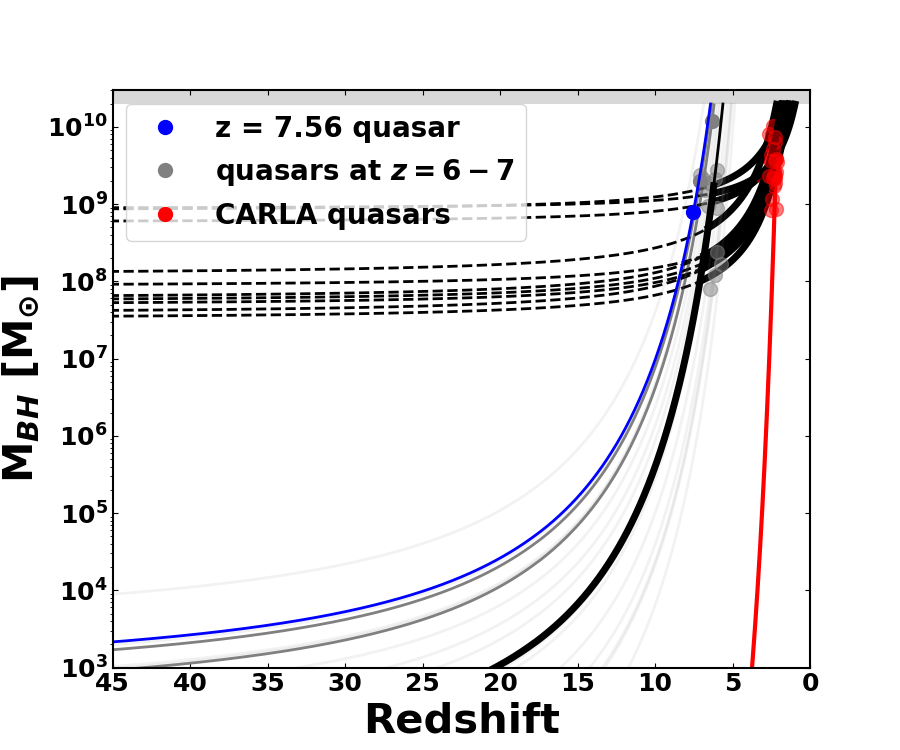}
\includegraphics[width=0.48\textwidth]{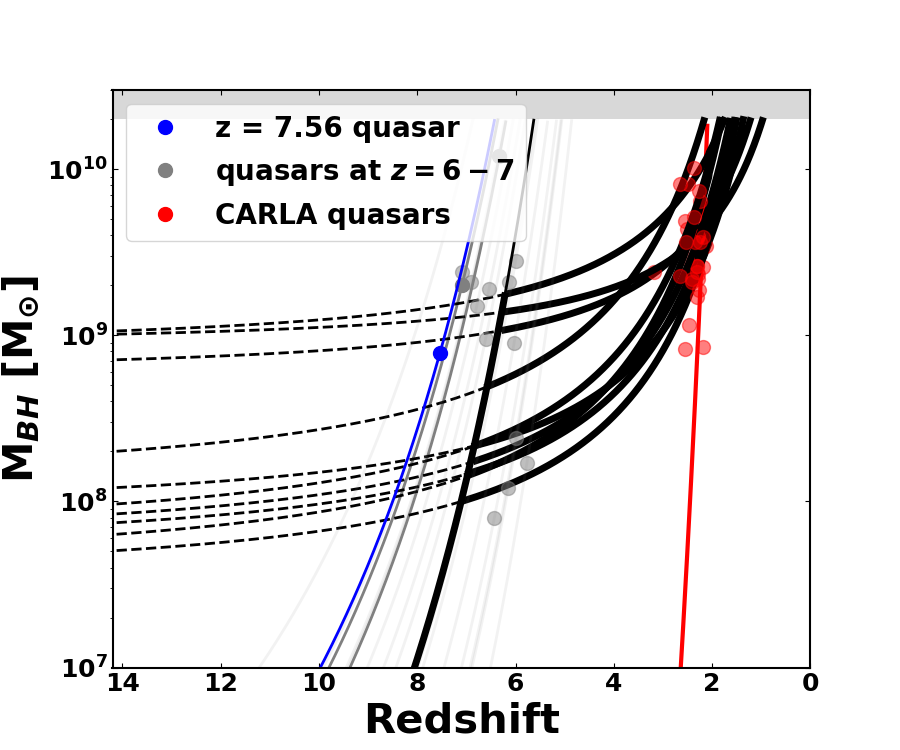}
\caption{BH masses and possible evolutionary tracks for high-redshift quasars in the $z=0-45$ (left-hand panel) and the $z=0-15$ redshift range (right-hand panel). Red filled circles at $z\simeq2-3$ show the CARLA quasars from this paper. The blue circle shows the quasar at $z=7.56$ from \citet{banados18}, and the grey circles a sample of quasars at $z=6-7$ \citep{gallerani17}. Lines trace the possible growth histories of these SMBHs. The blue line shows the Eddington-limited growth of a 1000 \msol\ seed at $z\gtrsim45$ that is consistent with the $z=7.56$ quasar (blue circle) from \citet{banados18}, while the grey lines show the same tracks for $z\sim6-7$ quasars (grey circles). Black dashed lines trace the masses of CARLA quasars back in time assuming they always grew at their observed Eddington ratios. These tracks require extremely large seed masses at high redshifts that are likely not physical (i.e., $>10^7$ \msol\ at $z\sim45$). The solid black lines assume Eddington-limited growth that can produce the population of $\gtrsim10^8$ \msol\ BHs observed at $z\simeq6-7$ (grey circles). Extending these curves to $z\sim2-3$ at the Eddington ratios measured for the CARLA quasars (red circles) can reproduce their measured BH masses (red circles). The red line shows a track of Eddington-limited growth starting from a $\sim1000$ \msol\ seed at $z<5$, which can also reproduce the BH masses of the CARLA quasars. The grey shaded region marks masses above the empirical limit of $\sim2\times10^{10}$ \msol.}
 \label{fig:growth-track}
\end{figure*}

\subsection{The Growth History of CARLA Quasars}
\label{sec:history}

In this work we obtained robust BH masses for 35 quasars at $z\sim2$, finding masses ranging from $\sim$10$^9$\,\msol\ to $\sim10^{10}$\,\msol. The results in Sections~4.2 and 4.3 show that these quasars are actively accreting, and have likely been growing for a substantial fraction of cosmic time. We have also shown that at fixed accretion rates, the growth times for radio-loud quasars tend to be longer than those of radio-quiet quasars, but this is mainly a consequence of the fact that the BH masses of radio-loud quasars are higher. In this section we explore a few simple evolutionary scenarios that may approximate the growth history of the CARLA quasars (Figure \ref{fig:growth-track}). 

The first scenario we consider is that the CARLA quasars have always accreted at their \led\ measured at $z\sim2-3$. These are indicated by the dashed tracks in Figure \ref{fig:growth-track}. These tracks are much too shallow (i.e., they rise too slowly with decreasing redshift), requiring the existence of seeds that are likely much too massive ($\gtrsim10^7$ \msol) too early ($z\gtrsim45$). This is a consequence of the long growth times we found for a significant fraction of the CARLA quasars. These growth times are often longer than the age of the universe (much longer in some cases). It is worth pointing out that this discrepancy is exacerbated by the fact that the values estimated here are actually a lower limit for the growth time. For example, an $f_{active}$ smaller than unity, a lower seed mass or a lower accretion rate would result in even longer growth times. Similar results were found by \citet{netzer07} for a significant fraction of quasars at $z=2.3-3.4$. 

The second, extreme scenario we consider is that the CARLA quasars grew at or near the Eddington limit for most of their lifetime (red solid track). This scenario easily reproduces the observed masses from relatively small seeds ($\sim1000$ \msol) at relatively late times ($z\lesssim6$), but probably is also not very realistic. First of all, it does not agree with the much lower values of \led\ measured for these CARLA quasars, but this is not such an important issue given that the accretion rate can easily fluctuate. A more serious objection to this scenario is that it assumes that $\sim1000$ \msol\ seeds are lingering at the centers of the quasar host galaxies as recent as $z\sim3$, even though those host galaxies themselves have already reached masses of the order of $\sim10^{10-12}$ \msol. 

The last scenario we consider here is that the CARLA quasars are direct descendants of the objects hosting the most distant SMBHs known at $z\simeq6-8$ \citep[e.g.,][]{mortlock11,wu15,banados18}. \citet{banados18} analysed the possible formation history of a quasar at $z=7.56$ with a SMBH of $\sim8\times10^{8}$ \msol\ (blue circle and blue line in Figure~\ref{fig:growth-track}). They presented an evolutionary track for a seed mass of 1000 \msol\ at $z=40$ growing at the Eddington limit with an efficiency of 10\%. In Figure \ref{fig:growth-track} we show this evolutionary track (blue line). Similarly, we indicate Eddington-limited tracks (grey lines) for a subset of other quasars at $z\sim6-7$ (grey circles) from \citet{gallerani17}. If we want to force a scenario in which the lower redshift SMBHs in CARLA quasars are the direct descendants of those in the higher redshift quasars, the latter must either drastically reduce their accretion rates after $z\sim6$ (assuming a constant efficiency, $\eta$) or grow more intermittently by decreasing their duty cycle \citep[see][]{trakhtenbrot12}. Interestingly, we can easily unite the two populations by assuming that the SMBHs first grow at or near the Eddington limit reaching about 10\% of their observed masses by $z\simeq6-8$, followed by a phase of slower accretion given by the actual Eddington ratios measured for the CARLA quasars at $z\sim2-3$ (thick black curves in Figure \ref{fig:growth-track}). Such a fast accretion phase at close or equal to the Eddington limit of BHs at early times, followed by a slower accretion phase at later times, is also supported by the results of \citet{marconi04}. This scenario is also consistent with the recent theoretical results by \citet{mcalpine18} based on the simulations of \citet{schaye15}. These authors showed that BHs only spend $\sim15\%$ ($\sim1.4Gyr$) of their active phase in a state of rapid accretion, which is close to the age of the universe at $z\sim6-7$. Most of their active phase is spent in the AGN-feedback regulated state with lower accretion rates. 

It is important to note that a substantial fraction of the CARLA quasars would still exceed the empirical upper limit on BH mass of $\sim2\times10^{10}$ \msol\ by $z\sim1-3$ even at their observed (sub-Eddington) accretion rates (black solid lines in Figure~\ref{fig:growth-track}). A further quenching of the BH activity will thus be required in order not to exceed this limit \citep[e.g., see][for a discussion of the limit black hole mass]{king16,inayoshi16}. The mechanism to accomplish this may be readily available in the CARLA objects, given that they all host very powerful radio jets, believed to be important agents that could provide such feedback. Our findings for the CARLA radio-loud quasars are also broadly consistent with the results found for the CARLA radio galaxies. \citet{falkendal19} derived BH accretion and star formation rates for radio galaxies at $z\sim1-5$ \citep[see also][]{drouart14}, concluding that while their star formation appears to have been quenched (i.e., they lie below the star-forming main sequence relation), their BHs are still rapidly accreting. Coupled with the high BH masses measured by \citet{nesvadba11} for a small subsample and the similarly high BH masses of \mbh$\sim10^{9-10}$ \msol\ assumed for the remainder of the sample of radio galaxies, this implies that both their BH accretion rates will have to come down fast while their host galaxies will still have to increase their stellar mass in order to not violate the local $M_{BH}-\sigma_*$ relation. This stellar mass growth could be accomplished mainly through dry mergers as this ensures that no further star formation or BH accretion occurs\footnote{BH growth may still occur through binary BH mergers} \citep{falkendal19}. 

Focusing again on the earliest, very rapid phase of the BH growth that occurred at $z\gtrsim6$, the calculations thus far considered assume that this growth is through efficient accretion only. However, theoretical modeling shows that, in fact, both BH mergers and radiatively (in)efficient accretion are necessary ingredients \citep{volonteri07,volonteri08}. Such models show that a $10^9$ \msol\ SMBH can be built in 100 Myr if the efficient accretion regime starts as soon as the seed reaches 10\%\ of the final desired mass of the \mbh. In this scenario, mergers could drive the BH growth up to the 10\% seed mass, after which accretion of gas associated with the mergers takes over the growth. However, it is still an open question whether the necessary number of BH-hosting halos indeed occurs at these high redshifts \citep{regan19}. 
In terms of the overall BH mass growth, this early-merging scenario is not very different quantitatively from the fast-followed-by-slow-accretion scenario considered above (black solid tracks in Figure \ref{fig:growth-track}). This scenario also agrees with the results of \citet{merloni04} and \citet{trakhtenbrot12}, who showed that \mbh$>10^9$ \msol\ BHs predominantly formed during high Eddington rate growth spurts at $z\gtrsim2$. These BHs are likely to be among the progenitors of the relic BHs found at the centers of local clusters \citep{McConnell11}. If the most massive SMBHs are furthermore preferentially located in the most massive halos (see Section 5.3), they may have experienced a relatively high merger rate at early times. These mergers would increase the BH spin \citep{orsi16}, which would favor both the accretion rate and the formation of radio jets \citep{cattaneo02,volonteri07}. Although the results shown in Figure~\ref{fig:led-corr} point to a general scenario where the more massive BHs in our sample typically have lower accretion rates than the less massive BHs, there is substantial scatter with some of the most massive BHs also having some of the highest accretion rates. These cases could be explained by massive BHs experiencing a secondary fast accretion event, perhaps as the result of a recent major merger in which the gas supply of a quasar or inactive BH is momentarily replenished. 

\begin{figure*}
\begin{center}
\includegraphics[width=0.45\textwidth]{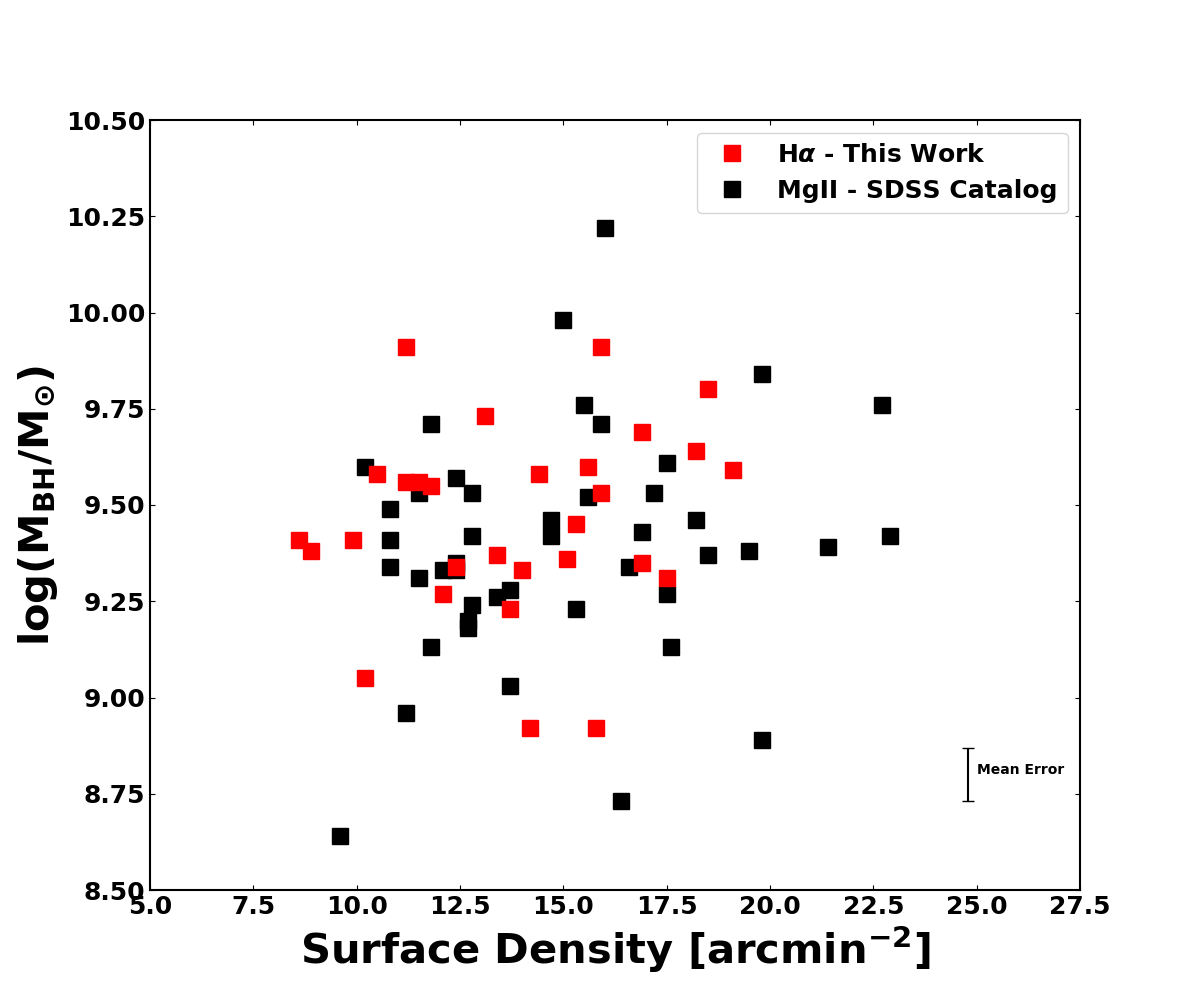}
\includegraphics[width=0.45\textwidth]{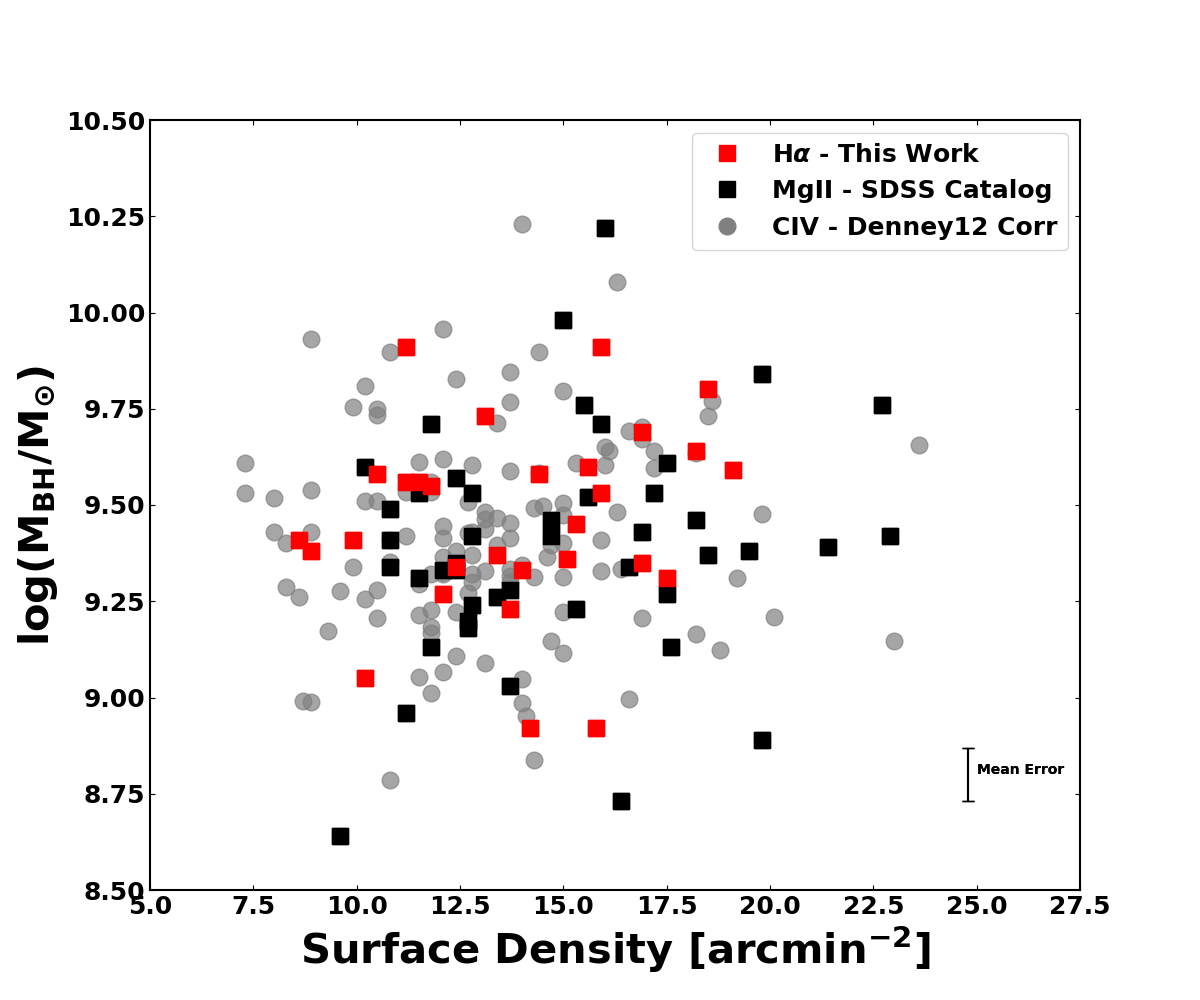}
\end{center}
\caption{Galaxy surface density as measured by \citet{wylezalek13} versus BH mass for the CARLA quasars studied in this paper. The left-hand panel shows the \ha-based \mbh\ obtained in this paper (red squares) complemented with CARLA quasars for which \mgii-based masses exist (black squares). In the right-hand panel we have added the \civ-based \mbh\ corrected following the \citet{denney12} rehabilitation method (see Section \ref{blackholes}) in order to obtain the largest sample of CARLA quasars having both accurate \mbh\ and density measurements.}
\label{fig:surface}
\end{figure*}

\subsection{Does Black Hole Mass Correlate with Environment?}

The main motivation for the CARLA project was to systematically study the environments of radio-loud AGN (radio galaxies and radio-loud quasars) at high redshifts \citep{galametz12,wylezalek13,hatch14}. One of the major results from this project was a clear excess in the number of (projected) galaxy counts for CARLA sources compared to carefully constructed control samples of radio-quiet quasars and inactive galaxies of similar stellar mass. \citet{hatch14} showed a $4\sigma$ significance effect that radio-loud AGN from CARLA reside in denser environments compared to the control samples. The environmental parameter used in this study is, by necessity, defined as the surface density of color-selected galaxies measured within a radius of 1 arcminute ($\sim$0.5 Mpc at $1.3<z<3.2$) away from the CARLA quasars. The redshift accuracy of the color-selection is not very high, but by analyzing a large sample the effects of for- and background interlopers are expected to average out. Moreover, \citet{wylezalek13} showed that the overdensities are centered on the radio-loud AGN, and that the overdensity distribution is similar for the type 1 (i.e, radio quasars) and type 2 (i.e, radio galaxies) radio-loud AGN. The fact that the environmental measure does not correlate with any of the radio properties of the sample (radio size, luminosity or spectral index), led \citet{hatch14} to the conclusion that it is the triggering of the radio jets that somehow depends on the environment of the host galaxies. It has been suggested that denser environments are conducive to higher merger rates between galaxies and thus of their BHs. This will in turn increase BH spins that are needed to launch powerful jets. At the same time, the enhanced mergers or gas inflow rates in denser environments lead to higher accretion rates, which boost the luminosity of these active BHs. One must be careful, however, in drawing such conclusion, as an alternative explanation could be that radio jets propagating through a denser medium have a higher working surface, and may therefore be more easily detected in radio surveys.

The existence of a strong $\mathrm{M_{BH}}-\sigma_*$ relation at the present-day demonstrates that at some point in their evolution, BH masses began to correlate with galaxy bulge masses. Because the most massive SMBHs locally are furthermore often found in the massive central galaxies in galaxy clusters, we also expect an (albeit indirect) correlation between environment and BH mass. The CARLA project traces some of the most massive SMBHs at high redshift that are found in, on average, high density environments suggestive of (proto)clusters of galaxies \citep{wylezalek13,overzier16}. This sample therefore offers a unique chance to investigate at what epochs the correlation between BH mass and environment may have been established. \citet{hatch14} tried to address this question by analysing if radio-loud quasars from CARLA with more massive SMBHs are found in denser environments. Although a weak correlation was found using the $\sim200$ quasars studied by CARLA (Spearman rank coefficient $\rho$ of 0.16 with $p$-value of 0.02), that study was based on \civ\ BH mass measurements. 

We have shown that \civ-based BH masses are less reliable than those based on \ha. In order to further investigate whether BH mass correlates with local galaxy density at $z\sim2-3$, we plot in the left-hand panel of Figure~\ref{fig:surface} the \ha-based \mbh\ determined in Section 4.1 against the galaxy surface densities for the 30 CARLA quasars observed as part of this work (red squares). In order to increase the statistics, we have also added those CARLA quasars for which reliable \mgii-based \mbh\ exists from SDSS (39 objects; black squares). The left panel shows that there is indeed a possible trend for more massive SMBHs to be found in denser environments, with a Spearman rank correlation of $\rho=0.16$ if we combine the \ha\ and \mgii\ data, but it is not significant ($p=0.16$). When we substitute the \ha-based masses for their original \civ-based masses used in \citet{hatch14}, the correlation coefficient is somewhat weaker than before, but again with a high $p=0.21$ (not shown). Despite the fact that we do not find conclusive evidence for a positive correlation between BH mass and environment, we have shown that the distribution of BH masses changes significantly between the \civ- and \ha-based measurements (Figure \ref{fig:BHM_ha}). This shows that this kind of analysis based on \ha\ should ideally be extended to the full CARLA sample before we can make a proper assessment of the question whether BH mass correlates with environment. Although the sample size of \ha-based BH masses for CARLA quasars is currently very limited, a possible improvement is to use the \civ-rehabilitation method by \citet{denney12} applied to the SDSS spectra of all CARLA quasars. As shown in Fig. \ref{fig:BHM_corr}), this offers a substantial improvement over the direct \civ-based measurements (0.24 dex with almost no systematic offset). We were able to apply the rehabilitation method to 125 additional CARLA quasars, and added the results to the right-hand panel of Figure~\ref{fig:surface}. Using this increased sample size of 199 quasars (\ha, \mgii, or \civ-rehabilitated \mbh), we find that the \citep{wylezalek13} surface densities and \mbh\ still correlate only weakly with $\rho=0.12$ and $p=0.1$. 

Despite the improvements in BH mass estimates that our analysis has been able to offer, a serious remaining limitation of our analysis is that the surface density parameters estimated from the CARLA data suffer from projection effects and large redshift uncertainties. Besides accurate BH masses, improved density measurements should thus also be obtained in order to perform the kind of analysis presented in \citet{hatch14} and Figure~\ref{fig:surface} in the most accurate way possible. This will be the subject of future work. 

\section{Conclusions}
\label{conclusions}

We analyzed a sample of 30 high-redshift quasars from the CARLA survey. We obtained spectroscopic observations with VLT/SINFONI in the $K$-band in order to estimate their BH masses and study the correlation between BH mass and other physical properties. We used \ha\ to obtain more accurate BH masses and to estimate accretion rates and BH growth times. We then studied whether the properties of the quasars correlate with radio power and with the large-scale environmennt. The results are summarized as follows:

\begin{enumerate}

	\item We used the \ha\ line (redshifted to the $K$-band) to estimate more accurate BH masses for the quasars in the CARLA sample. The results show a large scatter (0.35 dex) and systematic offset (0.2 dex) between the masses obtained using \ha\ and previously available \civ-based masses from SDSS spectra. Although the average \mbh\ increases when using \ha, it also reduced the number of objects having \mbh\ estimates of $\sim10^{10}$ \msol\ when using \civ. 

	\item We used various recent methods proposed for rehabilitating measurements based on \civ. We analyzed which best reproduce the BH masses obtained based on \ha. We found that the methods of \citet{denney12} and \citet{coatman17} reduce significantly the scatter and sytematic offsets between the \mbh\ determinations, while the methods of \citet{park17} and \citet{runnoe13} do not reproduce the \mbh(\ha) very well, resulting in a systematic offset.

	\item We calculated \led\ for our sample and find a range of 0.04-0.64, and used this to estimate the typical growth times for each quasar using basic assumptions for accretion efficiency, seed mass, and duty cycle. Although the median growth time of the sample is shorter than the age of the universe at the redshifts of the quasars, nearly half of the objects have growth times longer than the cosmic age. 

	\item We compared the radio power of our quasars with all measured quantities, i.e, \mbh, FHWM(\ha), \lopt, \led, \gt, and \civ\ blueshift. Our results show an anti-correlation between the \civ\ blueshift and radio power. This trend is consistent with the known effect that \civ\ blueshifts are larger for higher luminosity and higher \led\ quasars.  

	\item We explored various simple evolutionary tracks for the BH growth history of CARLA BHs, finding that a scenario in which the BH first accretes at or near the Eddington limit until $z\sim6-8$ and then switches to a slower accretion rate given by the \led\ measured at $z\sim2-3$ perhaps best explains the observations. In this scenario, the CARLA quasars could be direct descendants of the SMBHs observed in quasars at $z\sim6-8$. In order to be also consistent with the empirical upper limit of \mbh$\sim2\times10^{10}$ \msol, many of the CARLA quasars will have to lower their measured accretion rates further toward lower redshifts. In any case, given their high \mbh\ at $z\sim2-3$, we predict that their SMBHs will be found near the upper end of the local $M_{BH}-\sigma_*$ relation by $z\sim0$. 

	\item In order to shed new light on the origin of the local $M_{BH}-\sigma_*$ relation, we repeated the analysis of \citet{hatch14} to study whether BH masses are correlated with environment already at $z\sim2-3$. We use the \ha-based BH masses and galaxy surface density measurements from the CARLA survey as a proxy for environment. We find a weak correlation at very low significance. In future work, it will be important to extend the sample size of \ha-based \mbh, and reduce the uncertainty of the environmental measurements. 

\end{enumerate}

\section*{Acknowledgments}

We thank Andrew Humphrey, Nicole Nesvadba and the anonymous referee for very helpful comments. This study was financed in part by the Coordena\c{c}\~o de Aperfei\c{c}oamento de Pessoal de N\'ivel Superior - Brasil (CAPES) - Finance Code 001. We are grateful for financial support from the Conselho Nacional de Desenvolvimento Cient\'ifico e Tecnol\'ogico, CAPES (grant 88881.156185/2017-01) and the S\~ao Paulo Research Foundation (FAPESP; grants 2017/01461-2 and 2018/02444-7). YTL was supported by CNPq grant 400738/2014-7. The work of DS was carried out at the Jet Propulsion Laboratory, California Institute of Technology, under a contract with NASA. Based on observations collected at the European Organisation for Astronomical Research in the Southern Hemisphere under ESO programmes 089.B-0433(A), 090.B-0674(A), 091.B-0112(A), 092.B-0565(A), 093.B-0084(A), 094.B-0105(A), and 095.B-0323(A).

\appendix
\section{\ha\ and \civ\ emission line fits}

\begin{figure*}
   \includegraphics[width=1.0\textwidth]{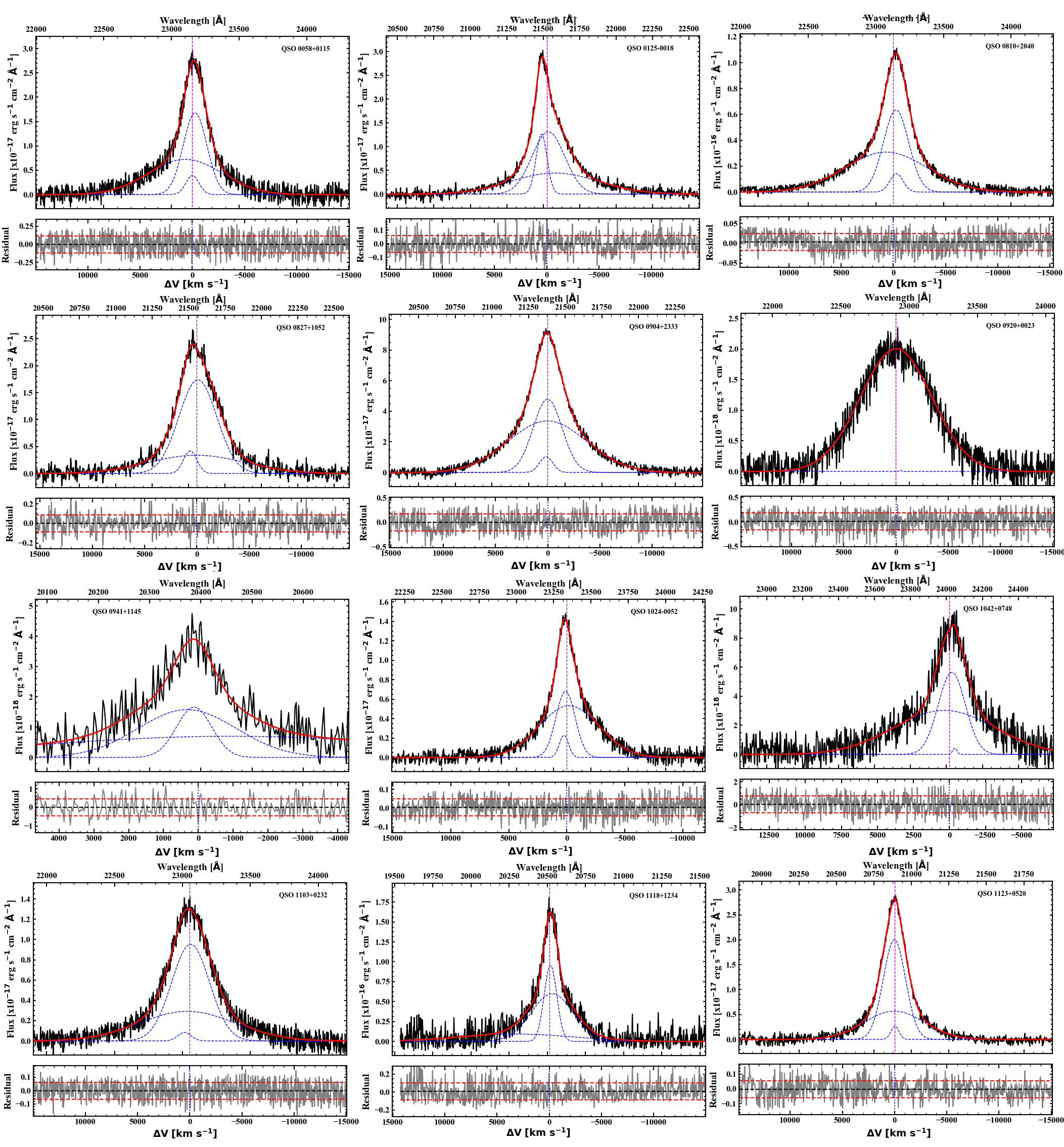}
    \caption{\ha\ fit for the entire CARLA sample. Top panels show the observed spectrum
    in black, the best fit components in dashed blue, and the sum of all components in red. 
    Bottom panels show the residual from the subtraction of the best fit from the observed spectrum
    in grey. The dashed black line shows the zero-level and the dashed red lines the standard deviation.
    Vertical dashed lines show the center of the line.}
   \label{fig:ha_fit1}
\end{figure*}

\begin{figure*}
   \includegraphics[width=1.0\textwidth]{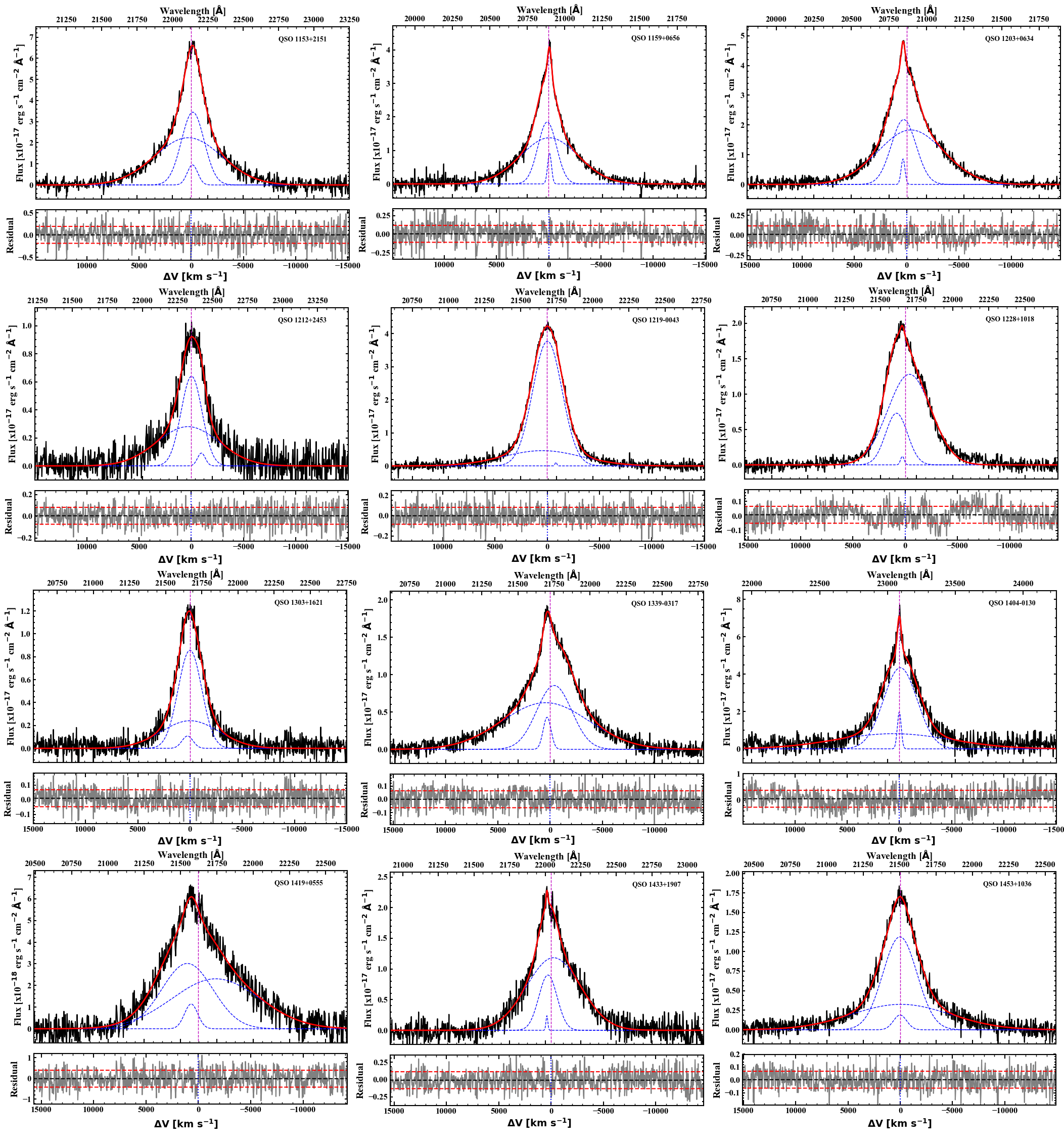}
    \caption{Continuation of Figure~\ref{fig:ha_fit1}.}
   \label{fig:ha_fit2}
\end{figure*}

\begin{figure*}
   \includegraphics[width=1.0\textwidth]{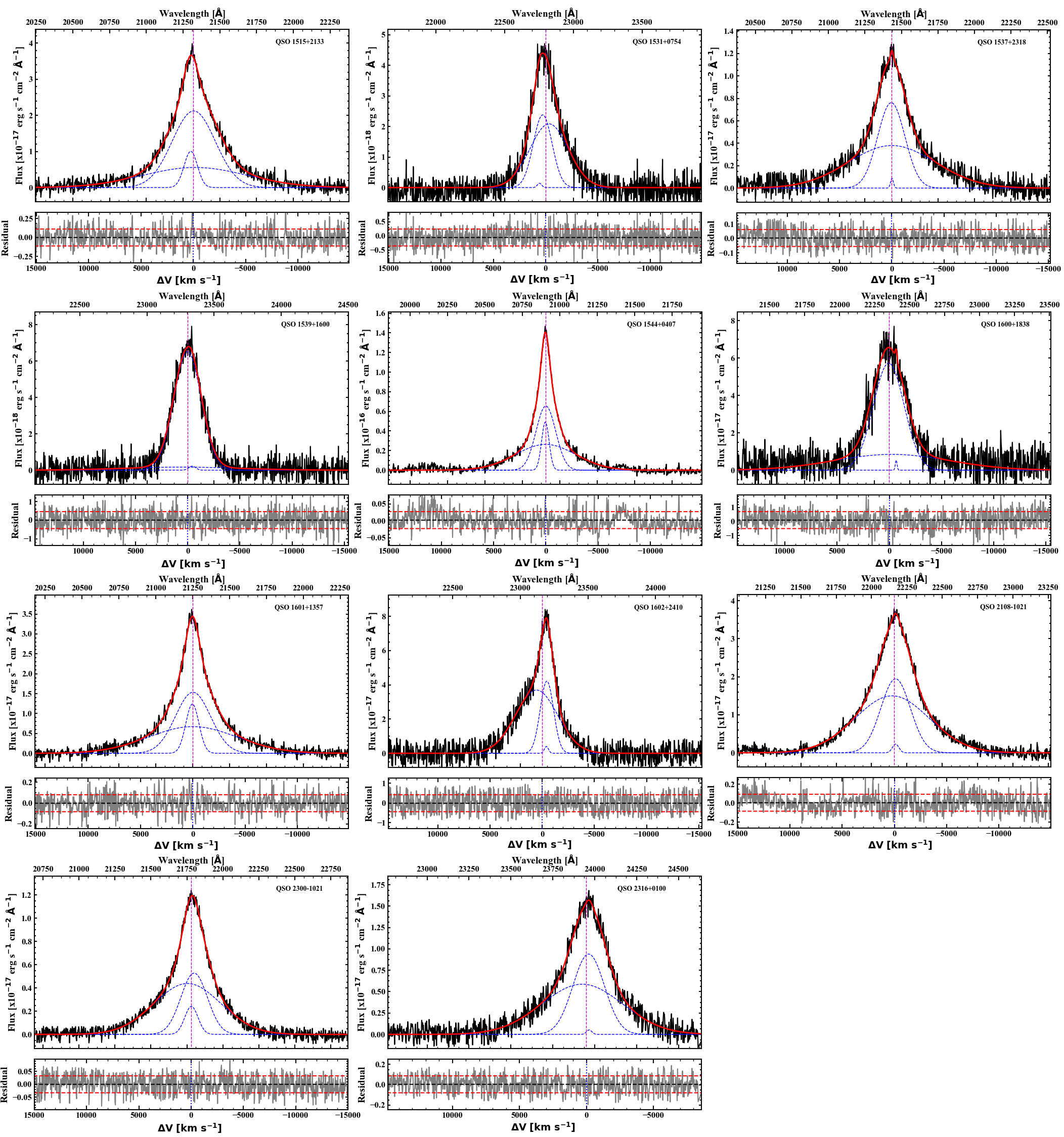}
    \caption{Continuation of Figure~\ref{fig:ha_fit1}.}
   \label{fig:ha_fit3}
\end{figure*}

\begin{figure*}
   \includegraphics[width=1.0\textwidth]{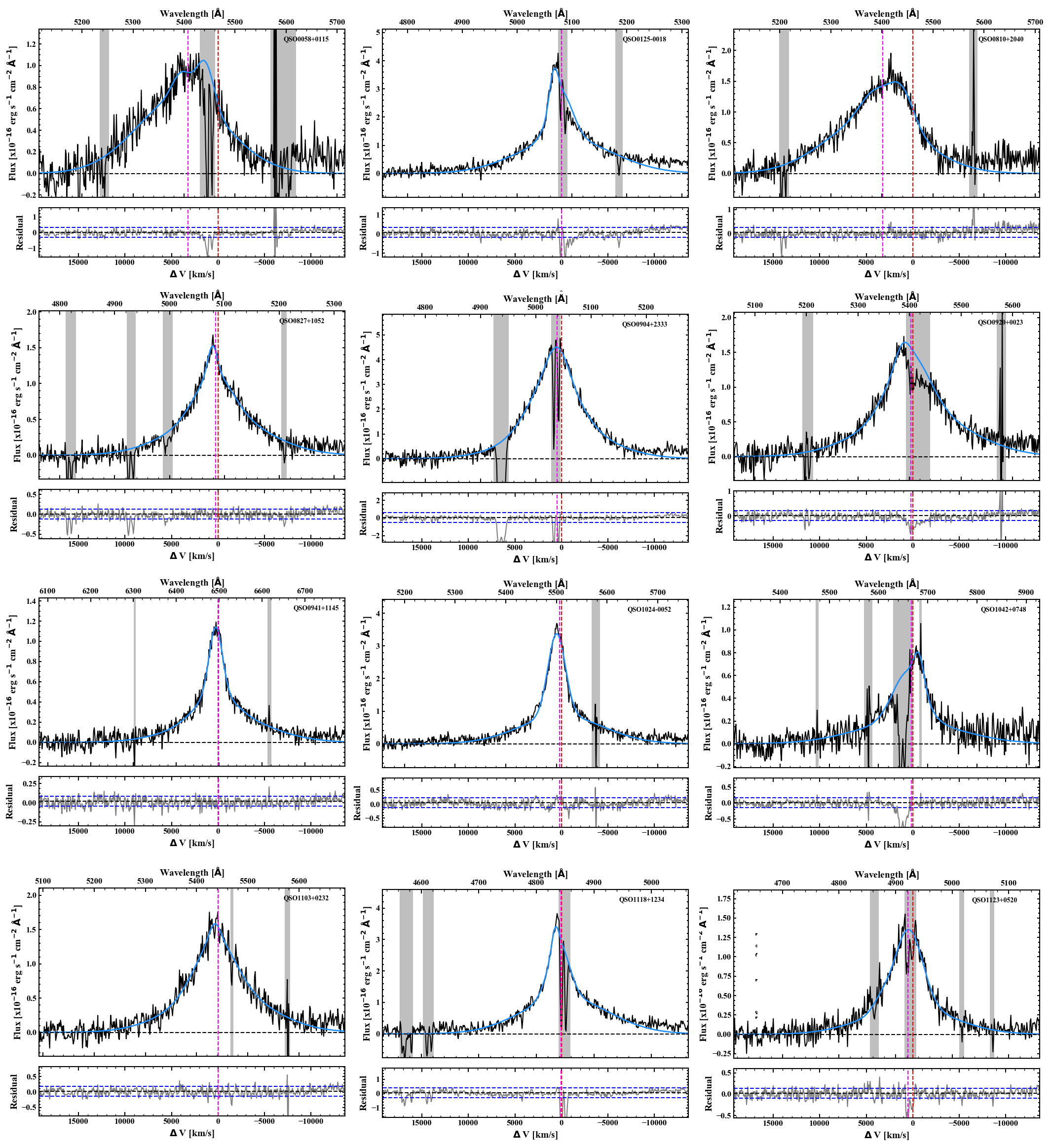}
    \caption{\civ\ fit for the entire CARLA sample. Top panels show the observed spectrum
    in black and the sum of all fitted components in blue. The thick horizontal blue line shows the FWHM of \civ.    
    Bottom panels show the residual from the subtraction of the best fit from the observed spectrum
    in grey. The dashed black line shows the zero-level and the dashed blue lines the standard deviation.
    In each panel, the vertical red dashed line shows the systematic velocity based on \ha, while the magenta dashed line shows the centroid of \civ. Vertical grey shaded regions indicate wavelength intervals that were masked in the fitting process.}
   \label{fig:civ_fit1}
\end{figure*}

\begin{figure*}
   \includegraphics[width=1.0\textwidth]{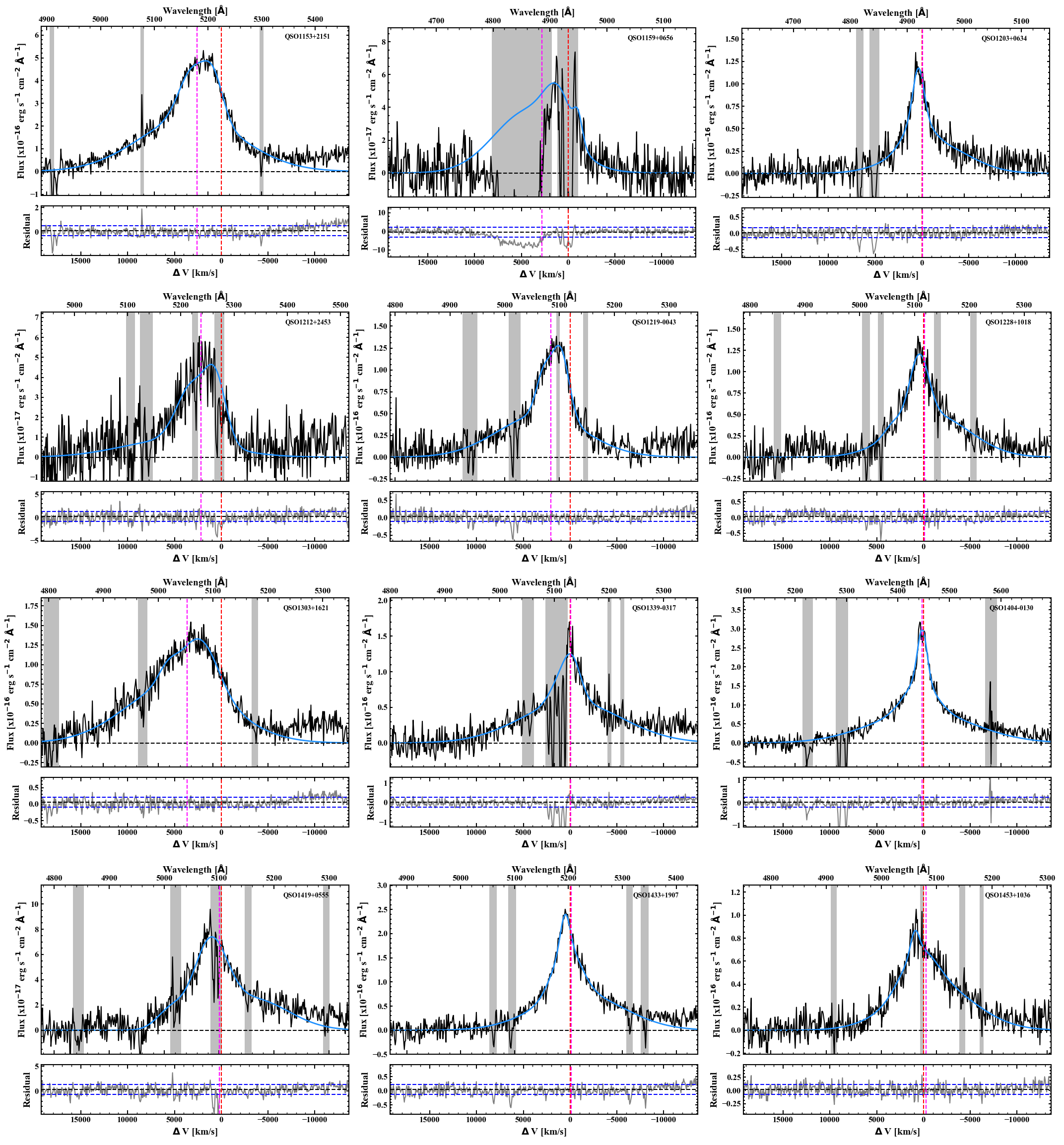}
    \caption{Continuation of Figure~\ref{fig:civ_fit1}.}
   \label{fig:civ_fit2}
\end{figure*}

\begin{figure*}
   \includegraphics[width=1.0\textwidth]{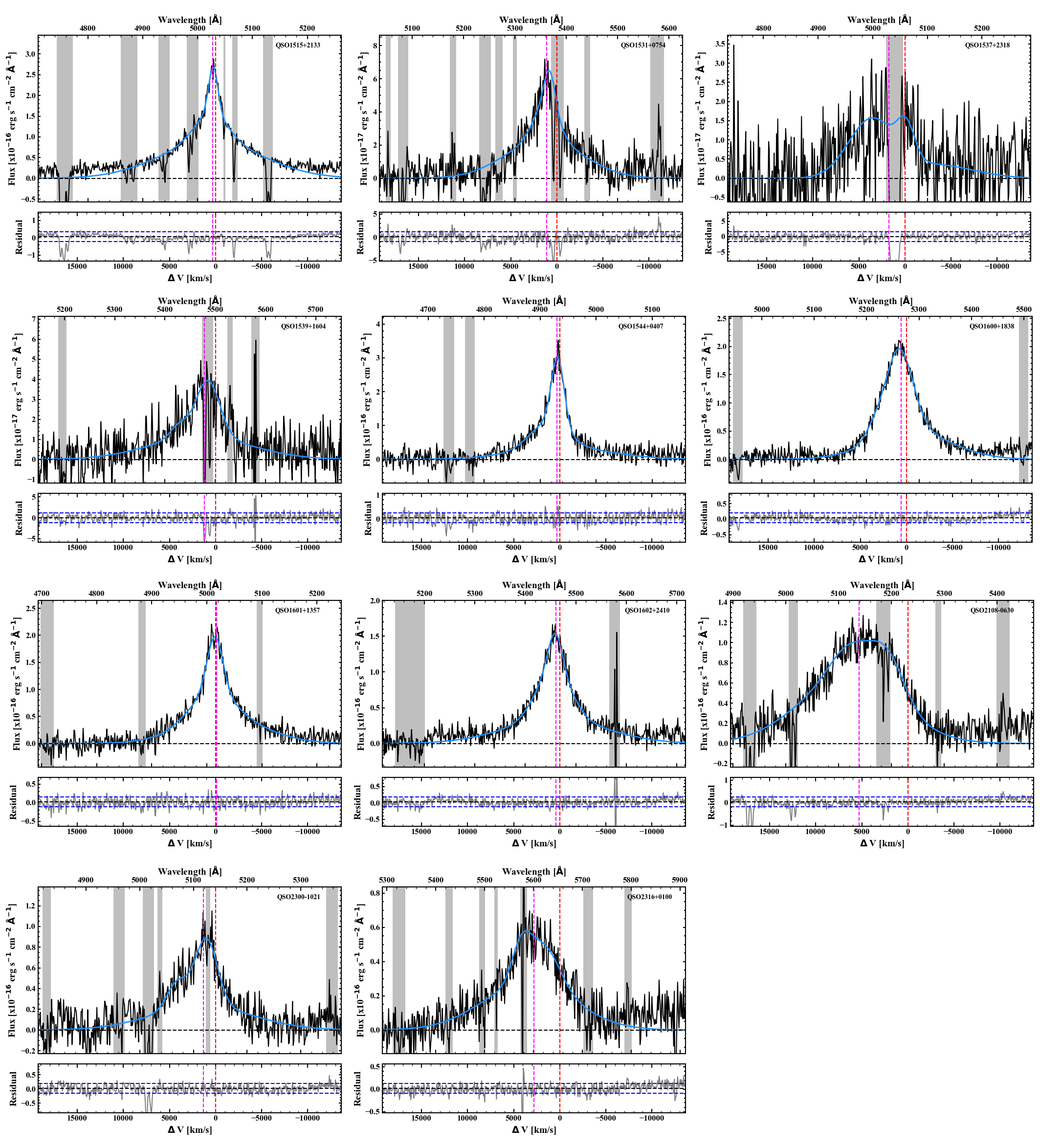}
    \caption{Continuation of Figure~\ref{fig:civ_fit1}.}
   \label{fig:civ_fit3}
\end{figure*}

\bsp	
\label{lastpage}
\end{document}